\documentclass[preprint,12pt]{elsarticle}

\usepackage{amssymb}
\usepackage[linesnumbered,ruled,vlined]{algorithm2e}%

\usepackage{url}
\usepackage[usenames]{color}
\definecolor{linkcolor}{cmyk}{0.98,0.13,0,0.43} % midnight blue

\graphicspath{{figures/}}
\DeclareGraphicsExtensions{.pdf}
\usepackage{comment}

\usepackage{lipsum} % For Tables
\usepackage{tabularx}
\usepackage{lipsum,amsmath,multicol}
\usepackage{multirow}
\usepackage{array,booktabs}
\usepackage{lipsum} % For Tables
\usepackage{tabularx}
\usepackage{adjustbox}
\usepackage{lipsum,amsmath,multicol}
\usepackage{multirow}
\usepackage{array,booktabs}
\usepackage{threeparttable} 
\usepackage{array}
\usepackage{makecell}

\usepackage{pbox}

\usepackage{csquotes}

\usepackage[caption=false,font=footnotesize]{subfig}
\usepackage{float}
\newboolean{printComments}
\setboolean{printComments}{true}

\newcommand{\yasr}[1] {{\ifthenelse {\boolean{printComments}} {\color{red}#1}{}}}
\newcommand{\yasb}[1] {{\ifthenelse {\boolean{printComments}} {\color{blue}#1}{}}}

\usepackage{geometry}
\journal{Journal of Systems and Software}

\begin{document}

\begin{frontmatter}

%% Title, authors and addresses

%% use the tnoteref command within \title for footnotes;
%% use the tnotetext command for theassociated footnote;
%% use the fnref command within \author or \address for footnotes;
%% use the fntext command for theassociated footnote;
%% use the corref command within \author for corresponding author footnotes;
%% use the cortext command for theassociated footnote;
%% use the ead command for the email address,
%% and the form \ead[url] for the home page:
%% \title{Title\tnoteref{label1}}
%% \tnotetext[label1]{}
%% \author{Name\corref{cor1}\fnref{label2}}
%% \ead{email address}
%% \ead[url]{home page}
%% \fntext[label2]{}
%% \cortext[cor1]{}
%% \address{Address\fnref{label3}}
%% \fntext[label3]{}

% Adel Nadjaran Toosi 

\title{The Impact of Distance on Performance and Scalability of Distributed Database Systems in Hybrid Clouds}

\author{Yaser Mansouri and M. Ali Babar}

\address{Centre for Research on Engineering Software Technologies (CREST) Lab.}
\address{School of Computer Science}
\address{The University of Adelaide, Adelaide, Australia}

\begin{abstract}
The increasing need for managing big data has led the emergence of advanced database management systems, e.g., NoSQL or SQL Clusters, which are usually deployed and operated on a cloud computing infrastructure. These database management systems are also called "big data management solutions", whose performance and scalability play a critical role in the operations of "big data applications". Recently, there has been increased efforts aimed at evaluating the performance and scalability of "big data management solutions" hosted by either private or public cloud datacenters. However, there has been little work on evaluating the performance and scalability of "big data management solutions" in hybrid cloud arrangement, where an enterprise leverages a public cloud along with it's own private cloud for it's processing or storage needs. In the hybrid cloud model, the \textit{distance} between private and public cloud datacenters can be one of the key factors that can affect the throughput performance (simply throughput) of "big data management solutions." Hence, any evaluation of the performance and scalability of "big data management solutions" in a hybrid cloud arrangement needs to consider the impact of \textit{distance} between a private cloud and a public cloud. 

In this article, we present a detailed evaluation of throughput performance, scalability, and VMs size vs. VMs number for six modern databases (MongoDB, Cassandra, Riak, CouchDB, Redis, and MySQL Cluster) in a hybrid cloud arrangement, consisting of a private cloud in Adelaide and Azure based public cloud datacenters located in Sydeny, Mumbai and Virginia regions. Based systematic and thorough evaluation, we make the following important observations. First, as the distance between private and public clouds increases, the throughput performance of most databases reduces. Second, MongoDB obtains the best throughput performance, followed by MySQL Cluster under the default setting in terms of replicas number and consistency mechanism.  Whilst Cassandra exposes the most fluctuation in through performance, the throughput of other databases initially reduces and then increases as the number of nodes burst into the public cloud increases. Third, vertical scalability improves the throughput of databases more than the horizontal scalability. Forth, exploiting bigger VMs (i.e., a VM with more cores) rather than more VMs with less cores can increase throughput performance for Cassandra by a factor of at most 800\%, for Riak by 30\%, and for Redis by 10\%.
\end{abstract}


\end{frontmatter}

\section{Introduction}\label{sec1}
For about half century, relational databases have been dominated solutions for storing, retrieving, and managing data \cite{orend2010}\cite{Chang2008}\cite{Leavitt2010}. However, due to essential requirements for high performance\footnote{In this paper, performance and throughput are used exchangeable; otherwise mentioned.} and scalability, NoSQL databases have emerged \cite{Schram2012}\cite{Kuznetsov2014}. The explosive growth in the usage of NoSQL databases has led to several efforts aimed at evaluating their performance and scalability. The performance of these databases depends on their features (data model, data replication strategy, consistency mechanism) as well as the hardware underpinning the computing and storage infrastructures utilized \cite{Floratou2012}\cite{Parker2013}\cite{Cooper2010}. Most of the industrial organizations deploy and operate "big data storage solutions" using cloud computing infrastructure since the emergence of the cloud computing paradigm. 

Cloud computing comes traditionally in three models \cite{Bokhari2018}\cite{Patidar2012}: \textit{public}, \textit{private} and \textit{hybrid}. A public cloud provides computing, storage and networking resources to the general public over Internet, while a private cloud provisions computing and storage resources from an organization''s own cloud infrastructure. A hybrid cloud is a seamless integration of public and private clouds to benefit from the best of both worlds \cite{Rimal2009}. It enables cloud bursting in which applications initially leverage private cloud and then can burst into a public cloud when private cloud's resources are not enough under spikes or increased workloads. A hybrid cloud offers its owner business opportunities in terms of security in compliance with the location of sensitive data, availability, reliability and monetary cost. 

 %MongoDB\footnote{MongoDB: \url{www.mongodb.com}}, Cassandra\footnote{Cassandra: \url{http://cassandra.apache.org/}}, Riak\footnote{Riak: \url{https://riak.com/}}, CouchDB\footnote{CouchDB: \url{https://couchdb.apache.org/}}, Redis\footnote{Redis:\url{https://redis.io/}}, and MySQL\footnote{HBase:\url{https://hbase.apache.org/}} https://www.project-voldemort.com/voldemort/

The performance of NoSQL and Relational databases deployed on public or private cloud datacenters varies as the hardware infrastructure resources and database configuration parameters may change. As indicated in the literature \cite{Klein2015}\cite{Kuhlenkamp2014}\cite{Rabl2012}, the more powerful resources, the higher throughput and less read and write latency are offered by using these databases. The value of these performance parameters is directly proportional to the power of hardware resources. There is a significant amount literature on the evaluation of NoSQL databases deployed on either public or private cloud datacenters. More specifically, Rabl et al. \cite{Rabl2012} evaluated the performance and sclability of four NoSQL databases (Voldermort\footnote{Voldermort: \url{https://www.project-voldemort.com/voldemort/}}, 
HBase \footnote{HBase:\url{https://hbase.apache.org/}}, Cassandra\footnote{Cassandra: \url{http://cassandra.apache.org/}}, and Redis\footnote{Redis: \url{https://redis.io/}}), two relational databases (MySQL\footnote{MySQL: \url{https://www.mysql.com/}} and VoltDB\footnote{VoltDB: \url{https://www.voltdb.com/}}). In this regard, Kuhlenkamp et al. \cite{Kuhlenkamp2014} conducted experiments on two NoSQL databases (i.e., Cassandra and HBase) from scalability and elasticity perspectives. The former work conducted on a private cloud, and the latter on a public cloud. Li et al. \cite{Li2018} investigated the performance of six distributed databases in terms of read, write, delete and instance initiate over a private cloud. Table \ref{tab:relatedwork} summarizes the studies on evaluation of NoSQL and relational databases deployed on cloud computing infrastructure.
%Some authors provide algorithms for  Bog of Task (BoT) applications in order to  optimize the execution time of tasks in constrained time, budget, or both over a simulated or implemented hybrid clouds \cite{Toosi]2018}\cite{Calheiros2012} \cite{Vecchiola2012}\cite{Tuli2020}\cite{zhou2019}. 

All the above-mentioned studies discuss throughput/performance, horizontal and vertical scalability of different NoSQL and relational databases in either private or public clouds, where the distance factor did not come into play. In contrast, \textit{distance} between private and public cloud datacenters in a hybrid cloud can be an important factor that should be considered on the evaluation of distributed databases in terms of performance and scalability. Recently, we evaluated the throughput performance of distributed databases running in hybrid clouds without considering the distance factor \cite{Mansouri2020}. 

Thus, the main research question arises here is \textit{to what extent distance between private and public cloud datacenters can impact on the performance and scalability of distributed  database systems?.} An obvious answer to this question is the smaller distance between public and private cloud datacenters, the smaller is the impact on the databases performance. This solution might not be desirable for all business companies due to differential in their needs, for example, privacy and monetary cost. As an instance, a business company might select a cheaper cloud datacenter in East USA region rather than the one in Sydney region to deploy its database. This makes a compromise between performance and hardware infrastructure cost. Furthermore, a wider selection of cloud datacenters across the world (e.g., 58 Azure regions worldwide\footnote{Azure regions: \url{https://azure.microsoft.com/en-us/global-infrastructure/regions/}}) provides more opportunities for business companies in public cloud datacenter selection in terms of distance to pair with their private cloud datacenter to build a hybrid cloud. Therefore, distance factor comes into play for building a hybrid cloud as an important research question stated above. 
%[???]

To answer the above question, we have exploited an automated hybrid cloud that enabled us to pair a public datacenter in different regions with a private one in a flexible way \cite{Mansouri2020}. This implementation of hybrid cloud also allowed to select VM size, VMs number, desired database installation, database cluster configuration and so on.  The selection of a public datacenter in a specific region depends on how much distance is desirable between private and public cloud datacenters. Since we intended to evaluate the effects of distance on performance of distributed databases, we chose public cloud datacenters in US East and Sydney regions as the closest and the farthest to a private cloud datacenter located in our Lab (Adelaide Uni), respectively. Moreover, we selected another public datacenter between the closest and the farthest datacanter from distance perspective, which led to the choice of a datacenter in West India (Mumbai) region.  

In addition to the above research question, we were also interested in investigating the impact of \textit{VM packing} on the throughput of distributed databases running in a hybrid cloud. VM packing means that we deploy fewer VMs with more cores instead of more VMs with less core so that the total cores in both deployment is an equal number (e.g., 2 VMs with 4 cores each instead of 4 VMs with 2 cores each). This investigation helped us to demonstrate how much latency between VMs in a cloud datacenter can impact the throughput of distributed databases. In other word, the latency between VMs in a single cloud datacenter reduces to the latency between cores in  a single VM, which might be effective means of a hybrid cloud deployment for distributed databases. 

There are several choices of distributed databases that can be evaluated for the impact of distance on their performance as deployed on a hybrid 
cloud \cite{Wu2013}. By thriving NoSQL in 2011 \cite{Lourenco2015}, currently these are more than 225 NoSQL databases\footnote{NoSQL databases: \url{http://nosql-database.org/}}; Among these databases, some are supported in a pre-installed and configured infrastructure component (e.g., MongoDB\footnote{How to install and configure MongoDB on a Linux VM: \url{https://docs.microsoft.com/en-us/azure/virtual-machines/linux/install-mongodb}} and Cassandra\footnote{Run Apache Cassandra on Azure VMs: \url{https://docs.microsoft.com/en-us/azure/architecture/best-practices/cassandra}} databases by the well-known cloud providers). Based on the widespread usage and popularity, we selected six databases to benchmark the impact of distance on their performance, scalability, and VM packing: MongoDB\footnote{MongoDB: \url{www.mongodb.com}}, Cassandra\footnote{Cassandra: \url{http://cassandra.apache.org/}}, Riak\footnote{Riak: \url{https://riak.com/}}, CouchDB\footnote{CouchDB: \url{https://couchdb.apache.org/}}, Redis\footnote{Redis:\url{https://redis.io/}}, and MySQL\footnote{MySQL:\url{https://www.mysql.com/}}. We have investigated the impact of distance between cloud datacenters making a hybrid cloud on the performance and scalability of these databases. In this investigation, we answer the following Research Questions (RQs):
%None of these studies have  evaluated the impact of distance as an imperative factor on the performance of distributed databases in a hybrid cloud. To fill this gap, we answer the following research questions:

\begin{itemize}
\item \textbf{RQ1}: What is the impact of distance  on the performance of widely used NoSQL and SQL databases in hybrid clouds?
\item \textbf{RQ2}: To what extent, NoSQL and SQL databases are scalable in hybrid clouds?
\item \textbf{RQ3}: What is the impact of VM packing on the performance of distributed databases in hybrid clouds?
\end{itemize}

The reminder of this paper is structured as follows. Section 2 reviews the literature with respect to the performance evaluation of NoSQL databases and data-intensive applications on different models of cloud computing. Section 3 discusses the background for the model of hybrid cloud implemented and deployed, as well as distributed databases under test. Section 4 presents the experimental setup, and performance evaluation results, and then discusses our findings. Finally, Section 5 draws some conclusions and identifies the future work. 

\begin{table*}[t]
	
	%\begin{threeparttable}
	\caption{Comparison of the relevant studies with our work}\label{tab:relatedwork}
	\centering
	%\vspace{-3mm}
	\tiny
	\begin{tabular}{p{2.5cm} p{4.5cm}p{1cm}p{1cm}p{1.2cm}p{4.2cm}}
		%\centering
		\hline
		              Paper                       &DBs name       &Cloud model     &Distance impact    &Cloud bursting   &Evaluation metrics\\\hline\hline
		Rabl et al.\cite{Rabl2012}                &Cassandra, Hbase, Redis, Voldermort, MySQL, VoltDB &Private         &No                 &No            & Throughput, horizontal and vertical scalability evaluation\\\hline 
		Kuhlenkamp et al.\cite{Kuhlenkamp2014}    &Cassandra, Habse  &Public  &No &No     &Scalability and elasticity evaluation \\\hline
		Li et al. \cite{Li2013}   &MongoDB, RavenDB, CouchDB, Cassandra, Hypertable, Couchbase, MySQL  &Private    &No &No    & Throughput, read, write, delete latency\\\hline
		Klein et al.\cite{Klein2015}              &MongoDB, Cassandra,Riak    &Public     &No  &No     & Throughput  evaluation for different consistency setting\\\hline
		Abramova and Bernardino\cite{Abramova2013}&MongoDB, Cassandra &Private     &No     &No    & A comparison between Cassandra and MongoDB in performance\\\hline
		Cooper et al. \cite{Cooper2008}           &Cassandra, Hbase, MySQL, PNUTS     & Private        &No    &No  &Throughput, scalability, read and write latency\\\hline
		Veen et al. \cite{vanderVeen2012}         &MongoDB, Cassandra,  PostgreSQL    &Private &No  &No     &Throughput evaluation \\\hline
		Bastiao et al. \cite{Bastiao2014}         &MongoDB, CouchDB, Lucene   &Private    &No  &No    &Retrieve and insert latency\\\hline
		Lourenco et al. \cite{Lourenco2015}      &Cassandra, CouchDB, MongoDB, MS SQL  &Private        &No    &No  &Throughput\\\hline
		Mansouri et al. \cite{Mansouri2020}               &MongoDB, Cassandra, Riak, Redis, Couchdb, MySQL   &Hybrid          &No          &Yes    & Throughput, read and write latency\\\hline      
		\textbf{This work}     &MongoDB, Cassandra, Riak, Redis, Couchdb, MySQL    &Hybrid      &Yes  &Yes    & Throughput, vertical and horizontal scalability, VM packing  \\\hline

	\end{tabular}
	%\begin{tablenotes}
    %  \small
    %  \item  Abbreviations: BoT: Bag of Task; OSN: Online Social Network; DDB: Distributed Databases; NSDB: No-SQL %Databases; NS: Not Specified;\\ Imp: Implementation; Sim: Simulation
   %\end{tablenotes}
	%\end{threeparttable}
	\vspace{-5mm}
\end{table*}

\section{Related Work}\label{sec3} 
We compare our work in this paper with the state-of-the-art studies in two categories: Performance evaluation of distributed databases deployed on cloud computing, and the impact of distance on data-intensive applications performance.

\textit{Performance evaluation of distributed databases on cloud computing:}
	By flourishing NoSQL databases in 2011, several of them including MongoDB, Cassandra, Riak, CouchDB, Riak, Redis, and Hbase are the center of studies \cite{Tudorica2011}\cite{JingHan2011}\cite{Davoudian2018}. As stated in the literature, these databases outperform the relational databases in performance and scalability, making them more popular for use within private or public cloud datacenters.

The performance evaluation of NoSQL databases is supported through using the Yahoo Cloud Serving Benchmark (YCSB) \cite{Cooper2010} that allows to measure throughput and latency of  read, write, insert, update, delete, and scan operations. Initially, Cooper et al. \cite{Cooper2010} used this benchmark to measure the performance of Cassandra, Hbase, MySQL, and PNUTS \cite{Cooper2008}.  Later, since 2012, researchers leveraged this benchmark to make comparison between NoSQL and relational databases deployed on cloud computing from performance and scalability perspectives \cite{Hecht2011}. 

Abramova et al. \cite{Abramova2013} compared  MongoDB and Cassandra in terms of their features and capabilities using YCSB. MongoDB is affected by high workloads, whereas Cassandra seemed to experience performance boosts with the increasing amounts of data. Also, Cassandra outperforms MongoDB in update operations. Veen et al. \cite{vanderVeen2012} made a comparison between MongoDB and Cassandra and concluded also that MongoDB provides high throughput as it is deployed in a single server. Klein et al. \cite{Klein2015} evaluated MongoDB and Cassandra with non-default setting parameters to measure throughput, latency for read and write operations. Rable et al. \cite{Rabl2012} conducted extensive experiments to evaluate the performance and scalability of Cassandra, Hbase, Redis, Voldemort, VoltDB, and MySQL. Kuhlenkamp et al. \cite{Kuhlenkamp2014} carried out a large experiment to measure scalability and elasticity of Cassandra and HBase deployed within a public cloud datacenter. As summarized in Table \ref{tab:relatedwork}, different from our work, all these studies (except in our work \cite{Mansouri2020}) evaluate the performance of distributed databases on either a private or a public cloud datacenter. Recently, we have evaluated throughput performance of distributed databases on a hybrid cloud without considering distance between private and public clouds involved in a hybrid cloud agreement model \cite{Mansouri2020}.

Several studies have used NoSQL databases to evaluate their applicability to different IT domains. Bastiao et al. \cite{Bastiao2014} leveraged MongoDB and CouchDB in the healthcare domain. They observed that there is no difference between these two databases in performance, and concluded NoSQL databases still should be improved. Lourenco et al. \cite{Lourenco2015a} evaluated Cassandra, CouchDB and MongoDB for a write-intensive application. The results revealed that Cassandra is better than the other NoSQL databases for a four-node setup, while a MS SQL Server running on a single node outperformed all NoSQL contenders for these specific settings. Rith et al. \cite{Rith2014} implemented a layer to translate SQL queries to NoSQL ones for MongoDB and Cassandra, moving from relational to non-relational databases. None of these studies has investigated the impact of distance between cloud datacenters making up a hybrid cloud for the performance, scalability and VM packing (i.e., VMs number vs. VMs flavour) of distributed databases.

Some researchers have evaluated their proposed algorithms, policies and methods through simulation and implementation for a hybrid cloud. Toosi et al.  \cite{Toosi]2018} has recently configured a hybrid cloud including Microsoft Azure and two PC workers to analyze the proposed resource provisioning algorithms. Using the same configuration for the hybrid cloud with different VM size though, Tuli et al. \cite{Tuli2020} evaluated  several resource provisioning and task scheduling algorithms. Calheiros et al. \cite{Calheiros2012} and Vecchiola et al. \cite{Vecchiola2012} provided almost similar setup to evaluate an algorithm leveraging dynamic resources to meet the deadline constraint for Bag-of-Tasks (BoTs). Li et al. \cite{Li2018} designed a cost-aware job scheduling approach based on the queuing theory in hybrid clouds. Differently, Loreti et al. \cite{Loreti2015} implemented a software layer on top of a hybrid cloud infrastructure to dynamically deploy and scale virtual clusters. Zhou et al. \cite{zhou2019} proposed an approach to optimize the monetary cost of workflow scheduling with constrained time. They extended this approach to minimize the execution time of tasks within constrained time and budget. Our work is different with these studies as they have implemented scheduling algorithms to complete tasks within a constrained time and budget for BoT applications in a simulated or implemented hybrid cloud environment. 
    
\textit{Impact of distance on data-intensive applications performance:}
User-perceived latency for database operations (e.g., read, write, delete, update or create, etc.) is a vital criterion at the database level. Latency for these operations can be impacted from network congestion, overloaded computing infrastructure, and the distance between the location of requests issued from, and the location of data stored. How to reduce network latency was well-studied in literature, and there are several simple solutions. They are varying from data replication \cite{Vulimiri2013}\cite{Nishtala2013}, requests iteration \cite{Wu2015}\cite{Dean2013} to more powerful hardware resources.  Obviously, these solutions are more effective if a public cloud datacenter with a proper distance is selected to be paired with the private one as experimentally observed in a study \cite{Wu2013}. All studies stated above did not investigate the impact of distance between cloud datacenters involving in the hybrid cloud on the performance and horizontal/vertical scalability of distributed databases. This investigation is our main contribution in this study.

%% \linenumber

\section{Distributed database systems evaluated}\label{sec2}
In this section, we  briefly provide an overview of the distributed databases under evaluation in this work, and then discuss how a hybrid cloud has been implemented to evaluate the performance and scalability of these databases.

\subsection{Distributed databases under evaluation}
In this section, we discuss five NoSQL and one relational databases. The criterion for the selection of these database for our study is their widespread adoption in the industry. In the following, we give more details for these databases to better understand the experimental results.

%The default setting of these databases can be found in [], and 
\textbf{MongoDB} is an open-source document-based database and supports horizontal scalability and automatic sharding \cite{Haughian2016}. It also provides full replication and the asynchronous master-slave model for consistency. This implies that writes are only made by the master node and reads can be conducted from both a master node and from one of the slave nodes. Writes are propagated to the slave nodes by reading from the master’s operation log \cite{Haughian2016}. Mongo DB offers different types of consistency models for clients by specifying whether reads are made from secondary nodes and how many nodes must confirm a read operation.

\textbf{Cassandra} is an open-source database based on the ideas behind Google BigTable \cite{Chang2006} and Amazon Dynamo \cite{Sivasubramanian2012}. It uses column-based data model in which each column consists of the name, value and timestamp; All of which are provided by a client. Consistency is highly tunable according to the requirements of the applications, making a trade-off between latency and consistency. Cassandra operates in a master-master mode \cite{Gudivada2014}, which makes easy for horizontal scalability to support \cite{Haughian2016}. This mode of operation implies no node is different from another leading to high write throughput operations \cite{Gudivada2014}\cite{Haughian2016} with the help of combining disk-persistence with in-memory caching of data. It also supports several partitioning and replication techniques.

\textbf{Riak} is  open-source and document-based NoSQL database. It supports master-less replication architecture without a single point of failure. Riak allows applications to define how many nodes are required to confirm read and write operations. This feature makes a trade-off between availability and consistency. It supports the options to choose between eventual (default option) and strong consistency for each data bucket.

\textbf{CouchDB} is an open-source database offering a document-oriented approach \cite{Kuznetsov2014} in JSON format. It provides ACID properties on the document level and no lock for read operations through  Multi-Version Concurrency Control (MVCC). CouchDB supports both master-master and master-slave replication between different CouchDB instances or on a single instance. It does not support sharding, while provides scaling by asynchronous data replication \cite{Kuznetsov2014}. CouchDB offers eventual consistency and performs conflict resolution through the most updated data. CouchDB works well if it can store the whole dataset in a RAM of cluster since it is essentially a RAM-based database.

\textbf{Reids} is an open source and in-memory data structure including strings, hashes, lists, sets, sorted sets.  It also uses master-slave asynchronous replication, where data can be replicated to multiple replica servers. This improves read performance (as requests can be split among the servers) and faster recovery when the primary server goes down. Redis offers a highly available in-memory cache to decrease data access latency, increase throughput and ease the load off NoSQL and relational databases.

\textbf{MySQL Cluster} provides shared-nothing clustering and auto-sharding for the MySQL database management system. It internally deploys synchronous replication through a two-phase commit mechanism in order to ensure that data is written to multiple nodes upon committing the data. MySQL Cluster automatically creates \textit{data node groups} from the number of replicas and data nodes specified by a user. Writes are synchronously replicated between nodes in a group to guarantee durability. However, it replicates data asynchronously between clusters to reduce the effects of network latency by locating data physically closer to a set of users.

\subsection{Hybrid Cloud implementation}
\begin{figure*}[h!]
	\centering
	%\subfloat[Read-intensive]{\label{figur:hyb-arc}\includegraphics[width=0.65\textwidth]{Fig/hybrid-arch.pdf}}
     \includegraphics[width=0.65\textwidth]{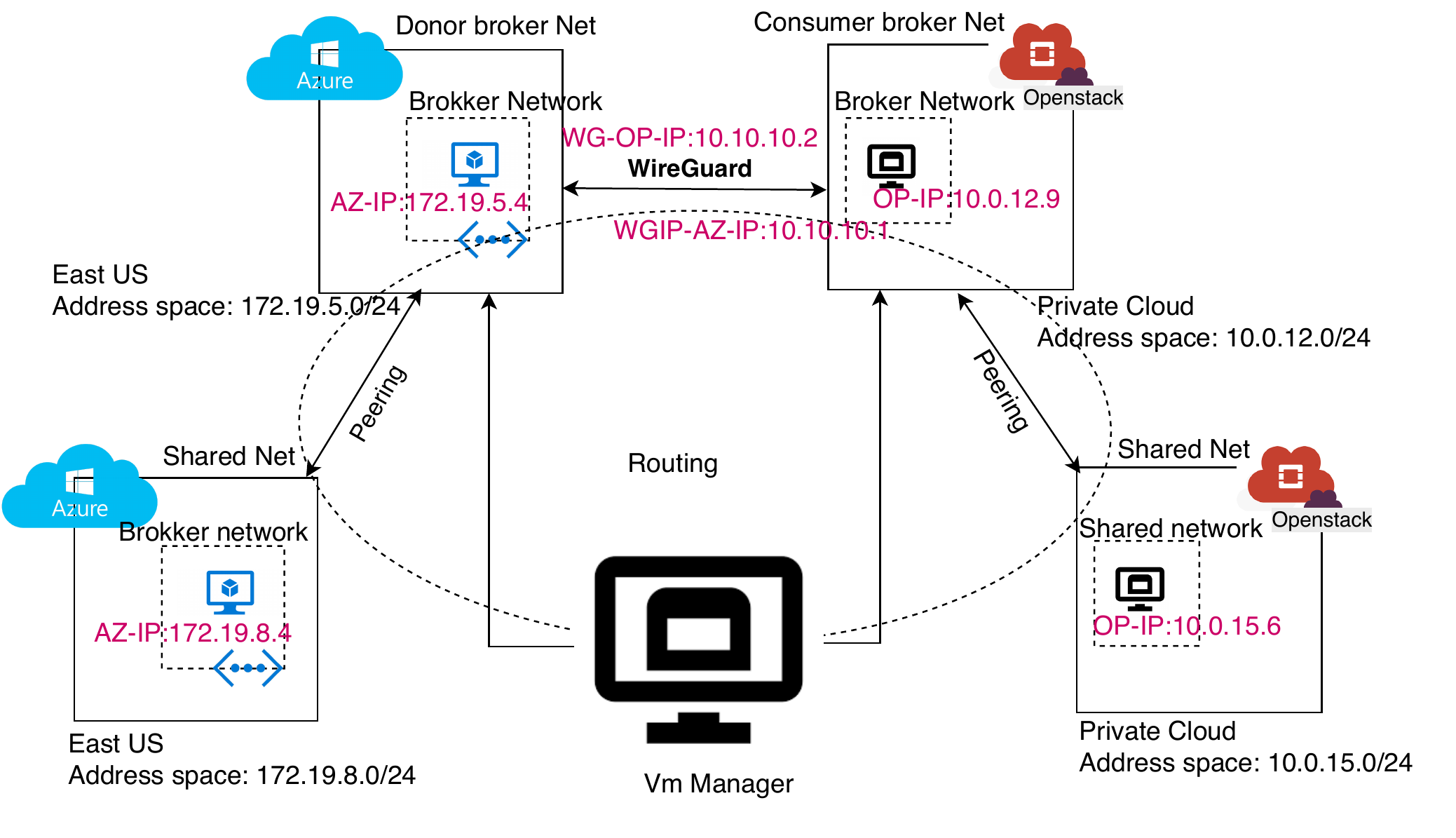}
	\caption{Hybrid cloud architecture spans over on-premises infrastructure resources and the public cloud datacenter in East US (Virginia) region by using WireGuard \cite{Mansouri2020}}
	\label{fig:hyb-arc}
\end{figure*}

We implemented an automated hybrid cloud across OpenStack\footnote{OpenStack: \url{https://www.openstack.org/}} and Microsoft Azure \footnote{Microsoft Azure \url{https://azure.microsoft.com/en-au/}}. In this implementation, we leveraged on-demand usage model (also called cloud bursting), where a data-intensive application running on a private cloud datacenter borrows resources (e.g., a VM instance) from a public cloud datacenter. One  of the main aspects of this implementation is a secure, robust, cost-free connection between private and public cloud datacenters. For this purpose, we used WireGuard\footnote{WireGuard: \url{https://www.wireguard.com/}} as a Linux kernel-based VPN tool recently released Version 1.0 as part of kernel 5.6 \footnote{WireGUard Version 1.0: \newline \url{https://www.archyde.com/wireguard-vpn-1-0-0-appears-in-linux-5-6-kernel-computer-news/}}. Using Wireguard rather than a public cloud VPN brings advantage in terms of security, reliability in connection, throughput, monetary cost and inter-portability \cite{Mansouri2020}. A schematic view of the hybrid cloud using WireGuard is illustrated in Fig. \ref{fig:hyb-arc}. 

As can be seen in Fig. \ref{fig:hyb-arc}, we initially implemented a client/consumer broker in OpenStack and exploited the on-demand model \cite{Mansouri2020} in which it might be required to expand the workload on Azure VM instances in the case of spiking workloads. Thus, we have implemented a server/donor broker in Microsoft Azure. Based on the environment specification, we need to create shared networks/sub-networks that can be connected to and disconnected from broker networks. Similarly, we might need shared networks in different regions in Microsoft Azure side. These shared networks should be able to connect/disconnect to/from  the broker network in Microsoft Azure side. We deployed our cluster nodes hosting distributed databases in the shared subnetworks. 

We used Terraform \footnote{Terraform: \url{https://www.terraform.io/}} as an open-source automation tool for provisioning and managing cloud infrastructure in an automated manner. This tool enabled us to define and execute the required infrastructure resources across OpenStack and Azure cloud datacenters in terms of quantity (e.g., VMs number) and specifications (e.g., VMs size). By using this tool, we installed six distributed databases and made cluster configuration between nodes hosting databases. Such automation of hybrid cloud implementation allows us to consistently re-produce experimental setup to evaluate distributed databases with minimal human interference. 

\section{Evaluation}\label{sec4}
We evaluated performance, horizontal scalability, vertical scalability and VM packing for six distributed databases deployed on a hybrid cloud that spans over a private cloud datacenter and  public cloud datacenters in Sydney, Mumbai, and Virginia regions\footnote{In this study, we reported only throughput for evaluation of performance, horizontal scalability, vertical scalability, and VM packing. We eliminated the latency results because the distance between private cloud and public cloud in three regions reflects latency. That is, the longer the distance between two clouds, the higher the latency for operations performed at databases.}. The aim of this evaluation is to investigate the impact of distance between cloud datacenters involved in a hybrid cloud on the performance of the most used and modern databases. In the following, we discuss specification and location of infrastructure resources, workload setup, experimental scenarios and results.

\subsection{Experiment Setup}

\textbf{Infrastructure resources specification:} We leveraged two clusters for our experiments as depicted in Fig. \ref{fig:hyb-arc}. The OpenStack cluster consists of Linux VMs, each equipped with 1 core CPU, 2 GB of RAM, and 10 GB disk. We also set up a cluster on Microsoft Azure consisting of Linux Standard\_B1ms with 1 vCPU, 2GB of RAM, and 4 GB SSD storage. The number of instances in both clusters is 8, where \textit{n$\ge$1} nodes/VMs in the private cloud/OpenStack and \textit{8-n} nodes in Azure cloud. Note that we consider at least one node running on OpenStack in all experiments to enforce compliance with the definition of hybrid clouds. A summary of deployed infrastructure is listed in Table \ref{tab:infras}.

\textbf{Infrastructure resources location:} The location of our private cloud infrastructure hosting OpenStack was CREST Lab in Adelaide University, and the locations of the used public cloud datacenters are Australia East (Sydney), India West (Mumbai) and US East (Virginia) regions. The reason behind the selection of these regions is  to evaluate the impact of \textit{distance} on the performance, scalability and VM packing of distributed databases. Thus,  we selected Sydney and Virginia as the closest and the furthest locations to our private infrastructure at a distance of 1374 KMs and 16671 Kms, respectively. Moreover, we selected one point in the middle of the closest and the furthest regions (i.e., (1374+16671)/2 =9022 KMs), which  led to the selection of the cloud datacenter in West India (Mumbai) region.   

\textbf{Benchmarking and system under test:} We used Yahoo Cloud Serving Benchmark (YCSB) that includes client component running a set of queries as core workload \cite{Cooper2010}. YCSB runs six different workloads as described in Table \ref{tab:coreworkload}; All workloads have a uniform requests distribution. For each experiment, we built cleaned VM instances and ran a YCSB load phase that inserts 10 K records in each cluster node. Each record consists of 10 fields with a length of 8 bytes. Thus, a record in the workload is 80 bytes. In all experiments, we used at least one instance in the shared network on OpenStack with the default number of threads (10 threads). 

The system under test consists of 6 distributed databases installed and configured on both private and public nodes as a single cluster based on the default settings \cite{Mansouri2020}. Cluster configuration for MySQL server is  different to NoSQL databases since it  requires three types of node. We ran \textit{manager} and \textit{mysql} server nodes on the same VM instance in OpenStack, and data nodes across the hybrid cloud. While NoSQL databases are configured in a master-slave model as needed; otherwise, all nodes were identical. 
As previously mentioned, we used Terraform to automate the deployment, destruction, installation and configuration of database cluster nodes with the least interference of human. This implementation allowed us to consistently and repeatedly evaluate distributed databases based on the desired configuration setting parameters in terms of hardware specification, cluster configuration and so on \cite{Mansouri2020}.

\textbf{Experimental Scenarios:}
We defined four different scenarios associated with RQs in the introduction section to evaluate the performance of 6 databases with different workloads as summarized in Table \ref{tab:expsce}. \textbf{RQ 1:} To evaluate the performance in terms of the number of operations per time unit (i.e., throughput), we considered all permutation of nodes that can be burst into the public cloud datacenter. Thus, we used the hybrid cluster configurations of (8\_0), (7\_1), …, (2\_6), and (1\_7), where the first and second element of each pair represents the number of nodes in the private and public cloud datacenter respectively. 
\textbf{RQ 2.1:} To assess the horizontal scalability of these databases across the hybrid cloud, we fixed one node in the private cloud and vary the number of nodes from 2 to 8 with a step of two in a public cloud\footnote{It is worth mentioning that this setup can be opposite, namely fixing one node in a public cloud and varying the number of nodes in a private cloud. We did not investigate the effect of the opposite configuration setup on the performance of databases since we intended to explore the impact of distance between two cloud datacenters, which were the same from a private cloud to a public cloud, and vice versa. }. 
\textbf{RQ 2.2:} To evaluate the vertical scalability of databases across the hybrid cloud, we considered two nodes in the private cloud and one node in the public cloud with a range of 2, 4, and 8 cores\footnote{Note that Azure cloud datacenters provide Bs-series VMs with 1, 2, 4, 8, 12,16, and 20 cores. We selected VMs with 2, 4, 8 cores to evaluated databases with respect to vertical scalability.  \url{https://azure.microsoft.com/en-us/pricing/details/virtual-machines/linux/\#a-series.}}. Note that we considered 3 nodes in this scenario because MySQL and Redis require at least three nodes for cluster configuration. 
\textbf{RQ 3:} In this scenario, we considered one node in the private cloud, and three sets of nodes with different cores number in the public cloud: 4 nodes with 2 cores (4X2), 2 nodes with 4 cores (2X4), and 1 node with 8 cores (1X8). The aim of this scenario is to evaluate the performance of databases under two cases: fewer VMs  with more cores for each and more VMs with less cores for each so that the overall cores in each case is the same (here the overall cores is 8).

\begin{table}[t]
	% \begin{threeparttable}
	\caption{A summary of infrastructure setup}\label{tab:infras}
	\centering
	\tiny
	%\vspace{-3mm}
	\begin{tabular}{p{2.7cm} p{5.5cm} p{5.5cm}}
		%\centering
		\hline
		System Setup      &Private Infrastructure (OpenStack)   &Public Infrastructure (Azure)    \\\hline\hline
		Instance type     & m1.small  & Standard\_1Bms  \\
		CPU               & 1 Core    & 1 Core  \\
		RAM               & 2 GB      & 2 GB    \\
		Disk              &10 GB      & 4GB SSD\\
		Location          &Adelaide   & Sydney, Mumbai, Virginia\\\hline
	\end{tabular}
	%\end{threeparttable}
	\vspace{-5mm}
\end{table}

\begin{table}[t]
	% \begin{threeparttable}
	\caption{Core workload in YCSB}\label{tab:coreworkload}
	\centering
	\tiny
	%\vspace{-3mm}
	\begin{tabular}{p{2.7cm} p{4.5cm} p{5.5cm}}
		%\centering
		\hline
		Workload type &         Operations &Label    \\\hline\hline
		Workload A     & 50\% Read + 50\% Update & Read-intensive  \\
		Workload B     & 95\% Read + 5\% Update  & Write-intensive  \\
		Workload C     & 100\% Read              & Read-only    \\\hline
		Workload D     &95\% Read + 5\% Insert   & Read-latest\\
		Workload E     &95\% Scan + 5\% Insert   & Scan\\
		Workload F     &50\% Read + 50\% update        &Read-Modify-Write(RMW) \\\hline
	\end{tabular}
	%\end{threeparttable}
	\vspace{-5mm}
\end{table}

\begin{table}[t]
	% \begin{threeparttable}
	\caption{A summary of experimental scenarios associated to RQs}\label{tab:expsce}
	\centering
	\tiny
	%\vspace{-3mm}
	\begin{tabular}{p{2.5cm} p{4cm} p{7.5cm}}
		%\centering
		\hline
		Research Question      &Description                          &Hardware setting    \\\hline\hline
		RQ1    &Throughput evaluation                & Up to 8 VMs with one core across the hybrid cloud          \\
		RQ2.1    &Horizontal scalability               & one VM in OpenStack and up to 8 VMs with one core in Azure  \\
		RQ2.2    & Vertical scalability                &2 VMs in OpenStack and one node with 2, 4, and 8 cores in Azure     \\
		RQ4    &VMs number vs. VMs size                & one VM in OpenStack, and 4X2, 2X4, and 1X8 VMs in Azure\\\hline
   \end{tabular}
	\vspace{-5mm}
\end{table}

%\subsection{Performance Metrics}

\subsection{Experiment Results}
In this section, we report results for research questions indicated in Introduction Section.

\subsubsection{Performance Evaluation}

\begin{figure*}[h!]
	\centering
	\subfloat[Read-intensive]{\label{figur:mongo-perf-a}\includegraphics[width=0.33\textwidth]{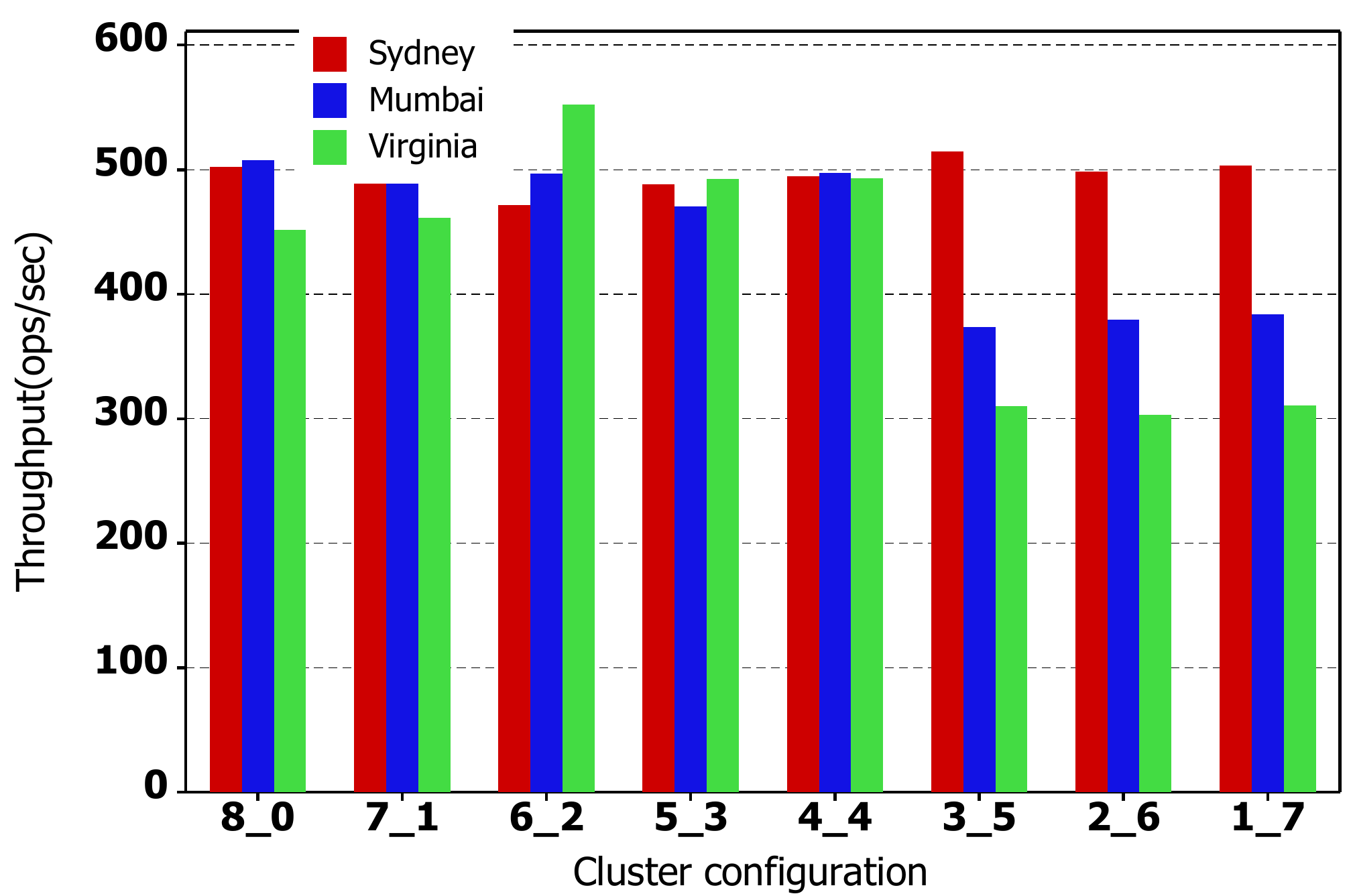}}
	\subfloat[Write-intensive]{\label{figur:mongo-perf-b}\includegraphics[width=0.33\textwidth]{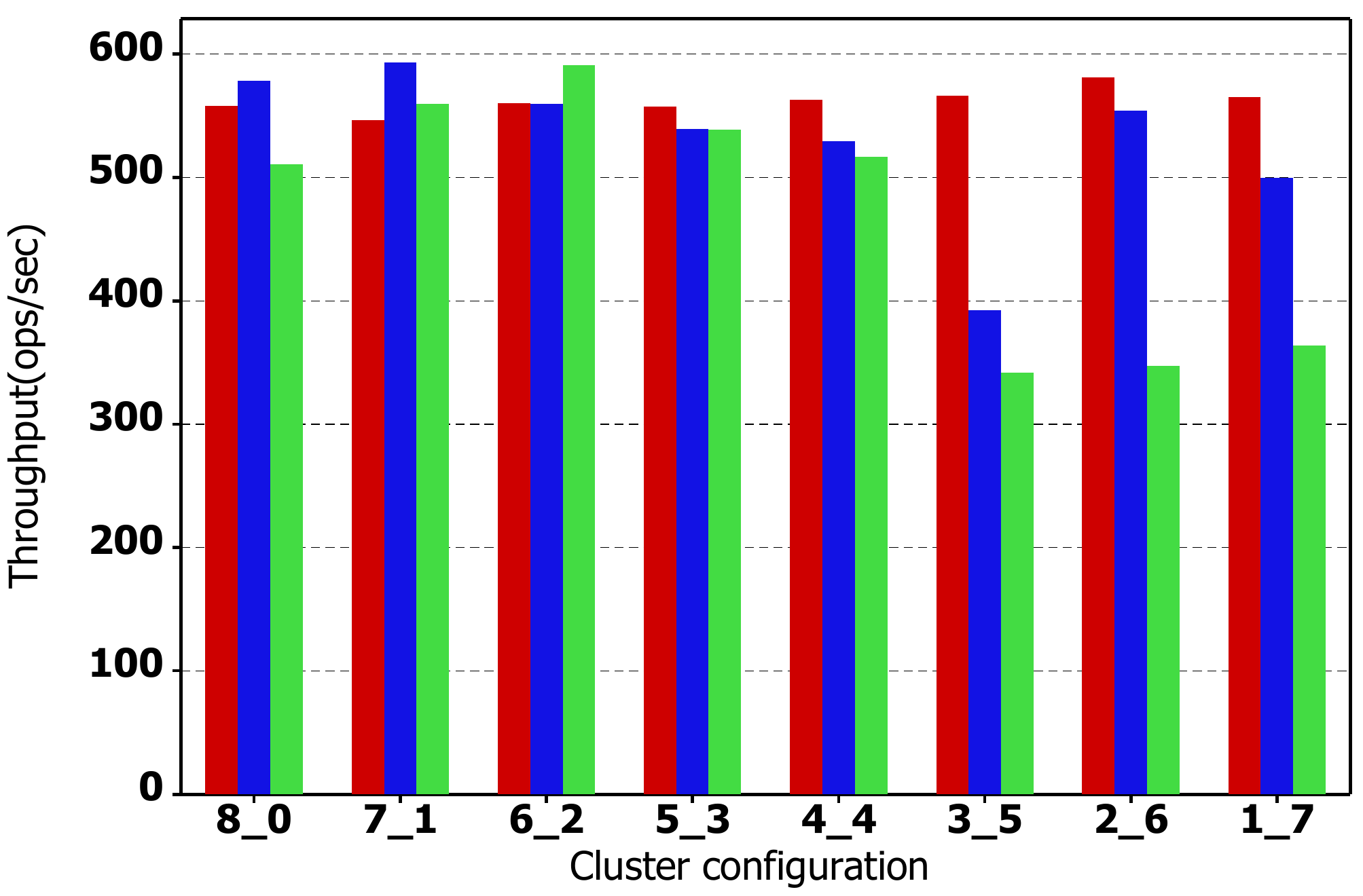}}
	\subfloat[Read-only]{\label{figur:mongo-perf-c}\includegraphics[width=0.33\textwidth]{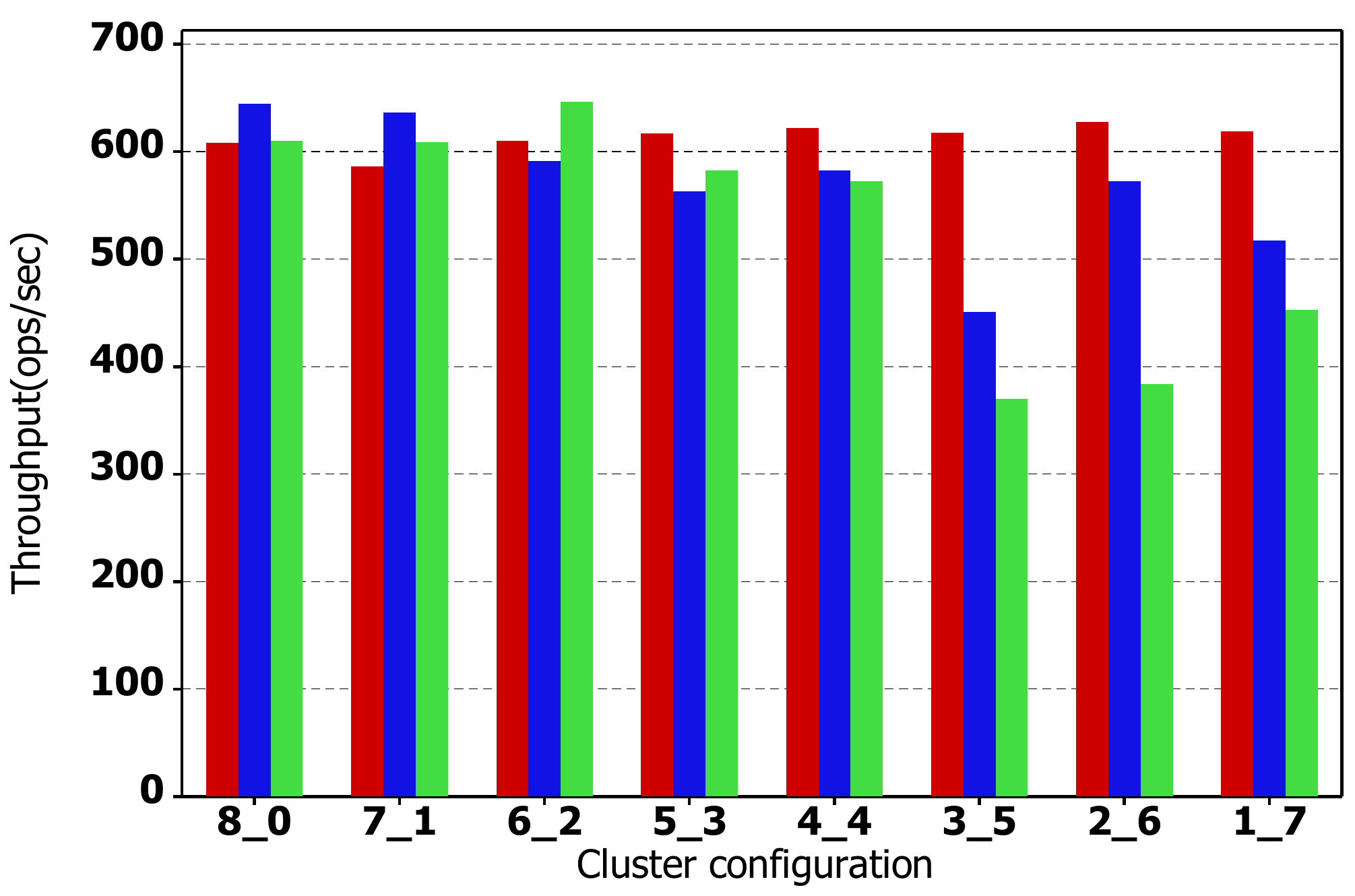}}\\
	\subfloat[Read-latest]{\label{figur:mongo-perf-d}\includegraphics[width=0.33\textwidth]{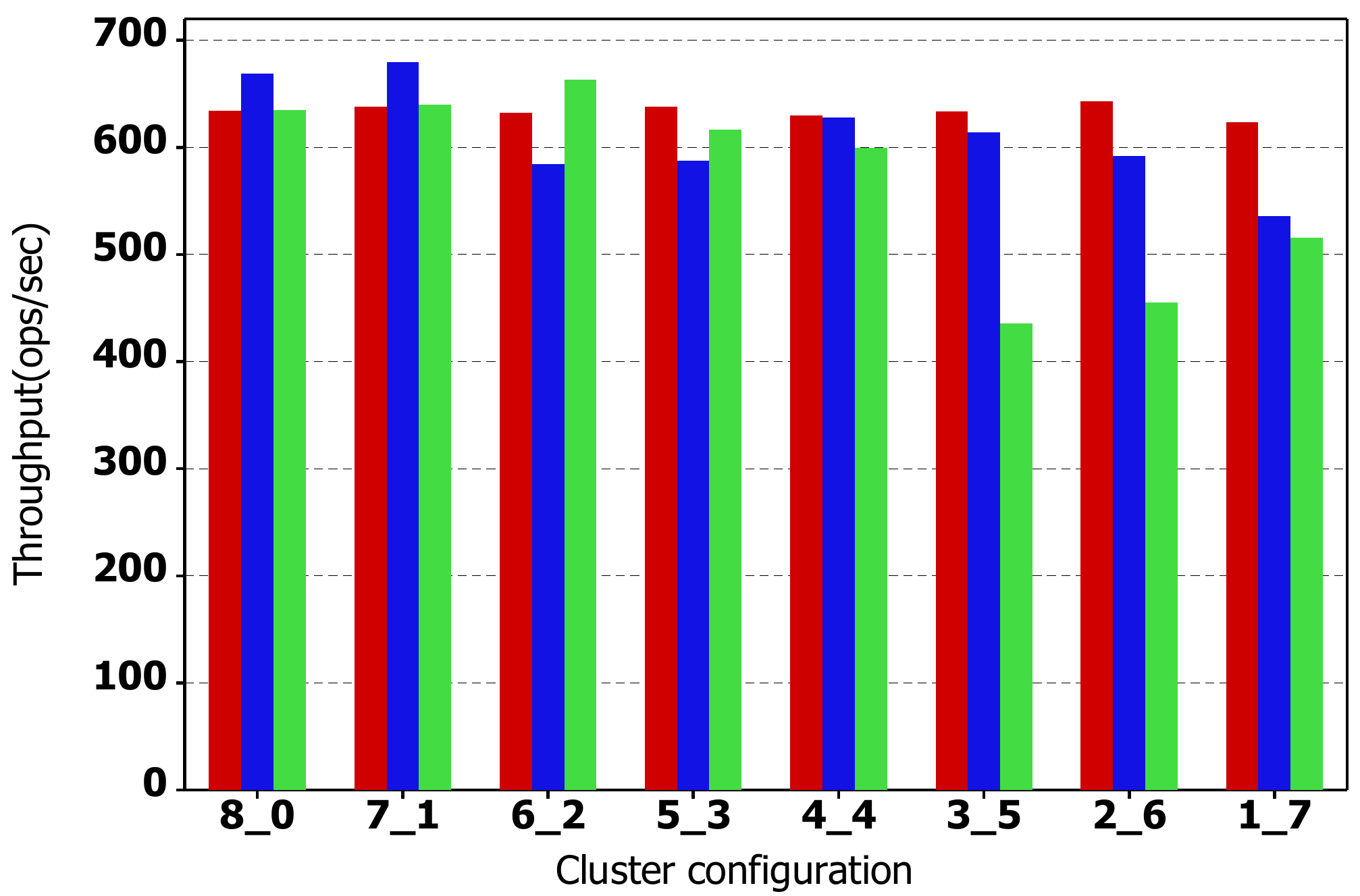}}
	\subfloat[Scan]{\label{figur:mongo-perf-e}\includegraphics[width=0.33\textwidth]{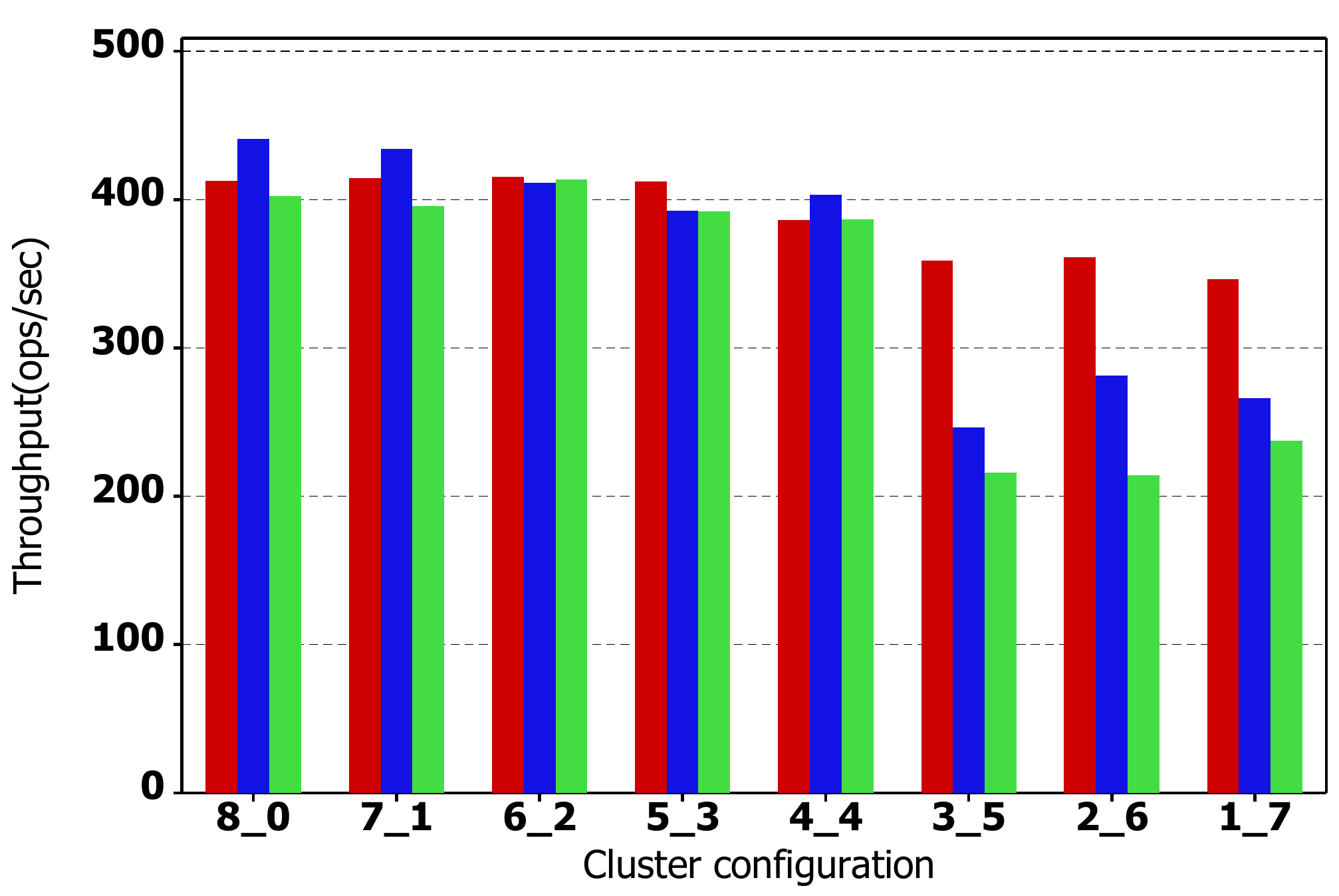}}
	\subfloat[Read-Modify-Write(RMW)]{\label{figur:mongo-perf-f}\includegraphics[width=0.33\textwidth]{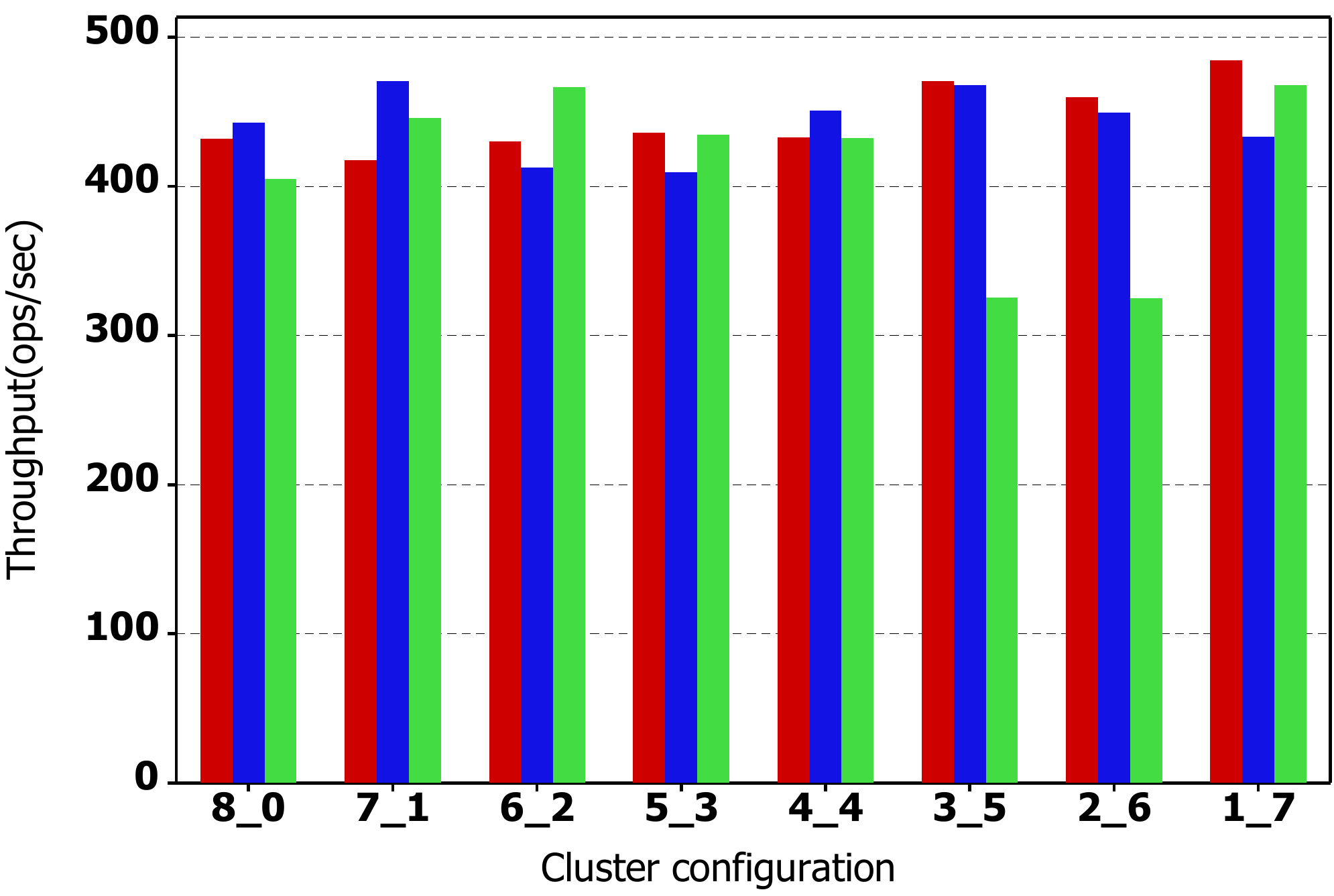}}
	\caption{Throughput for \textbf{MongDB} in Sydney, Mumbai and Virginia regions. Value $n\_m$  in axis X represents that the hybrid cloud consists of  $n$ nodes in the private cloud and $m$ nodes in the public cloud.}
	\label{fig:mong-perf}
\end{figure*}

\begin{figure*}[h!]
	\centering
	\subfloat[Read-intensive]{\label{figur:cass-perf-a}\includegraphics[width=0.33\textwidth]{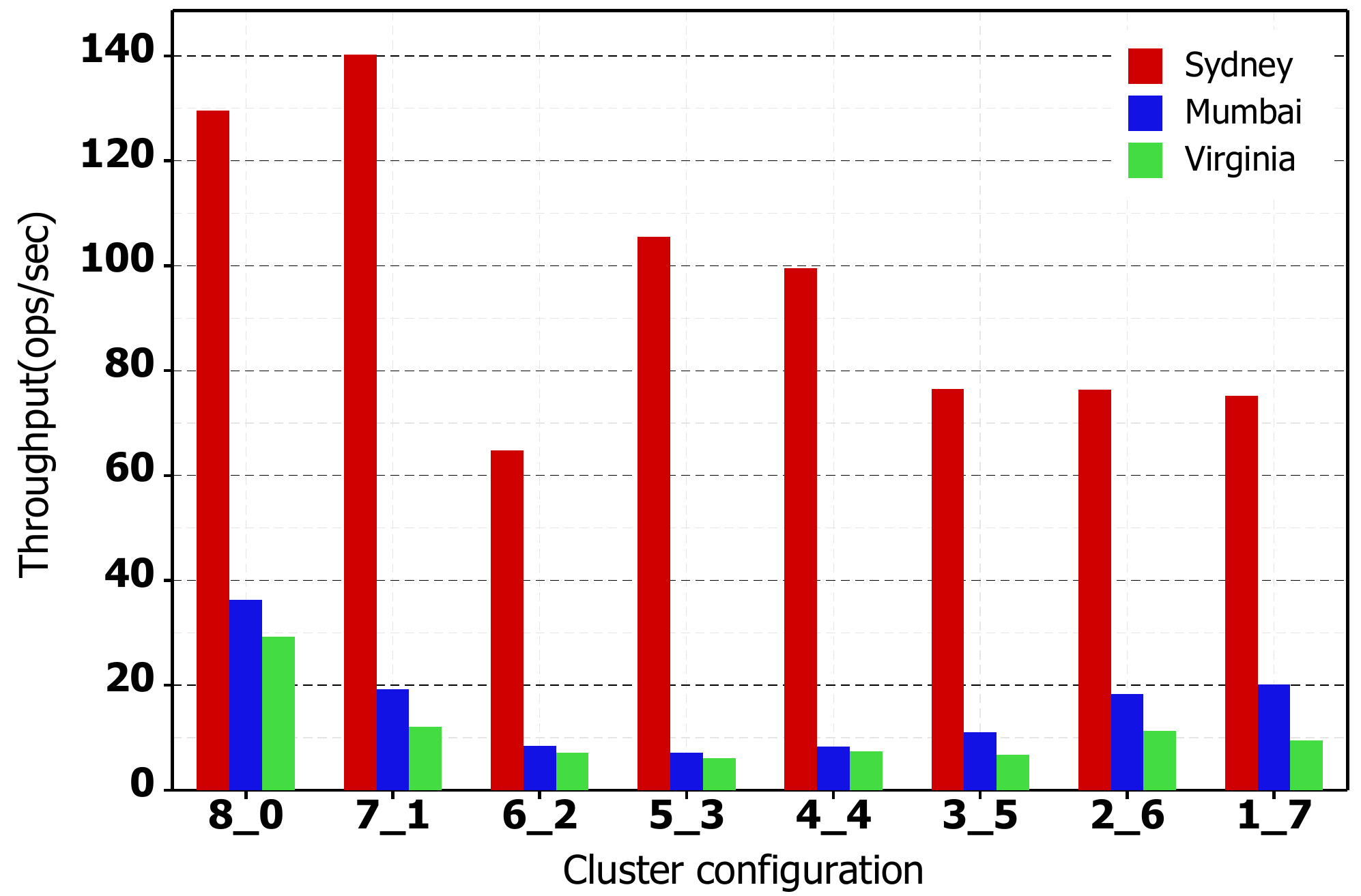}}
	\subfloat[Write-intensive]{\label{figur:cass-perf-b}\includegraphics[width=0.33\textwidth]{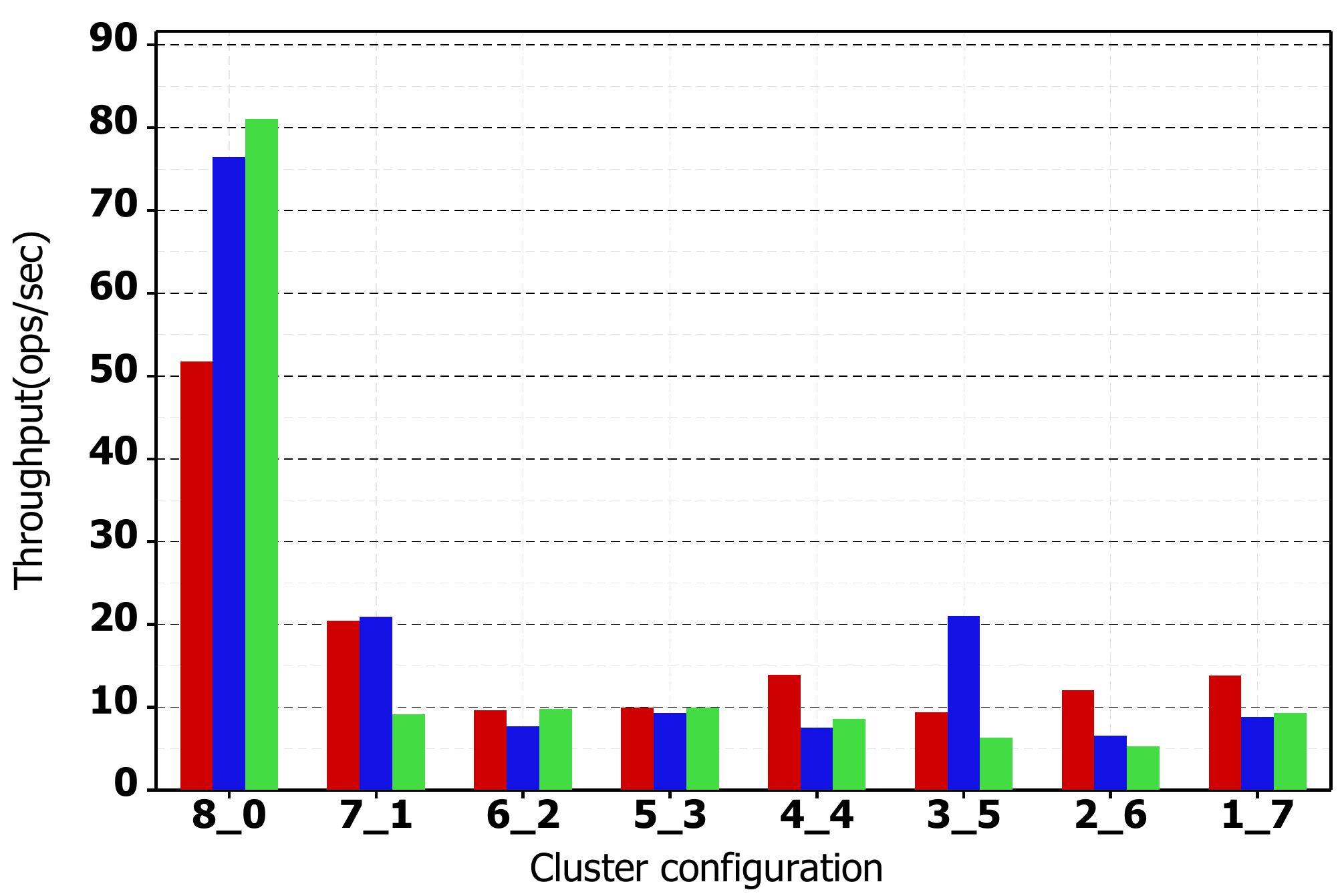}}
	\subfloat[Read-only]{\label{figur:cass-perf-c}\includegraphics[width=0.33\textwidth]{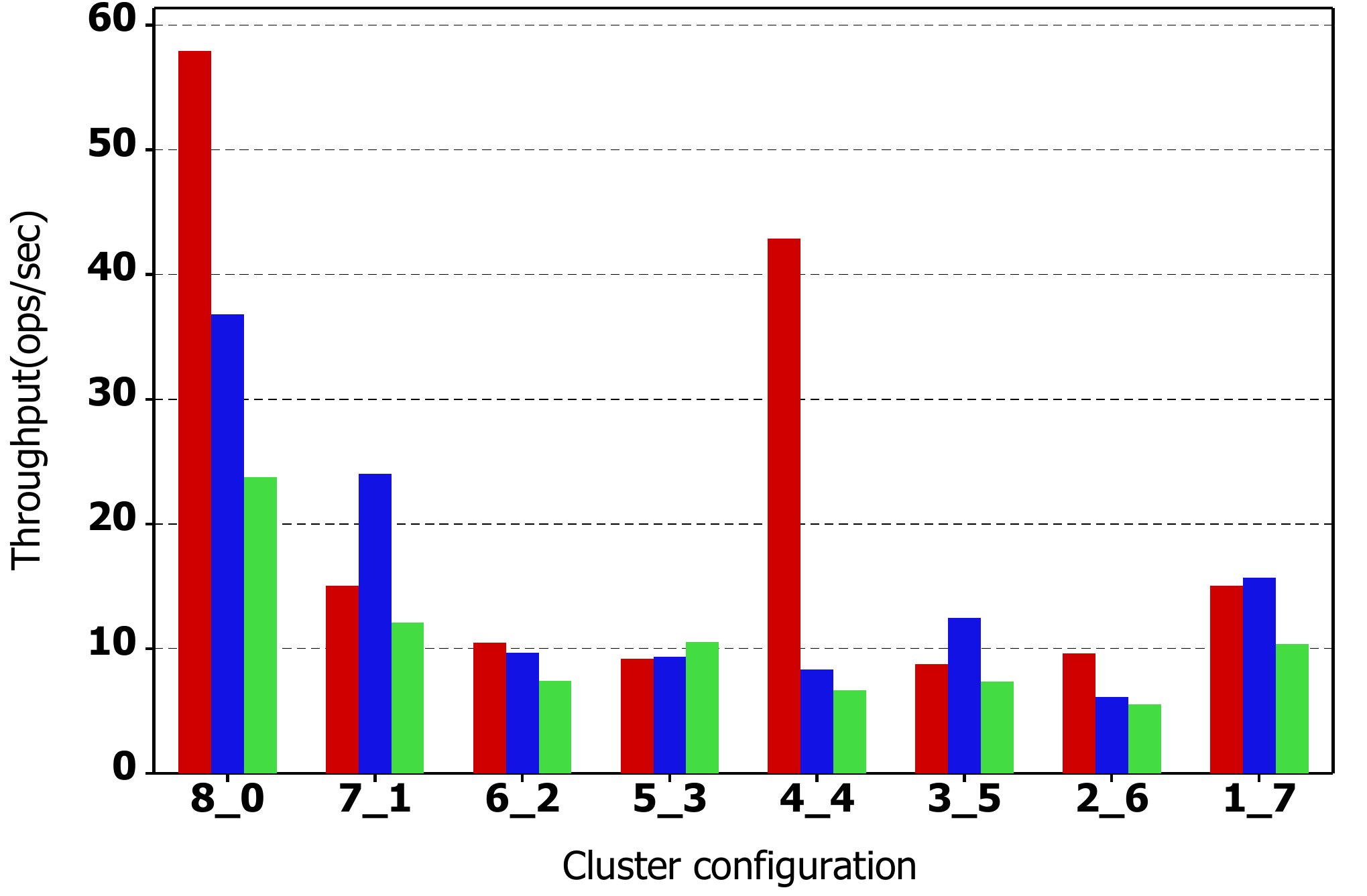}}\\
	\subfloat[Read-latest]{\label{figur:cass-perf-d}\includegraphics[width=0.33\textwidth]{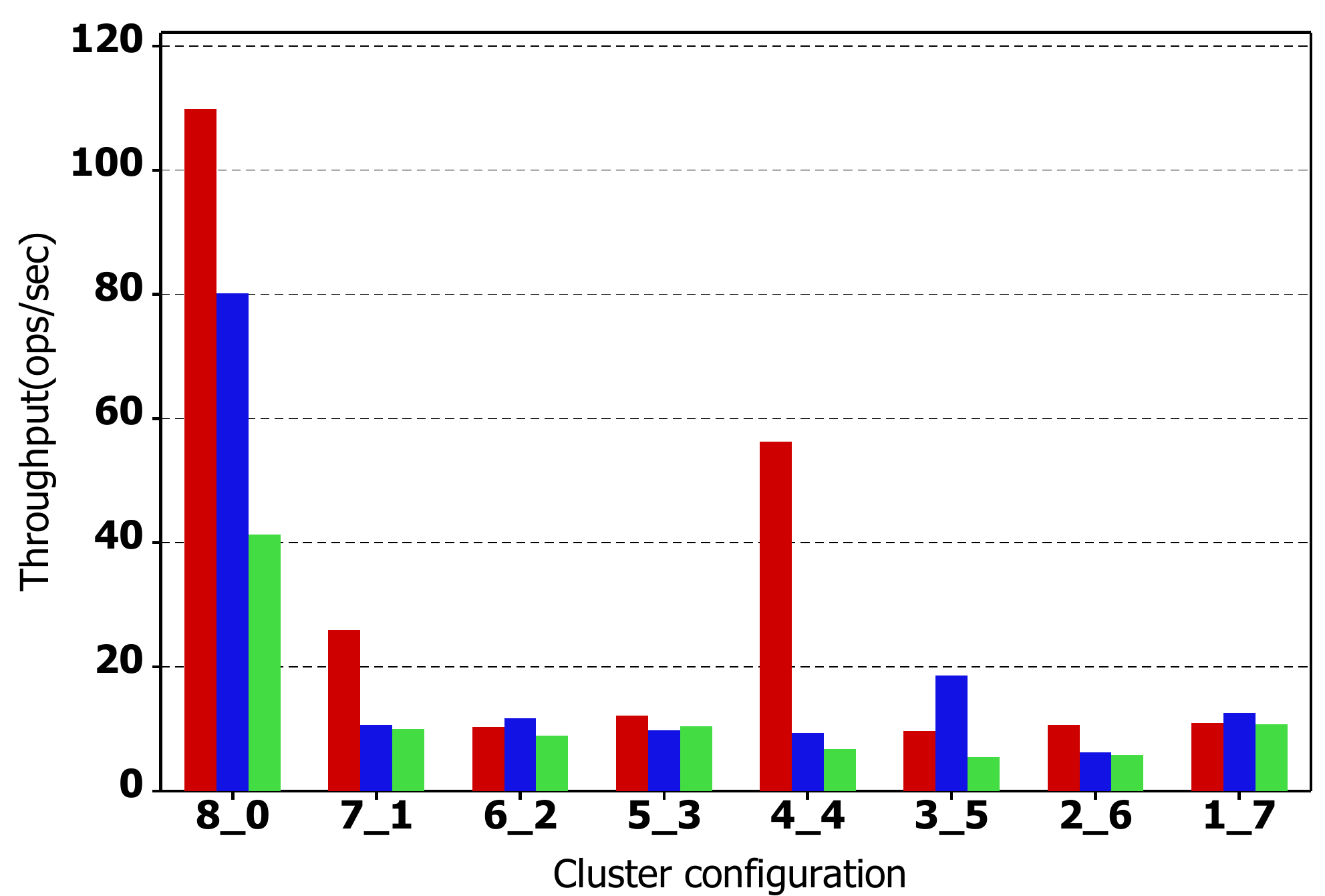}}
	\subfloat[Scan]{\label{figur:cass-perf-e}\includegraphics[width=0.33\textwidth]{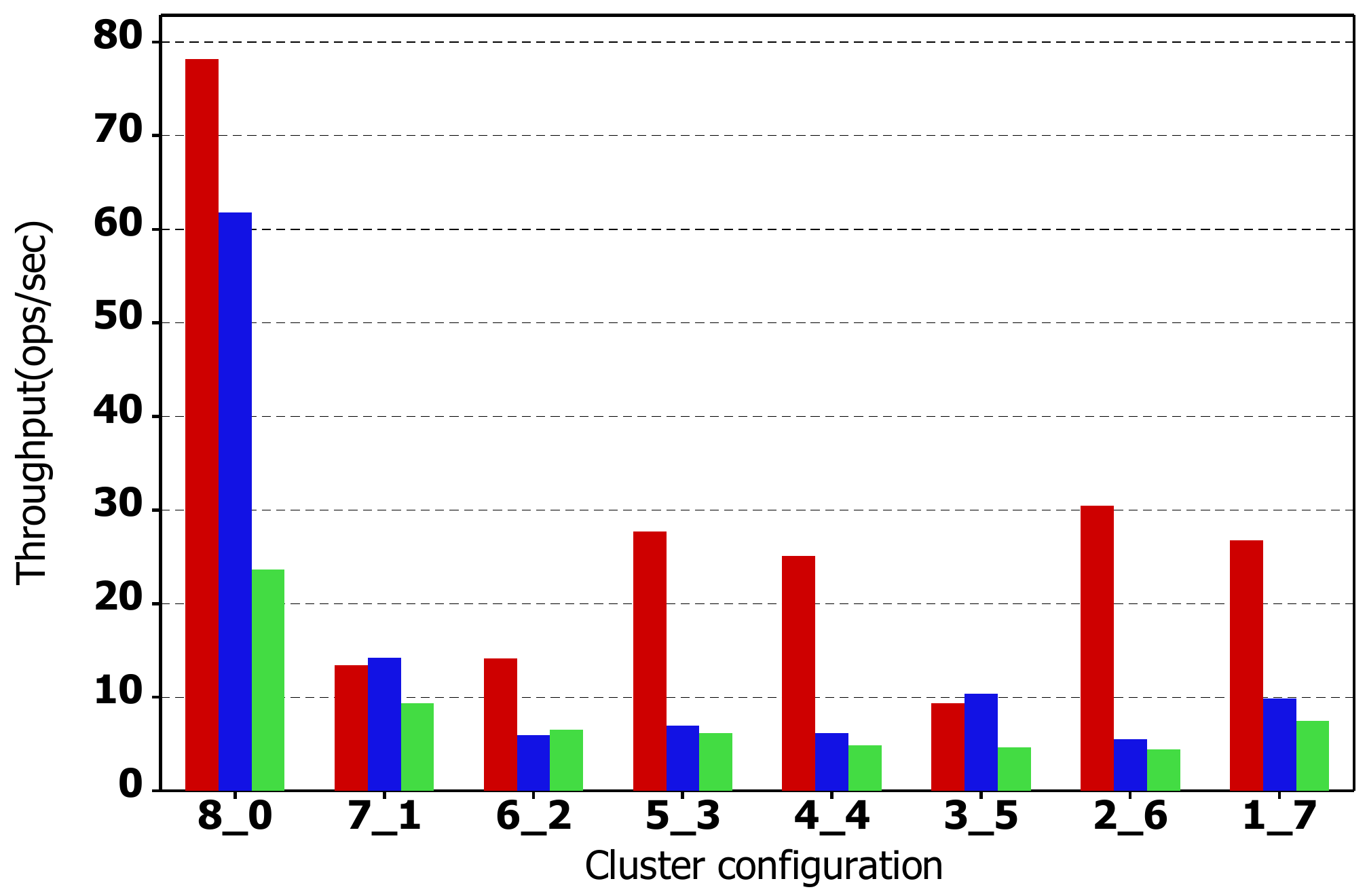}}
	\subfloat[Read-Modify-Write(RMW)]{\label{figur:cass-perf-f}\includegraphics[width=0.33\textwidth]{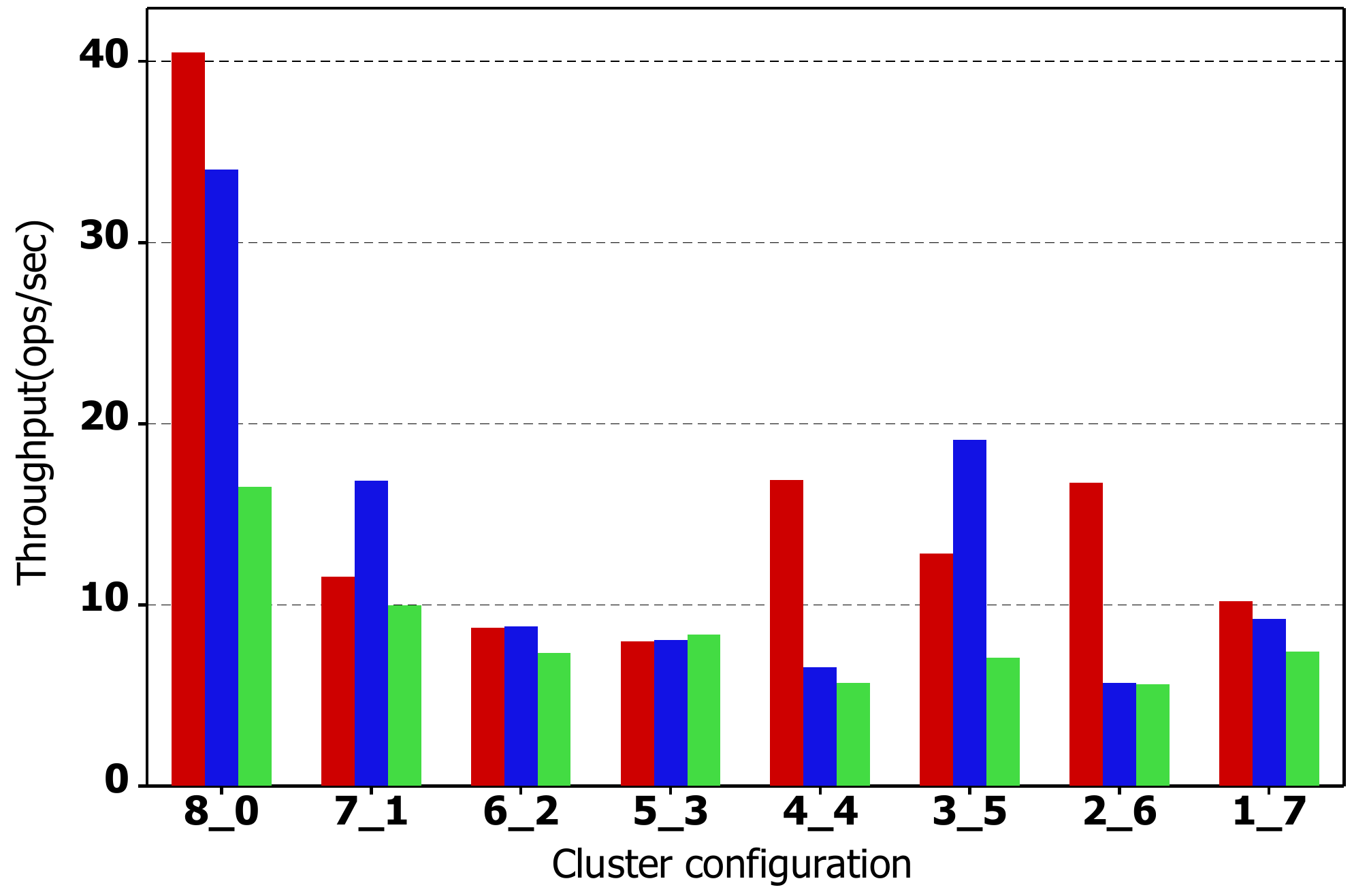}}
	\caption{Throughput for \textbf{Cassandra} in Sydney, Mumbai and Virginia regions. Value $n\_m$  in axis X represents that the hybrid cloud consists of  $n$ nodes in the private cloud and $m$ nodes in the public cloud.}
	\label{fig:cass-perf}
\end{figure*}

\begin{figure*}[h!]
	\centering
	\subfloat[Read-intensive]{\label{figur:riak-perf-a}\includegraphics[width=0.33\textwidth]{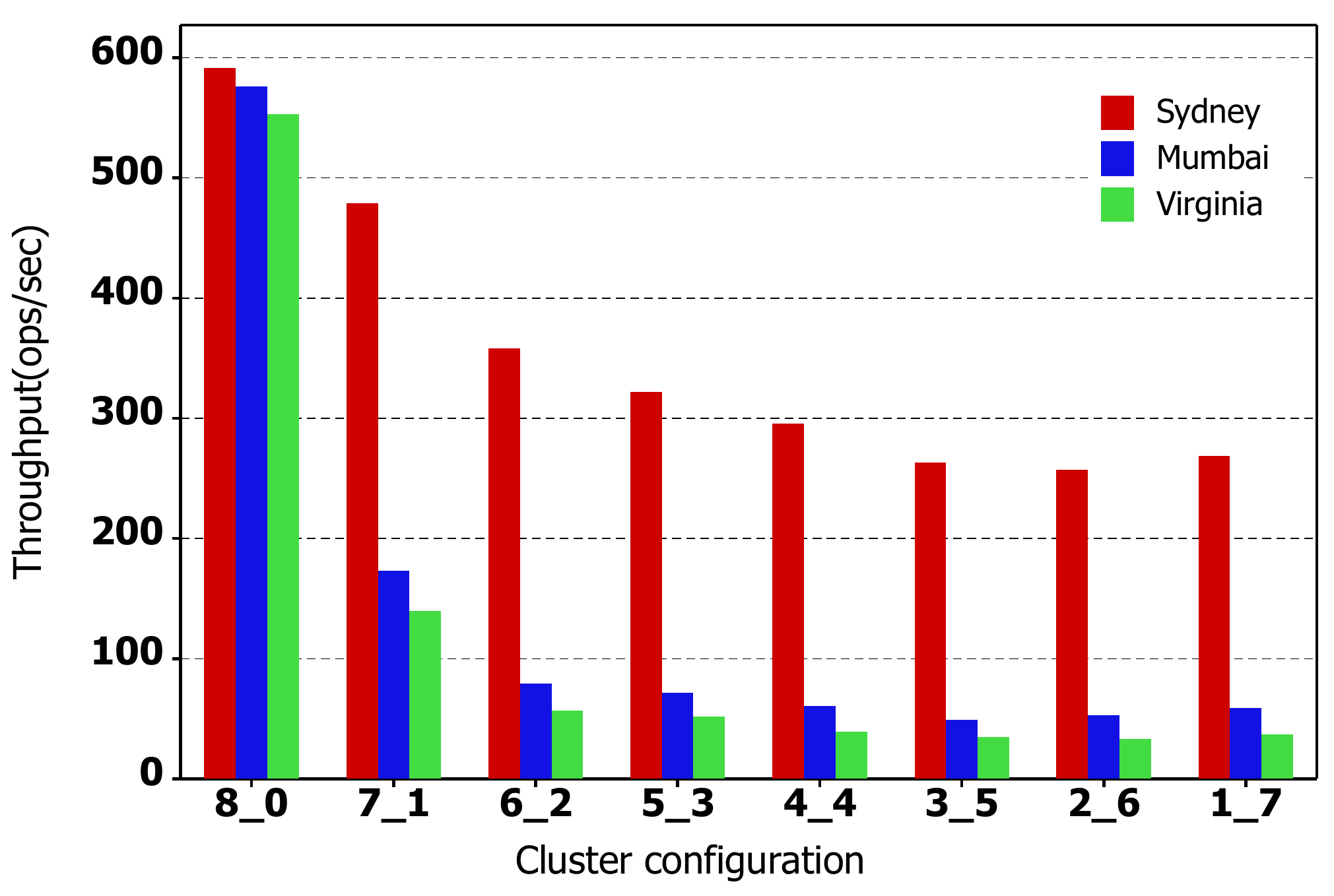}}
	\subfloat[Write-intensive]{\label{figur:riak-perf-b}\includegraphics[width=0.33\textwidth]{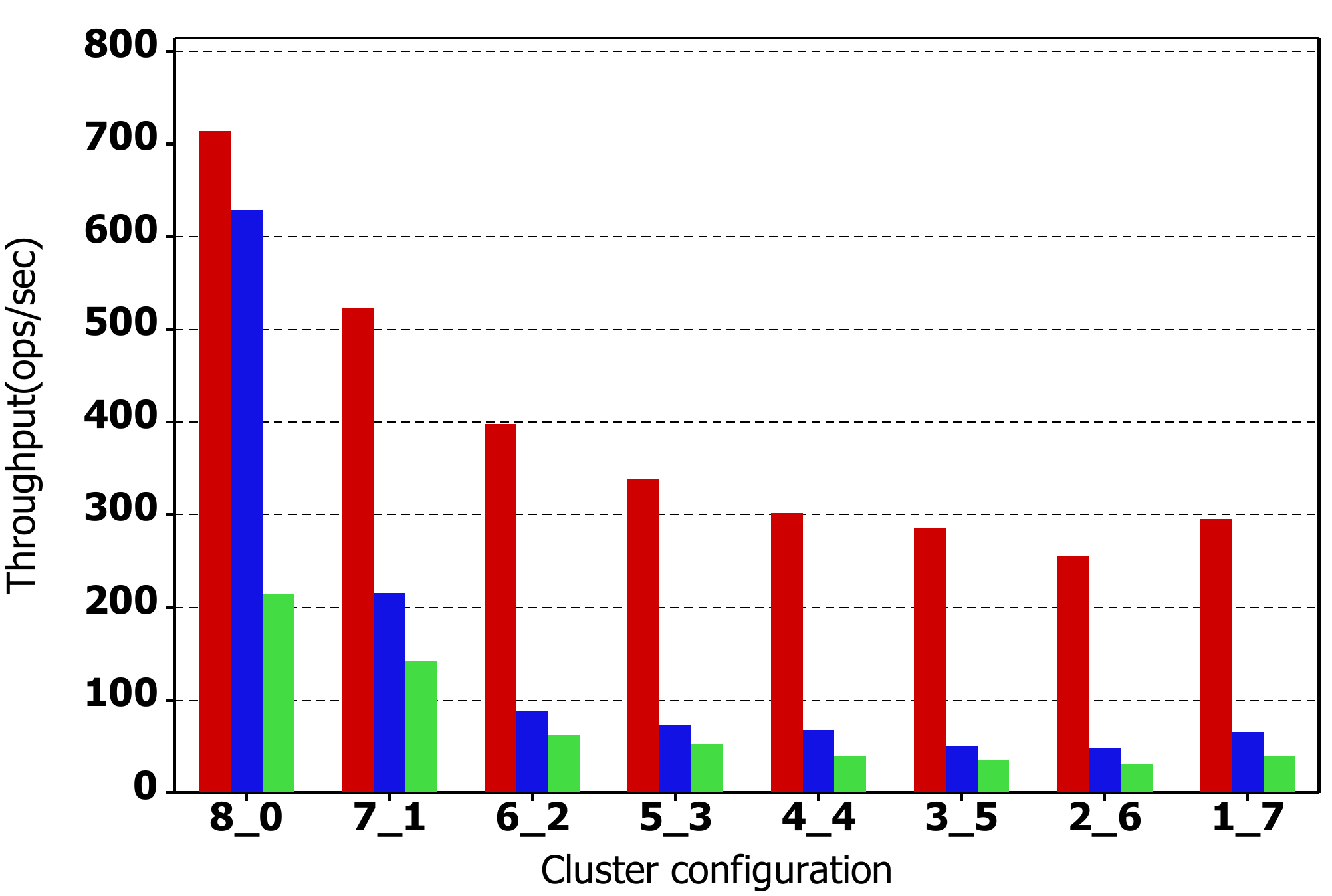}}
	\subfloat[Read-only]{\label{figur:riak-perf-c}\includegraphics[width=0.33\textwidth]{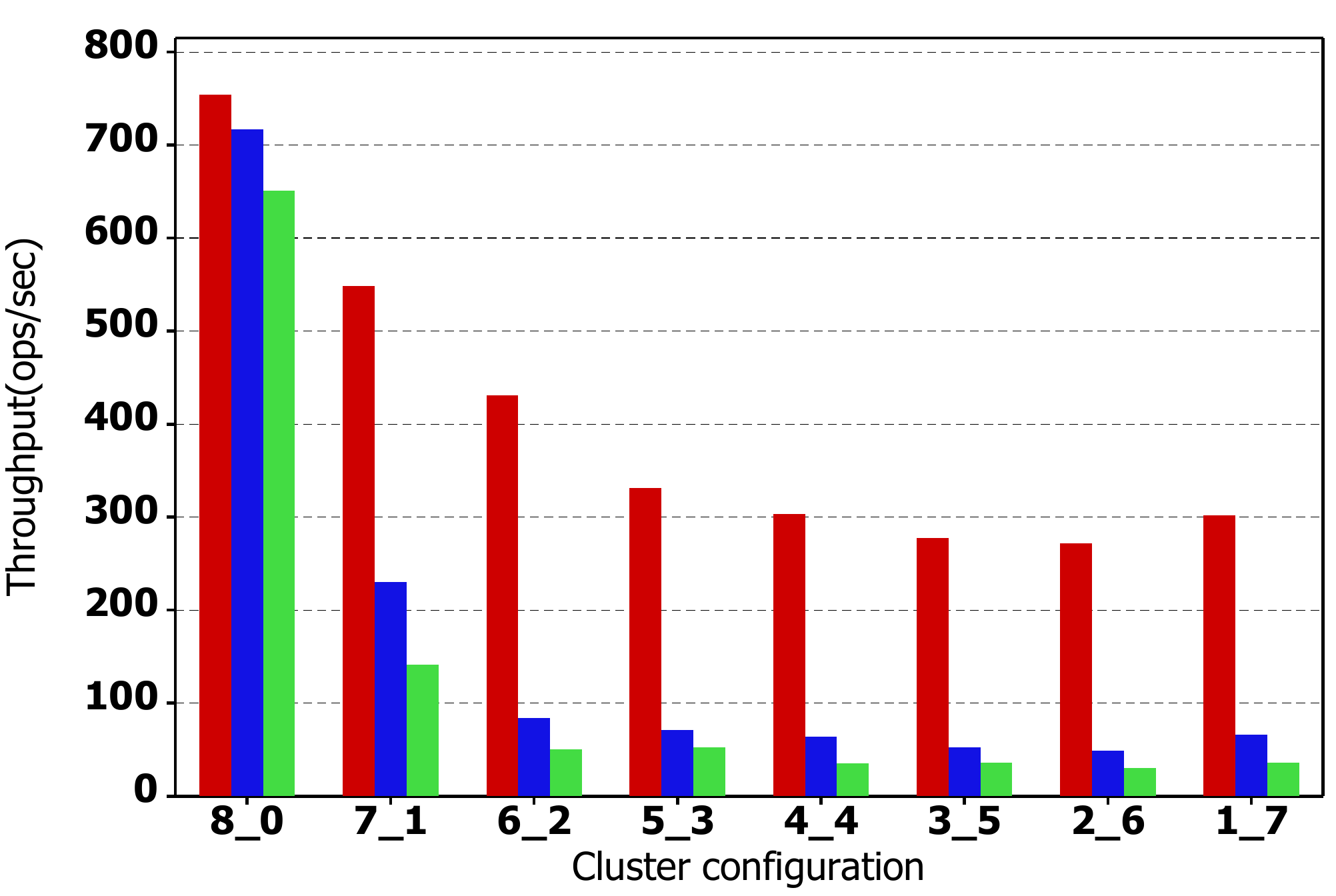}}\\
	\subfloat[Read-latest]{\label{figur:riak-perf-d}\includegraphics[width=0.33\textwidth]{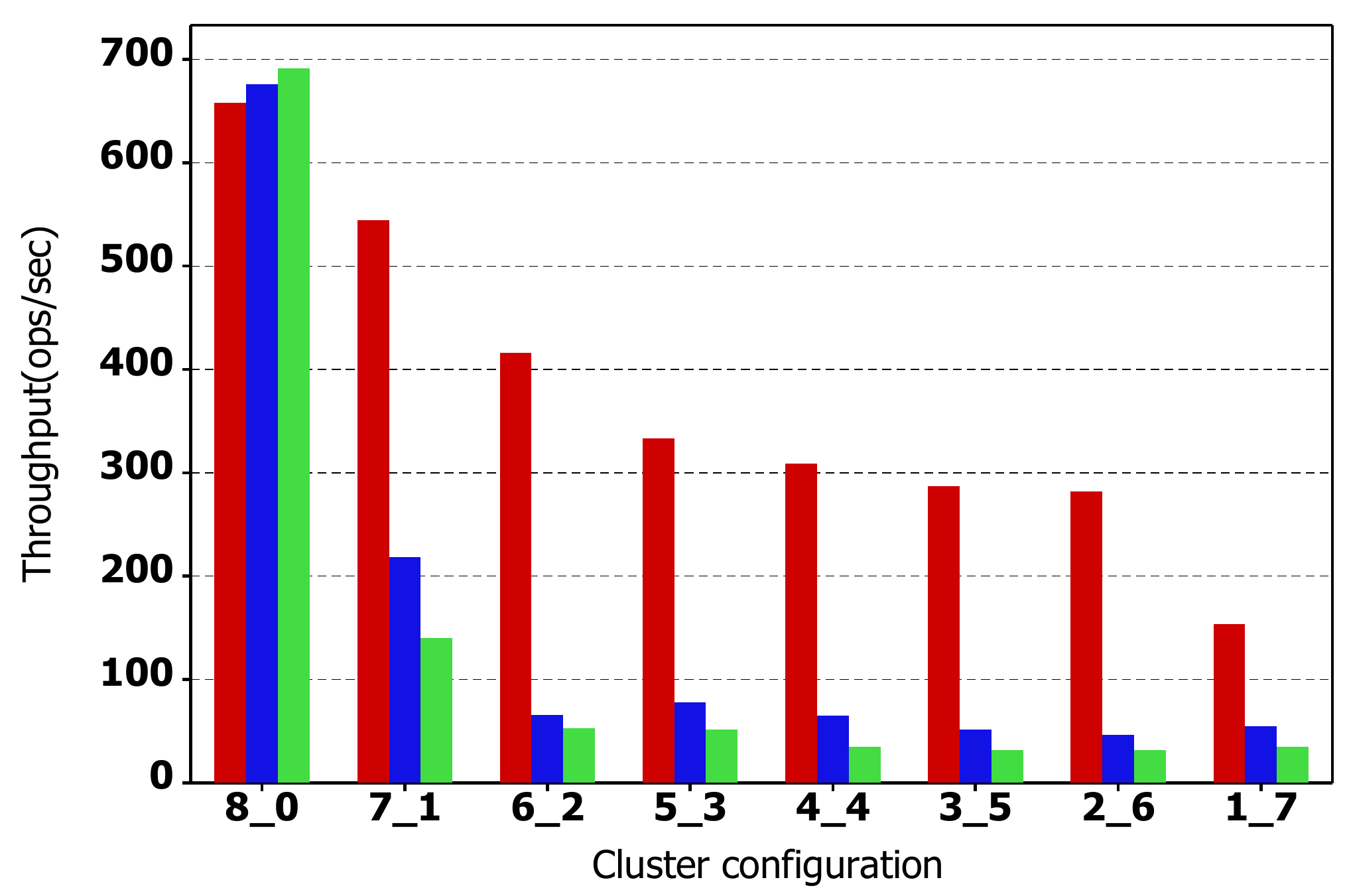}}
	%\subfloat[Scan]{\label{figur:redis-thr}\includegraphics[width=0.33\textwidth]{Fig/riak-perf-e.pdf}}
	\subfloat[Read-Modify-Write(RMW)]{\label{figur:riak-perf-f}\includegraphics[width=0.33\textwidth]{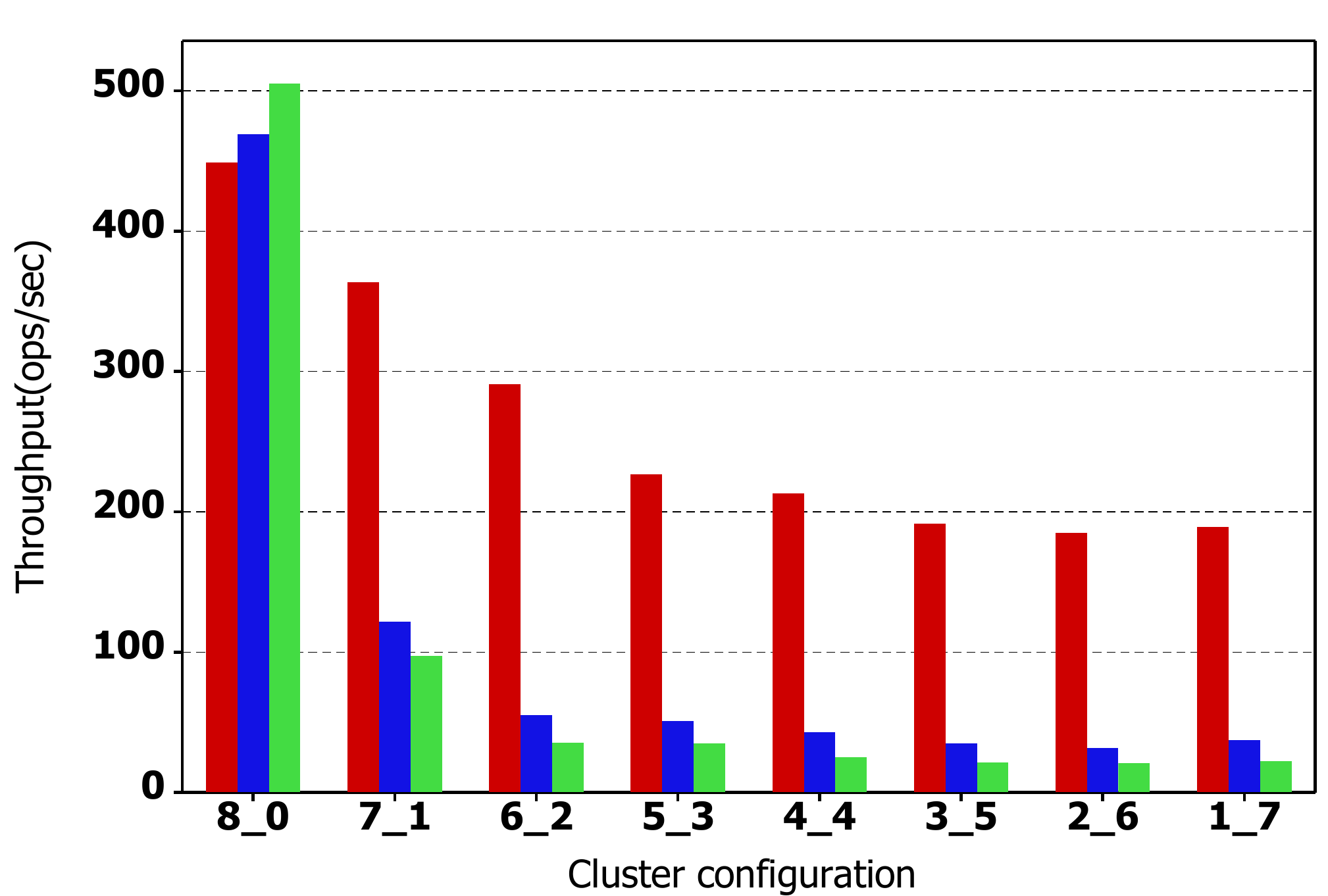}}
	\caption{Throughput for \textbf{Riak} in Sydney, Mumbai and Virginia regions. Value $n\_m$  in axis X represents that the hybrid cloud consists of  $n$ nodes in a private cloud and $m$ nodes in a public cloud.}
	\label{fig:riak-perf}
\end{figure*}

\begin{figure*}[h!]
	\centering
	\subfloat[Read-intensive]{\label{figur:mongo-thr}\includegraphics[width=0.33\textwidth]{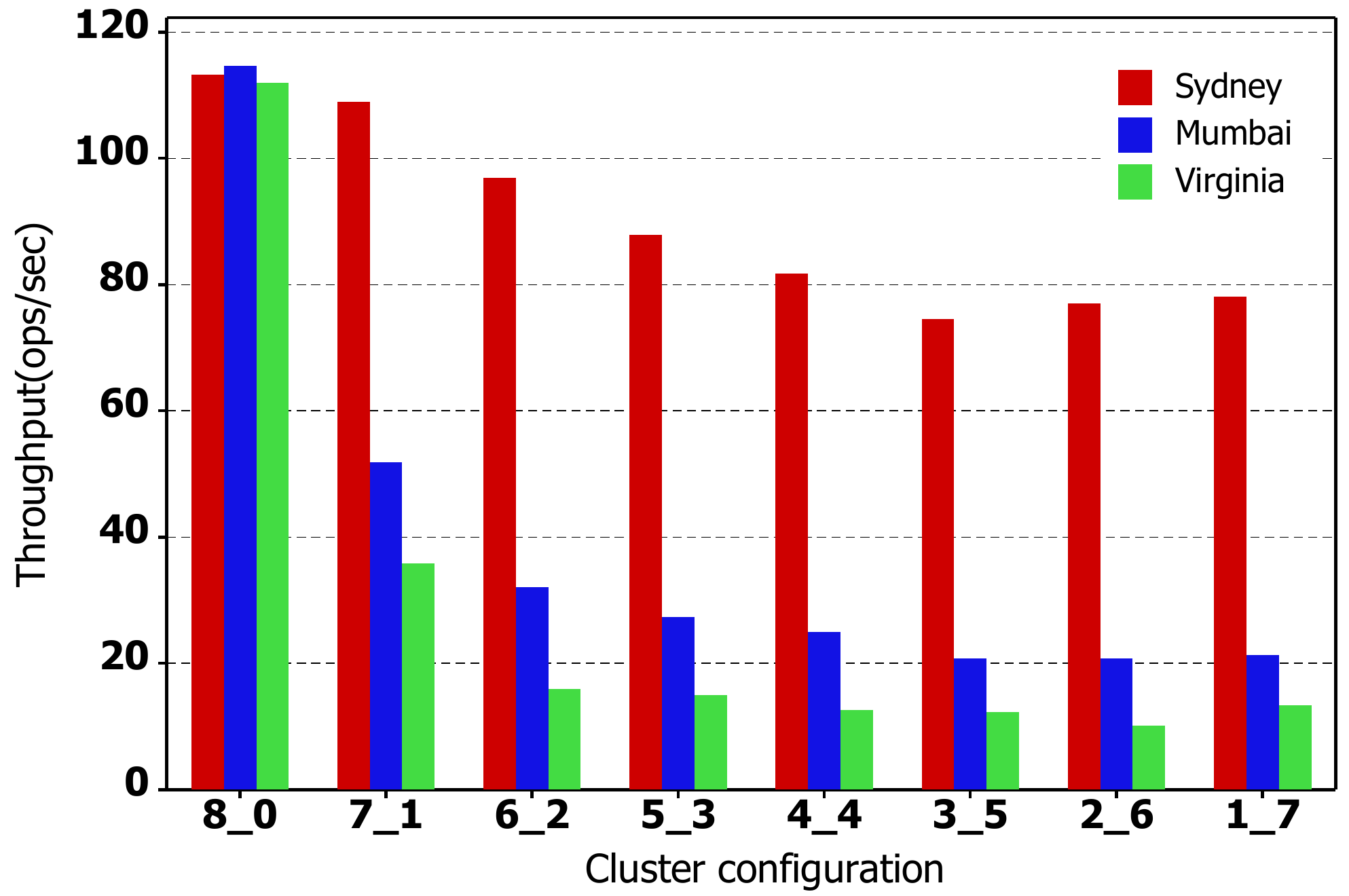}}
	\subfloat[Write-intensive]{\label{figur:cassandra-thr}\includegraphics[width=0.33\textwidth]{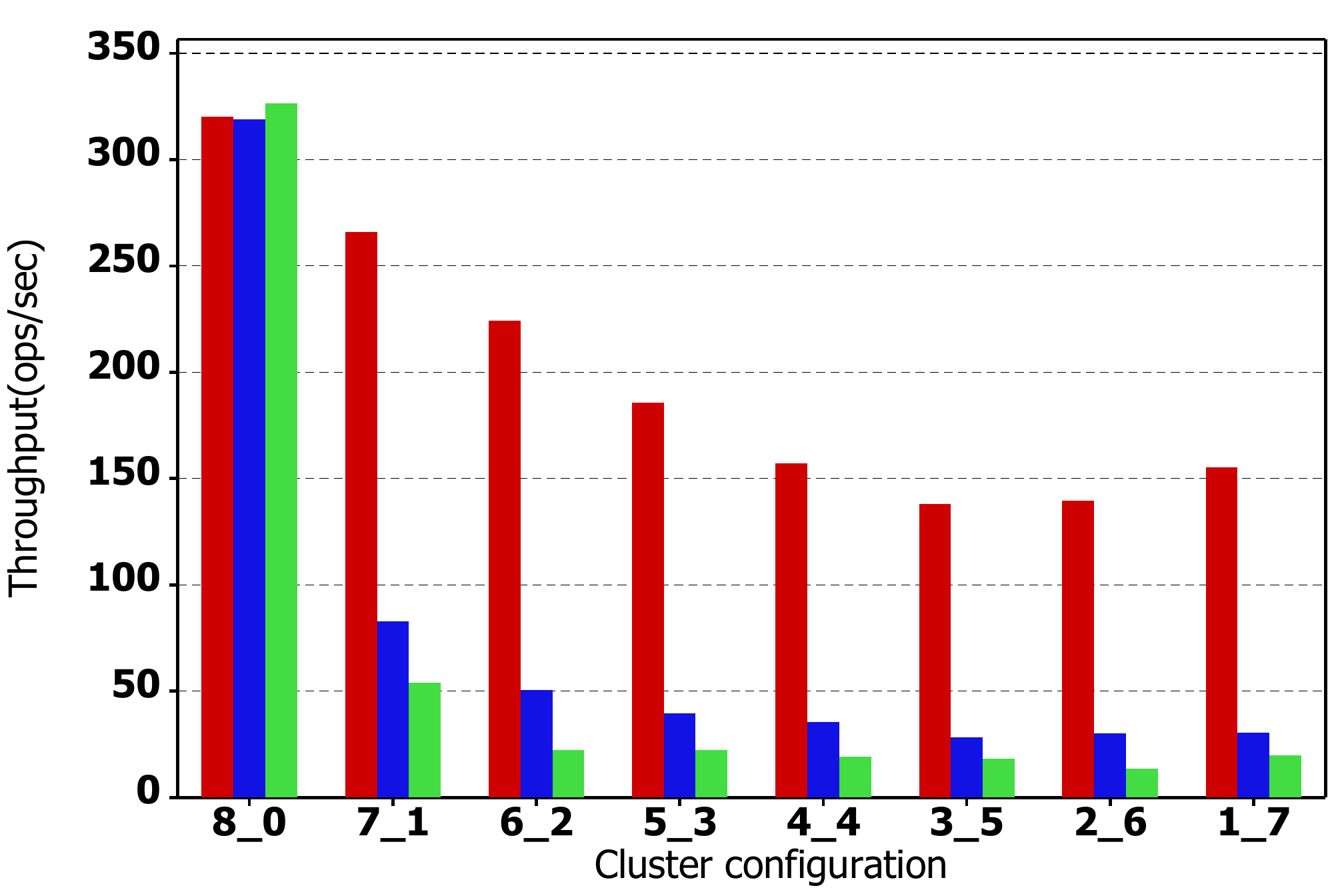}}
	\subfloat[Read-only]{\label{figur:riak-thr}\includegraphics[width=0.33\textwidth]{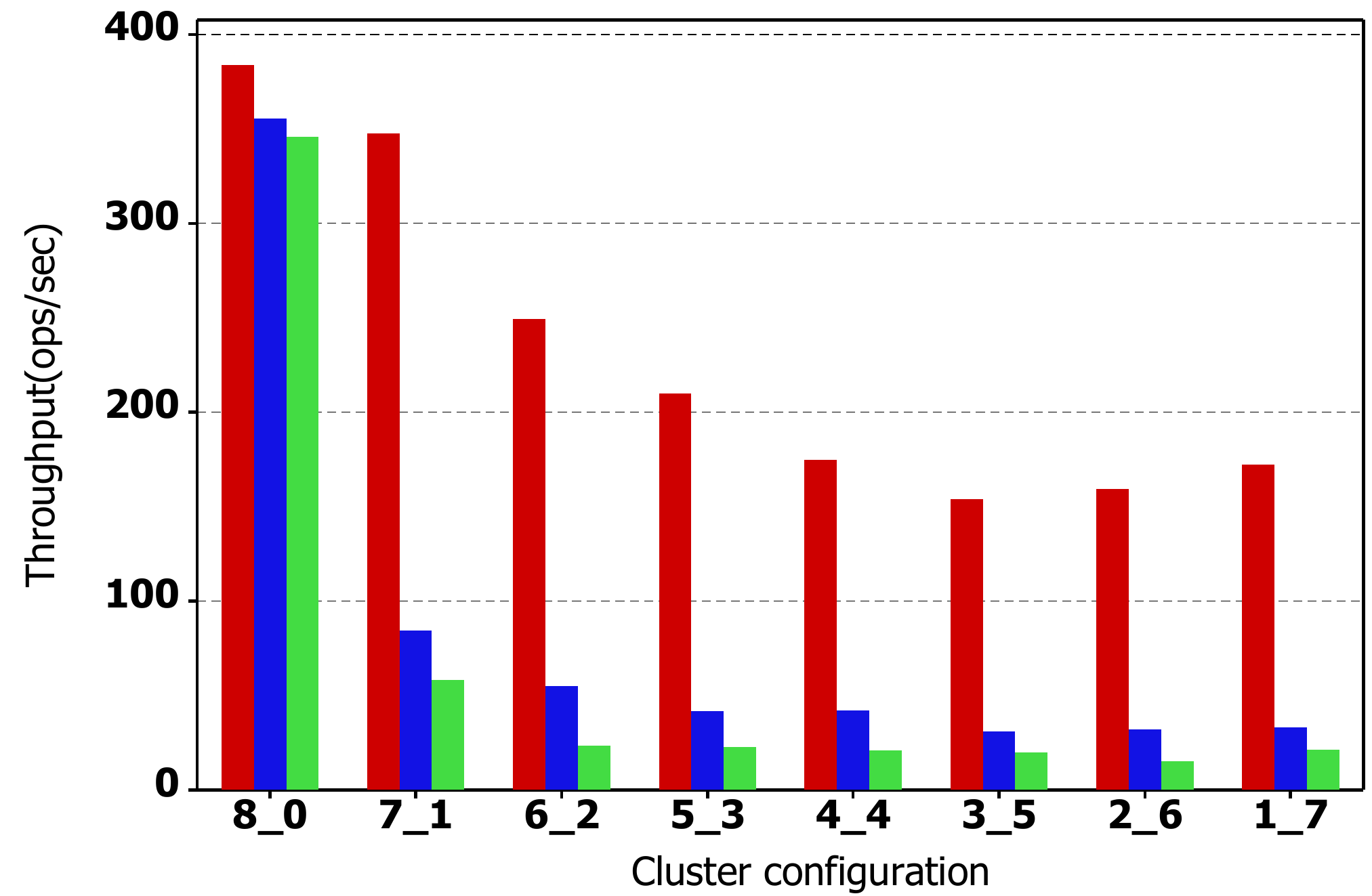}}\\
	\subfloat[Read-latest]{\label{figur:couchdb-thr}\includegraphics[width=0.33\textwidth]{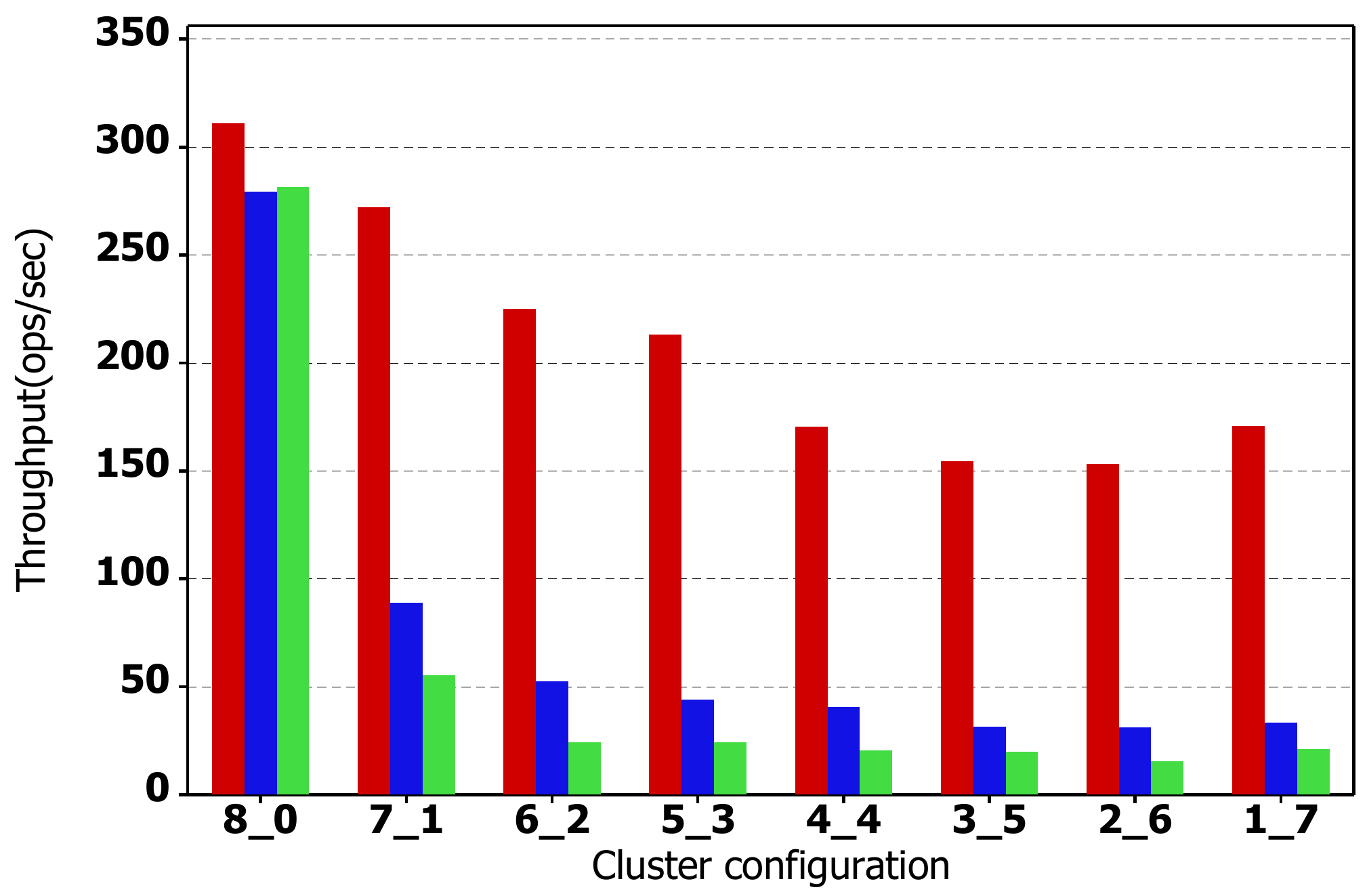}}
	\subfloat[Scan]{\label{figur:redis-thr}\includegraphics[width=0.33\textwidth]{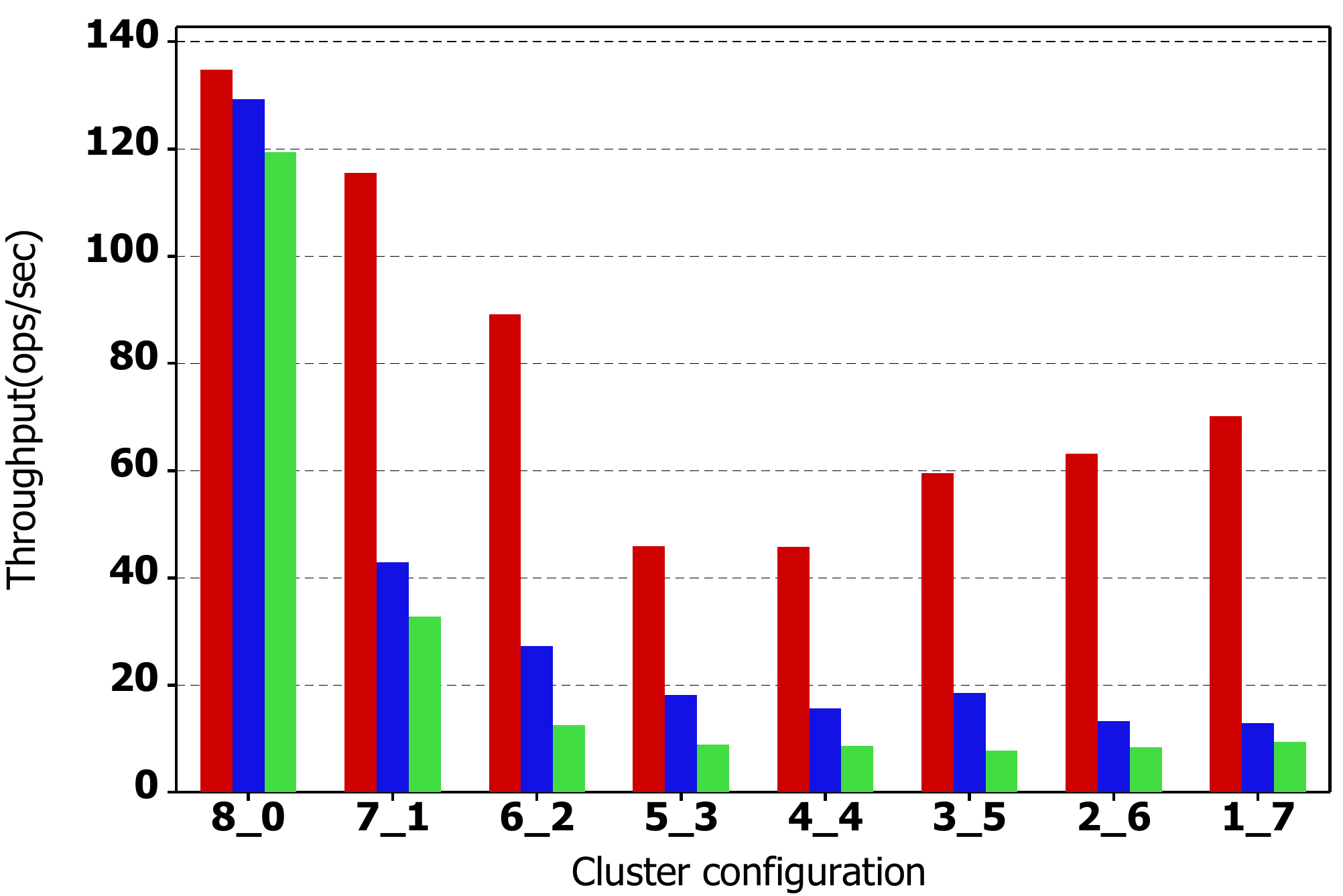}}
	\subfloat[Read-Modify-Write(RMW)]{\label{figur:mysql-thr}\includegraphics[width=0.33\textwidth]{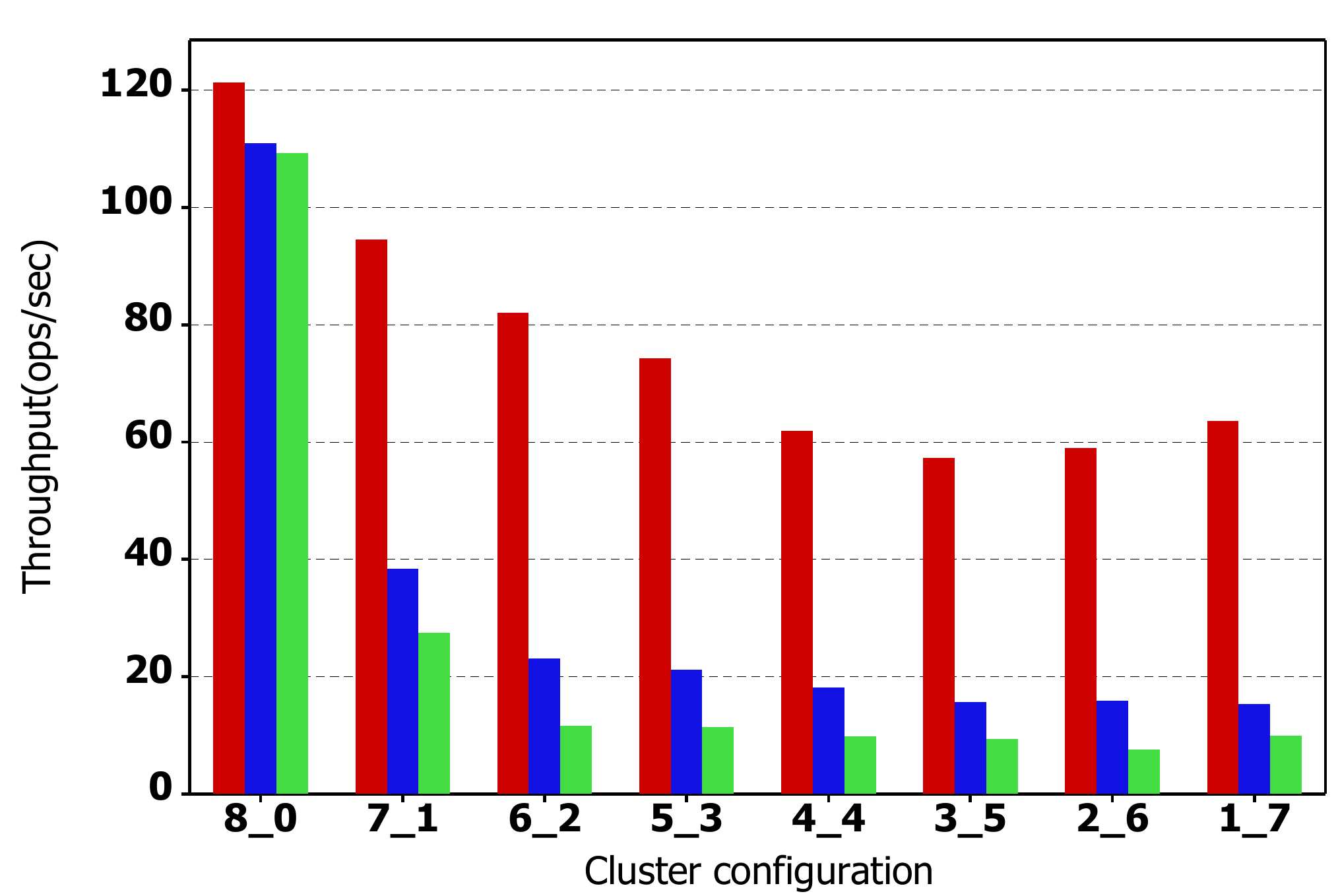}}
	\caption{Throughput for \textbf{Couchdb} in Sydney, Mumbai and Virginia regions. Value $n\_m$  in axis X represents that the hybrid cloud consists of  $n$ nodes in a private cloud and $m$ nodes in a public cloud.}
	\label{fig:couchdb-perf}
\end{figure*}

\begin{figure*}[h!]
	\centering
	\subfloat[Read-intensive]{\label{figur:redis-perf-a}\includegraphics[width=0.33\textwidth]{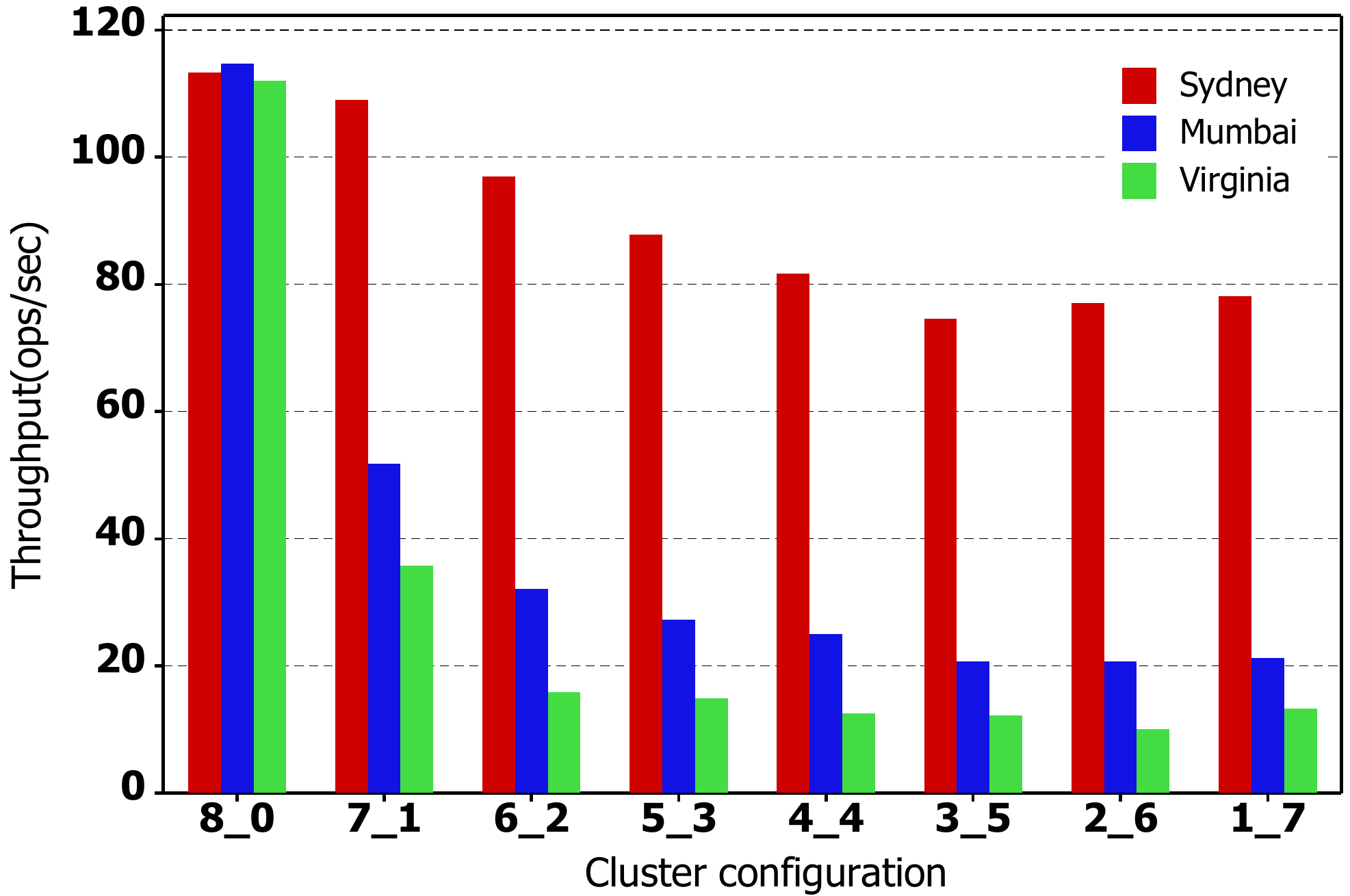}}
	\subfloat[Write-intensive]{\label{figur:redis-perf-b}\includegraphics[width=0.33\textwidth]{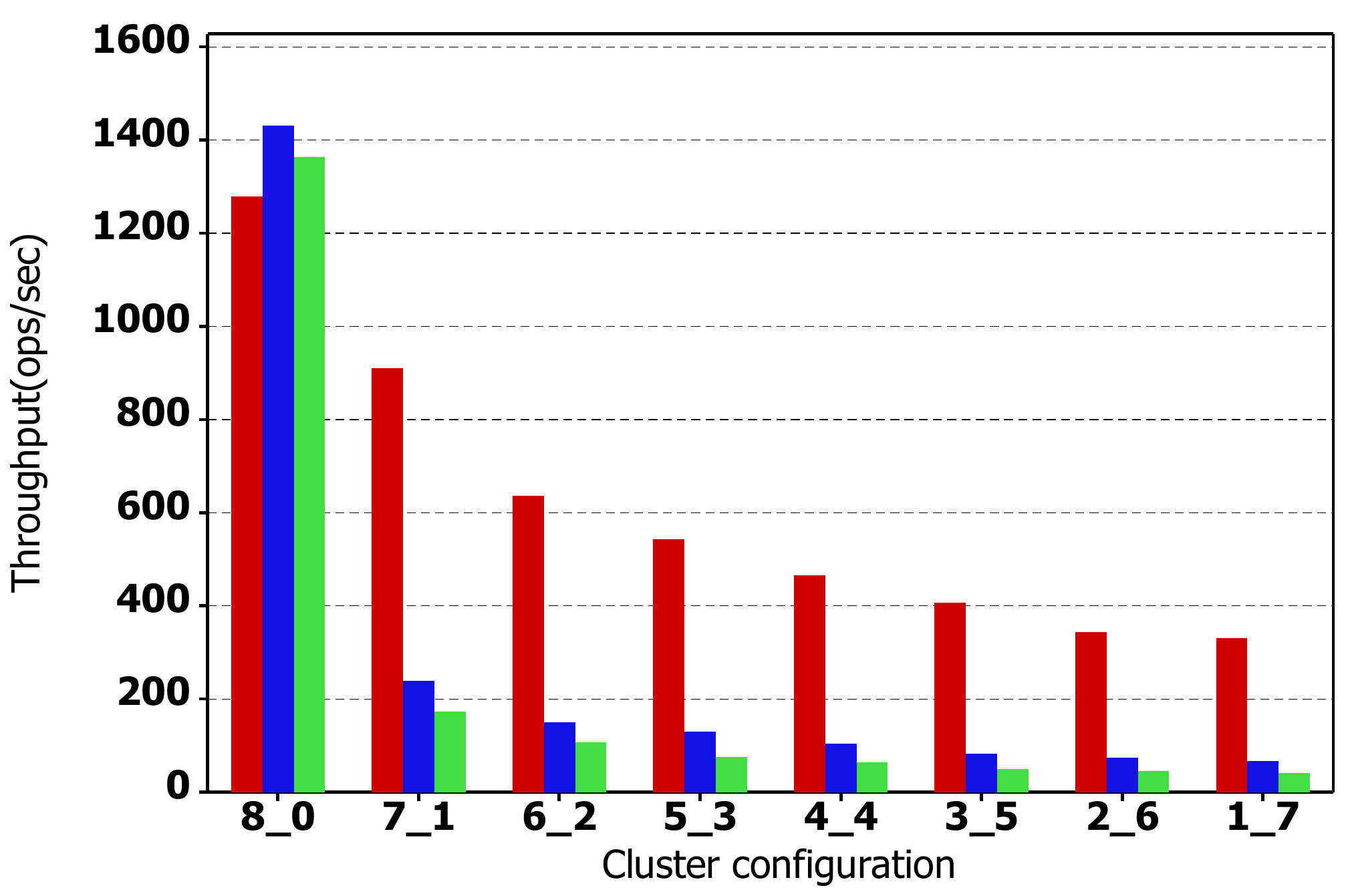}}
	\subfloat[Read-only]{\label{figur:redis-perf-c}\includegraphics[width=0.33\textwidth]{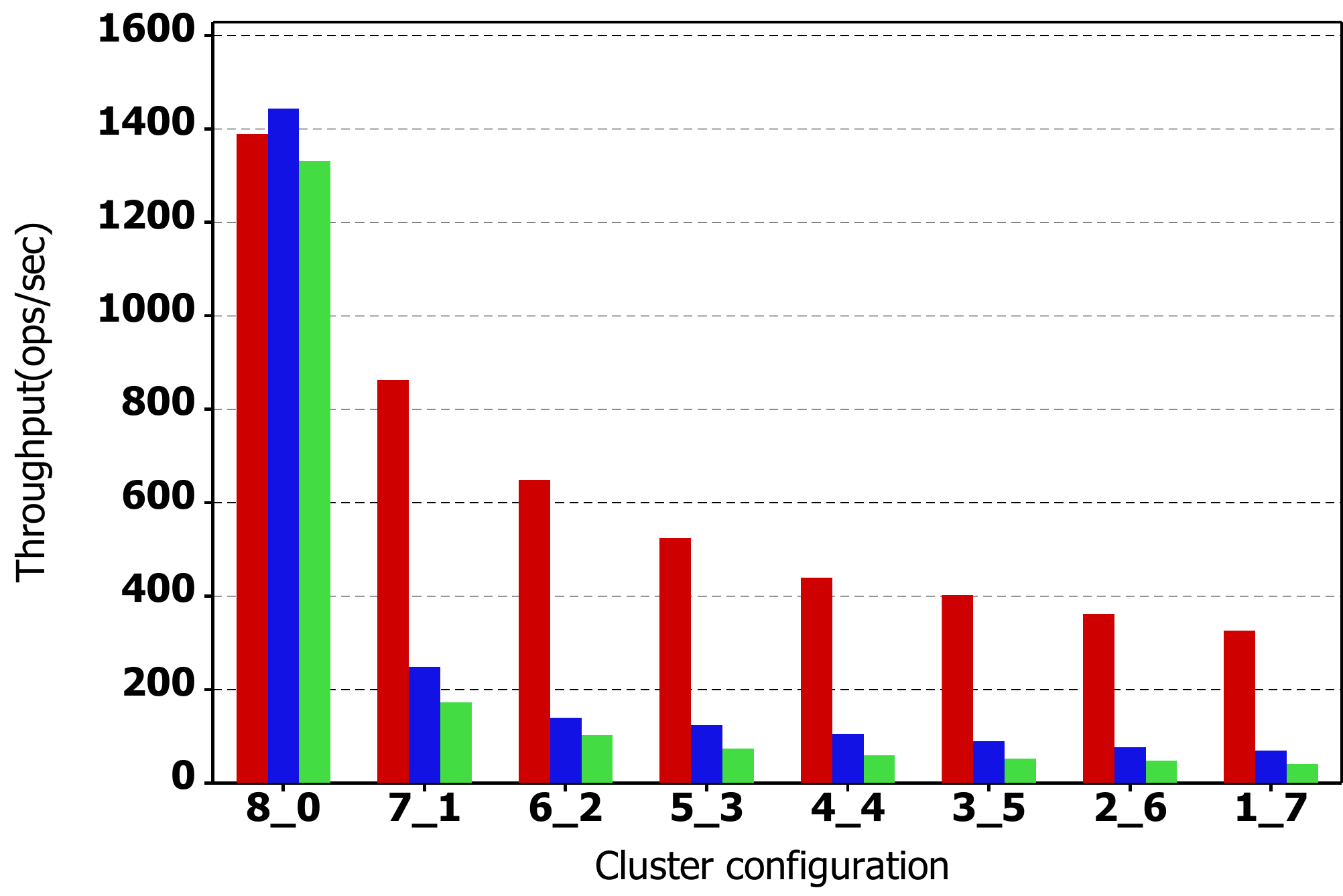}}\\
	\subfloat[Read-latest]{\label{figur:redis-perf-d}\includegraphics[width=0.33\textwidth]{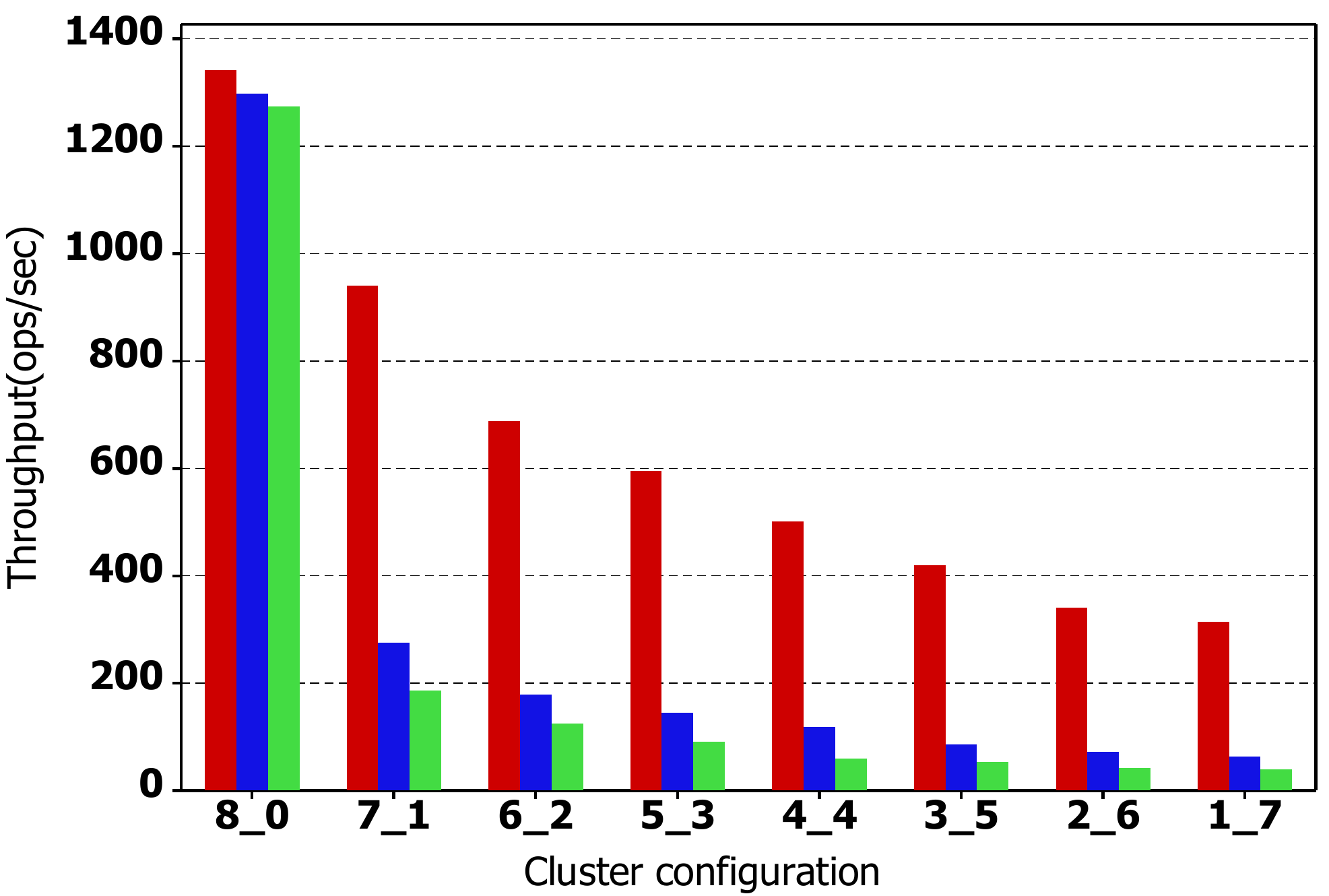}}
	\subfloat[Scan]{\label{figur:redis-perf-e}\includegraphics[width=0.33\textwidth]{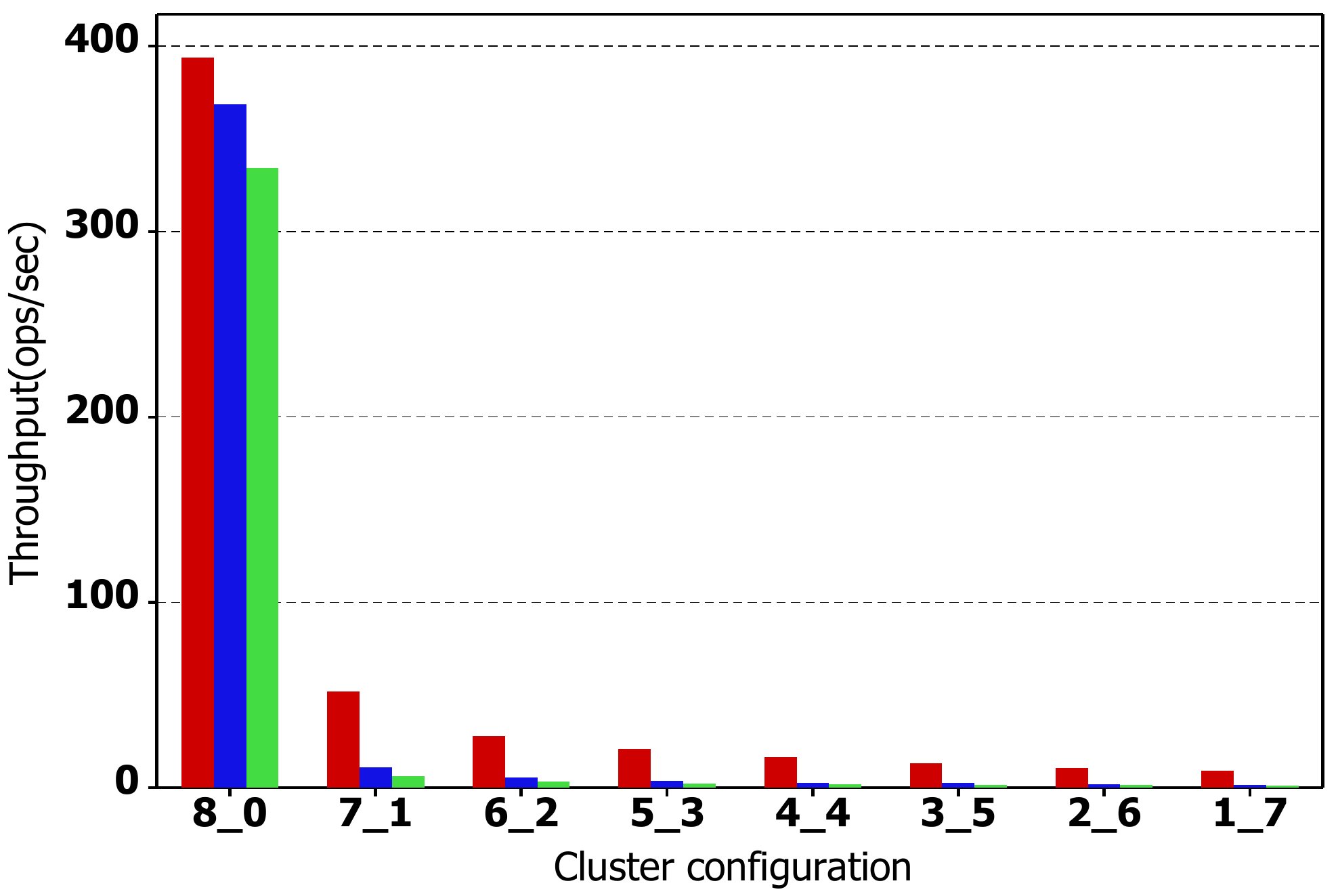}}
	\subfloat[Read-Modify-Write(RMW)]{\label{figur:redis-perf-f}\includegraphics[width=0.33\textwidth]{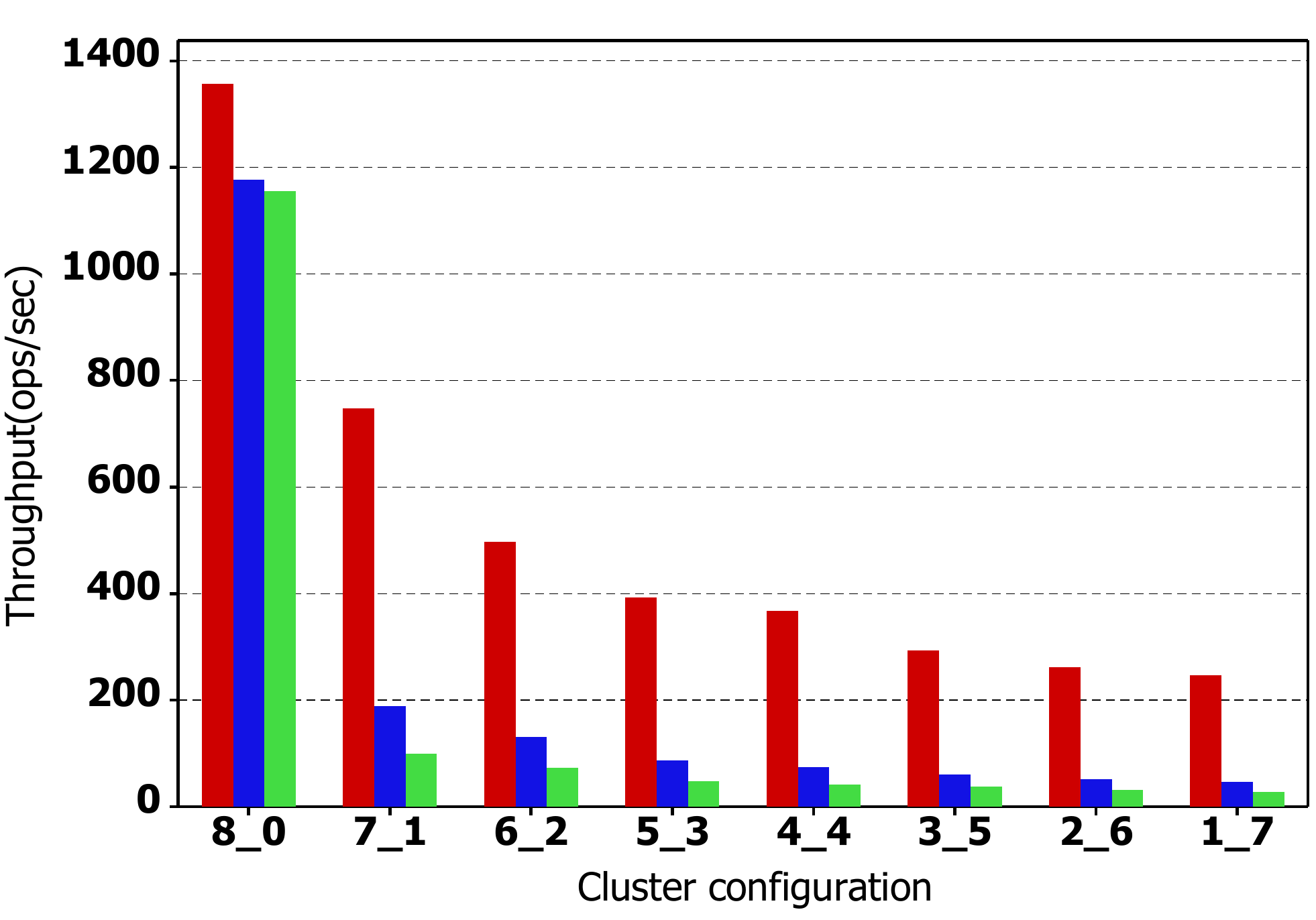}}
	\caption{Throughput for \textbf{Redis} in Sydney, Mumbai and Virginia regions. Value $n\_m$  in axis X represent $n$ nodes that the hybrid  cloud consists of  $n$ nodes in a private cloud and $m$ nodes in a public cloud.}
	\label{fig:redis-perf}
\end{figure*}

\begin{figure*}[h!]
	\centering
	\subfloat[Read-intensive]{\label{figur:mysql-perf-a}\includegraphics[width=0.33\textwidth]{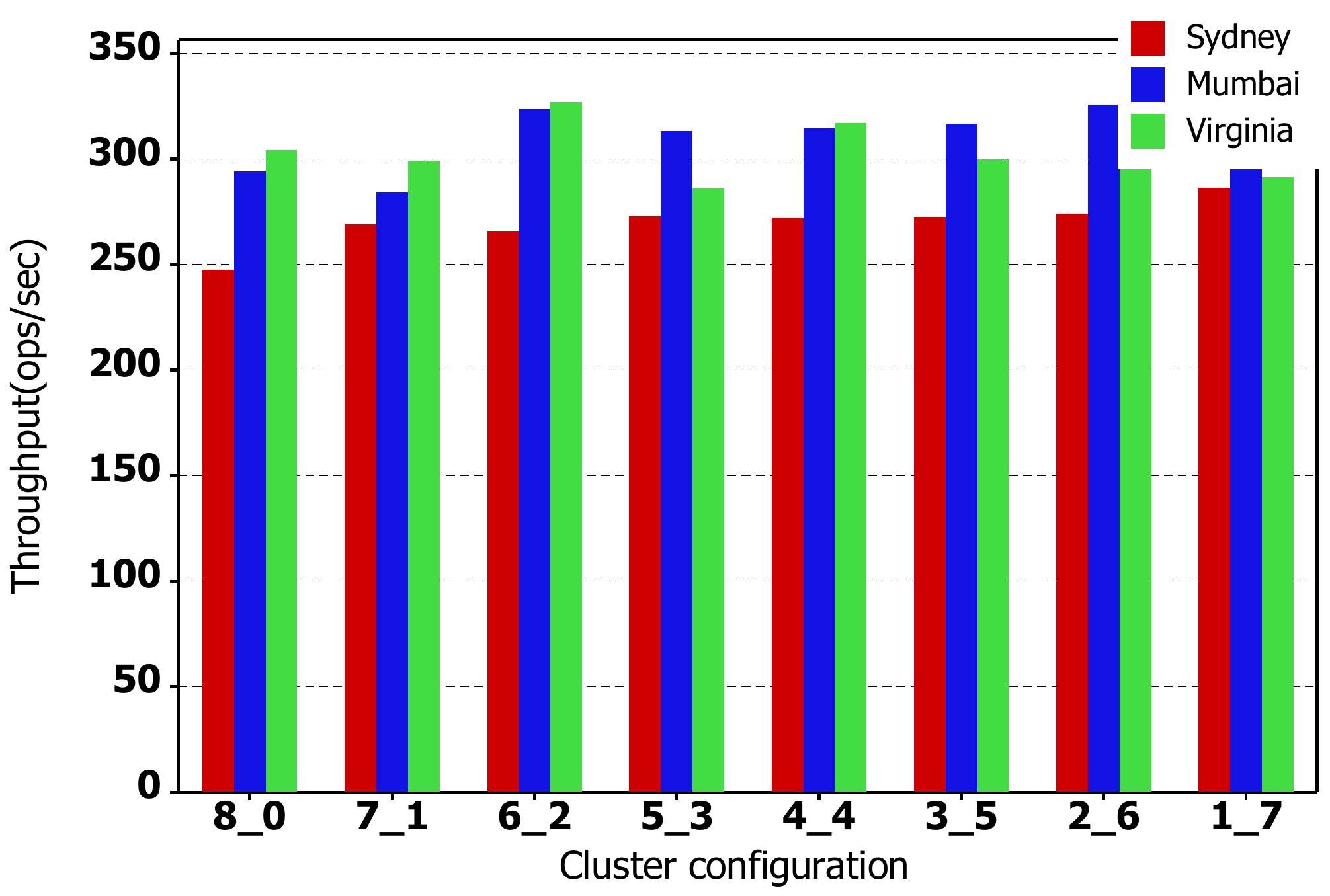}}
	\subfloat[Write-intensive]{\label{figur:mysql-perf-b}\includegraphics[width=0.33\textwidth]{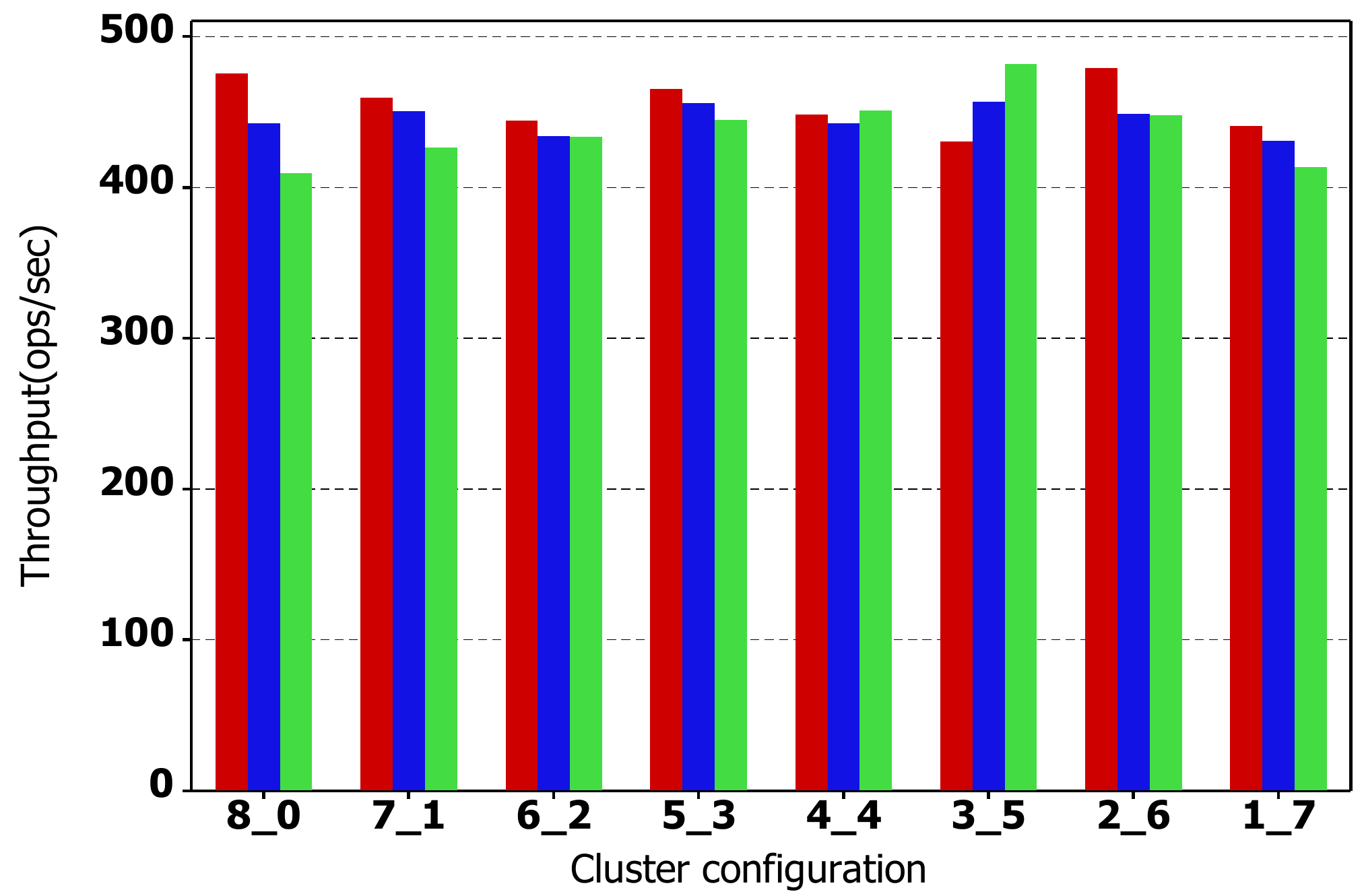}}
	\subfloat[Read-only]{\label{figur:mysql-perf-c}\includegraphics[width=0.33\textwidth]{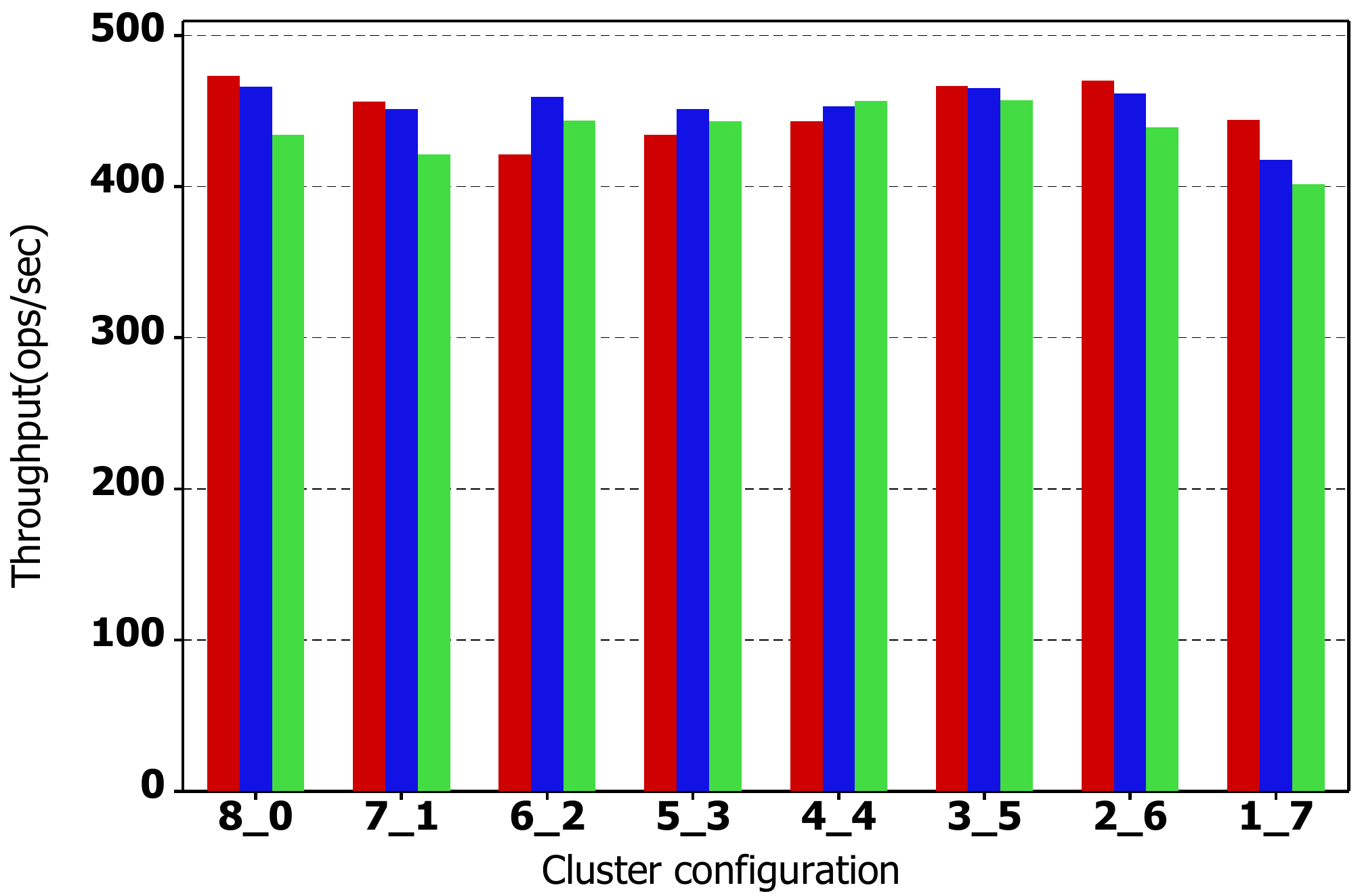}}\\
	\subfloat[Read-latest]{\label{figur:mysql-perf-d}\includegraphics[width=0.33\textwidth]{Fig/mysql-perf-c.pdf}}
	\subfloat[Scan]{\label{figur:mysql-perf-e}\includegraphics[width=0.33\textwidth]{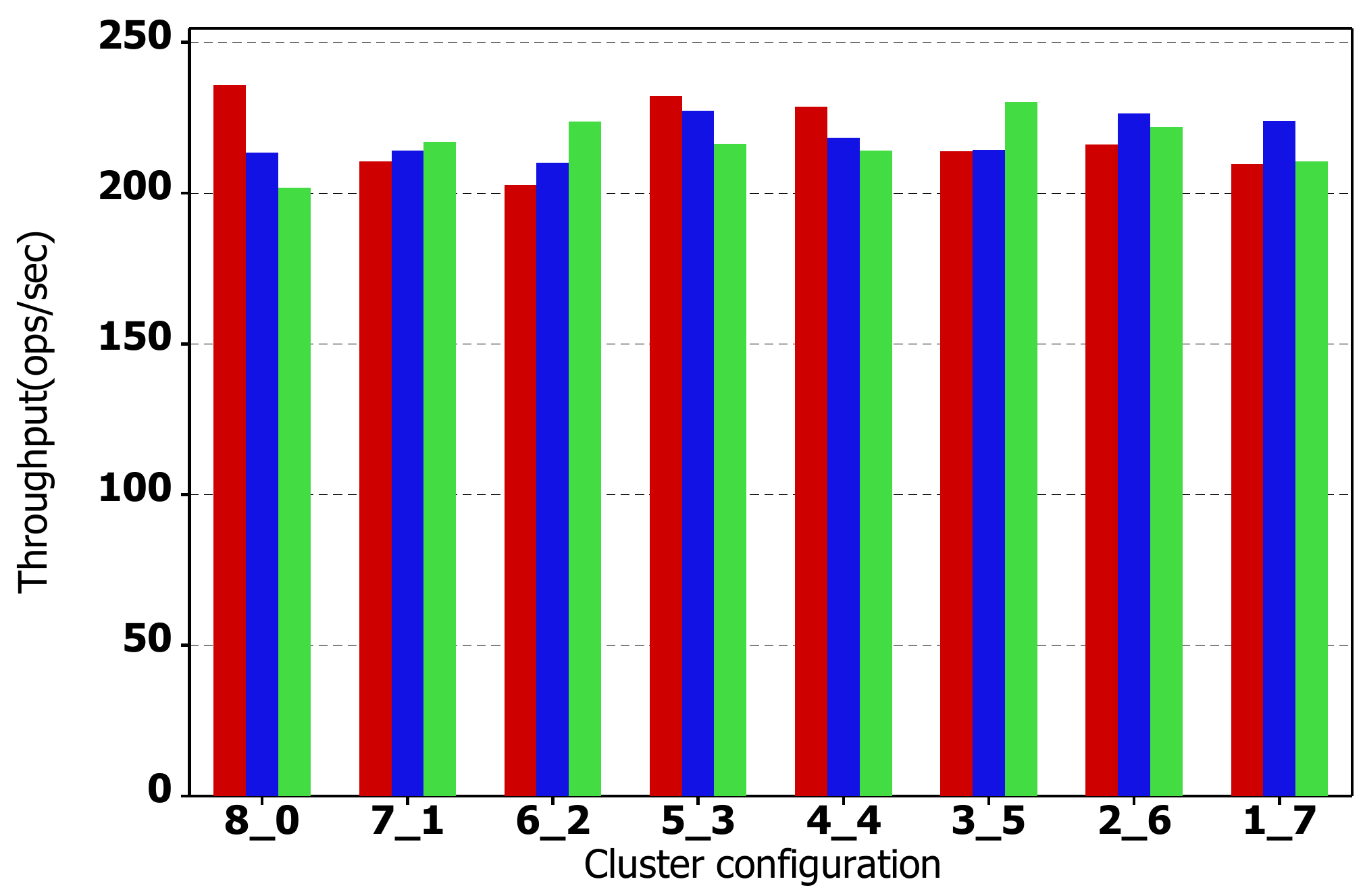}}
	\subfloat[Read-Modify-Write(RMW)]{\label{figur:mysql-perf-f}\includegraphics[width=0.33\textwidth]{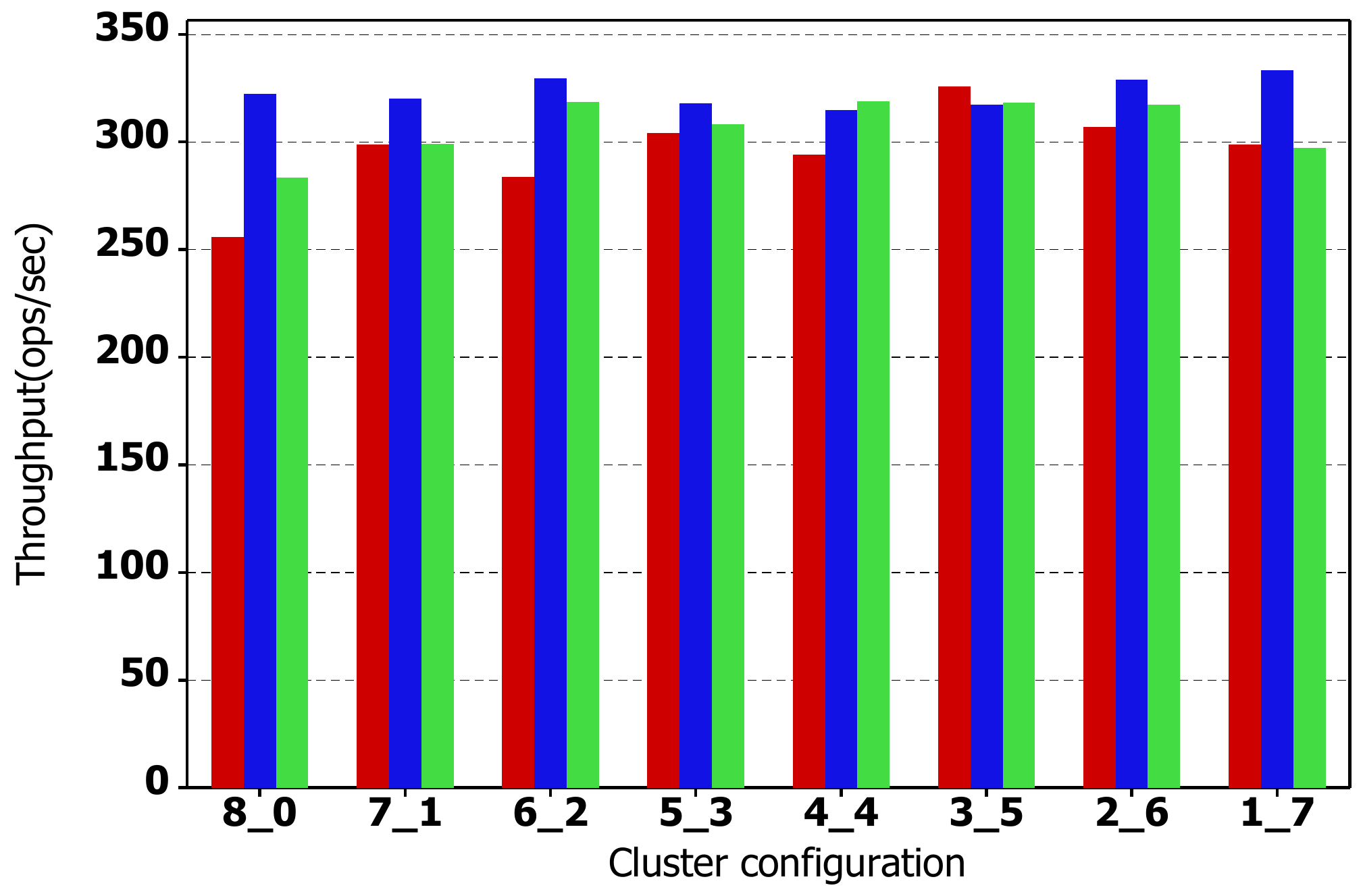}}
	\caption{Throughput for \textbf{MySQL Cluster} in Sydney, Mumbai and Virginia regions. Value $n\_m$  in axis X represents that the hybrid cloud consists of  $n$ nodes in a private cloud and $m$ nodes in a public cloud.}
	\label{fig:mysql-perf}
\end{figure*}

In this section, we intend to answer RQ1: \textit{What is the impact of distance between private and public cloud datacenters on the performance and scalability of distributed databases running  on a hybrid cloud?} In order to reflect distance impact, as we discussed before, we consider three regions hosting public cloud datacenter: Sydney, Virginia and Mumbai. 
Figs. \ref{fig:mong-perf}-\ref{fig:mysql-perf} illustrate the throughput of the six distributed databases against hybrid cluster configuration labelled with pairs of $(n\_m)$, where $m$ and $n$ respectively are the number of VM instances exploited in private and public cloud datacenters. For each database and cluster configuration, we used a freshly installed and established database cluster and loaded data. We refer to cluster configuration with pairs $(8\_0)$ and $(1\_7)$ as \textit{non-bursting} and \textit{full-bursting} respectively. All pairs, except (8\_0), are referred to as hybrid cluster configurations. It should be noted that in the full-bursting setting, we still exploit one VM instance in a private cloud datacenter due to keeping the definition of hybrid cloud.     

Fig. \ref{fig:mong-perf} shows the throughput for MongoDB in three regions. For all six workloads, the distance\footnote{In this context, ``the impact of distance'' implies the impact of  distance between a private cloud datacenter and a public one in all regions (Sydney, Mumbai, Virginia) on the performance. } has slight impact on the throughput when at least half of the VMs hosted in a private cloud (i.e., cluster configuration of $(8\_0)$, $(7\_1)$, $(6\_2)$,  $(5\_3)$, and $(4\_4)$). For these hybrid cluster configurations, MongoDB obtained the best throughput for read-only and read-latest workloads (about 600 ops/sec), followed by read- and write-intensive workloads (between 500-600 ops/sec). In contrast, as the number of nodes bursting into a public cloud increases (i.e., cluster configurations of $(3\_5)$, $(2\_6)$, $(1\_7)$), the distance affects the throughput of MongoDB for almost all workloads in Virginia and Mumbai regions. For read-intensive, write-intensive, and read-only workloads with these cluster configurations, the throughput decreases by 30\%-40\%  and 5\%-20\% in  Virginia and Mumbai respectively. The throughput for the scan workload reduces even more (around 45\%) especially in Virginia. For two other workloads (Read-latest and RMW), distance has less impact on the throughput reduction specifically in Mumbai region.

In Fig. \ref{fig:cass-perf}, the throughput for Cassandra in three regions is summarized. 
%In the hybrid cloud  with all cluster configurations, as the distance between the private cloud and the public cloud increases\footnote{Increment in the distance means the deployment of hybrid cloud in Mumbai  rather than Sydney, or the deployment of the hybrid cloud in Virginia instead of Mumbai/Sydney.}, the throughput significantly decreases. 
For the non-bursting hybrid cluster configuration (i.e., $8\_0$), Cassandra exhibits different values for throughput in three regions although all nodes are hosted in a private cloud. This implies that a fluctuation in the latency between the broker and shared sub-networks and the latency between shared VMs hosting Cassandra database nodes.  
When hybrid cluster configuration changes from non-bursting (i.e., $8\_0$) to bursting ($7\_1$, ...,$1\_7$), the distance has substantial effect on throughput especially for read-intensive workload (Fig. \ref{figur:cass-perf-a}). As the distance between private and public cloud datacenters increases\footnote{Increment in the distance means the deployment of hybrid cloud in Mumbai rather than Sydney, or the deployment of the hybrid cloud in Virginia instead of Mumbai/Sydney.}, the throughput significantly decreases. 
For other workloads, although cloud bursting reduces throughput for all hybrid cluster configurations, the distance has less impact on the performance of Cassandra. It is worth mentioning that Cassandra exposes better performance with hybrid cluster configuration of ($4\_4$) compared to other hybrid cluster configurations (i..e, $7\_1$,...,$1\_7$) especially for Read-only and Read-latest workloads in Sydney. The reason might be the VMs and consequently the data are placed almost equally on two cloud datacenters, which, in turn, the throughput for read operations based on the default setting (three replicas and quorum-based consistency) increases.

The throughput for Riak \footnote{Riak does not support workload E.} is captured in Fig. \ref{fig:riak-perf}. In comparison to MongoDB and Cassandra, Riak exposes more stable in performance trend. As the distance increases and the number of nodes bursting into a public cloud datacenter rises, the throughput of Riak decreases. For the read-intensive workload, the throughput reduces by half in Sydney region when the cluster configuration changes from non-bursting ($8\_0$) to full-bursting ($1\_7$); Whilst for two others regions, the throughput decreases by a factor of about 5. For the remaining workloads (Figs. \ref{figur:riak-perf-a}-\ref{figur:riak-perf-f}), the throughput closely follows the one for read-intensive workload in the reduction trend. In summary, as long as the distance between private and public cloud datacenters is effectively close (less than 1370 KM-- distance between Sydney and Adelaide), the performance of Riak is effective if more than half of Riak database nodes are hosted by a private cloud. 

In Fig. \ref{fig:couchdb-perf}, the throughput results for Couchdb can be seen.
Like Riak, Couchdb demonstrates stability in throughput for all workloads, though this performance metric decreases by 25\% for the RMW workload - 80\% for the read-intensive workload as the hybrid cluster configuration changes from ($8\_0$) to ($1\_7$) in Sydney region.
%with non-bursting cluster configuration in Sydney region. 
In this region, Couchdb's throughput also exposes a decrement of 23\% for the read-only workload - 50\% for the Read-latest and RMW workloads as the cluster configurations vary from non-bursting to full-bursting. In Mumbai region, upon cluster configuration changes from non-bursting to bursting ($7\_1$), Couchdb obtains the least reduction in throughput for the read-intensive workload by a factor of about 2.2; Likewise, the most reduction for the read-only workload by a factor of 4. For both Sydney and Mumbai regions, the throughput of Couchdb initially reduces as half of VMs burst into the public cloud datacenter, and then it gradually increases or stays at a constant level when more than half of VMs are exploited in the public cloud.   

%as the number of nodes bursting into the public cloud, generally the degradation in throughput becomes less or even stays at a fixed level. In summary, cloud bursting for Couchdb can be reasonable if the distance between public and private cloud datacenters is close. 

Fig. \ref{fig:redis-perf} plots the throughput for Redis. Compared to Riak and Couchdb, it has the same stability in performance as distance between private and public cloud datacenters increases. However, for non-bursting cluster configuration in Sydney region, Redis outperforms both Riak and Couchdb in throughput for all workloads (apart from the read-intensive and scan workloads). As hybrid cluster configuration changes from non-bursting to bursting ($7\_1$), Redis's throughput in Sydney region is 4 times compared to the one in Mumbai region for all workloads except for the scan workload. This increment for the hybrid cloud configuration in Sydney raises even by a factor of 5 compared to this configuration in Virginia region. This increment trend for throughput in Sydney region remains fairly stable in comparison to the one that a hybrid cloud obtains in Mumbai and Virginia regions. 

Fig. \ref{fig:mysql-perf} illustrates throughput performance for MySQL Cluster with default setting, where MySQL provides strong consistency among replicas in each \textit{node group cluster}. The size of data node group equals to the number of replicas, which is two as a default in our experiment. With this default setting \cite{Mansouri2020} and the amount of data uploaded to the data node groups, MySQL achieves high throughput (more than 300 ops/sec) for all workloads except the scan workload in all regions. Thus, MySQL holds the second rank among the six investigated databases (after MongoDB) in term of throughput. In respect to MySQL default setting and performance, it is worth mentioning that the following remarks. (i) As we observed, MySQL smartly clusters data nodes in each \textit{data node group} based on the distance between nodes, and (ii) for non-bursting cluster configuration, MySQL's throughput is lower than the throughput of Riak and Redis (See Figs. \ref{fig:riak-perf}, \ref{fig:redis-perf}, and \ref{fig:mysql-perf}).    

\subsubsection{Horizontal Scalability Evaluation} 

\begin{figure*}[h!]
	\centering
	\subfloat[Cassandra]{\label{figur:cass-hsca}\includegraphics[width=0.5\textwidth]{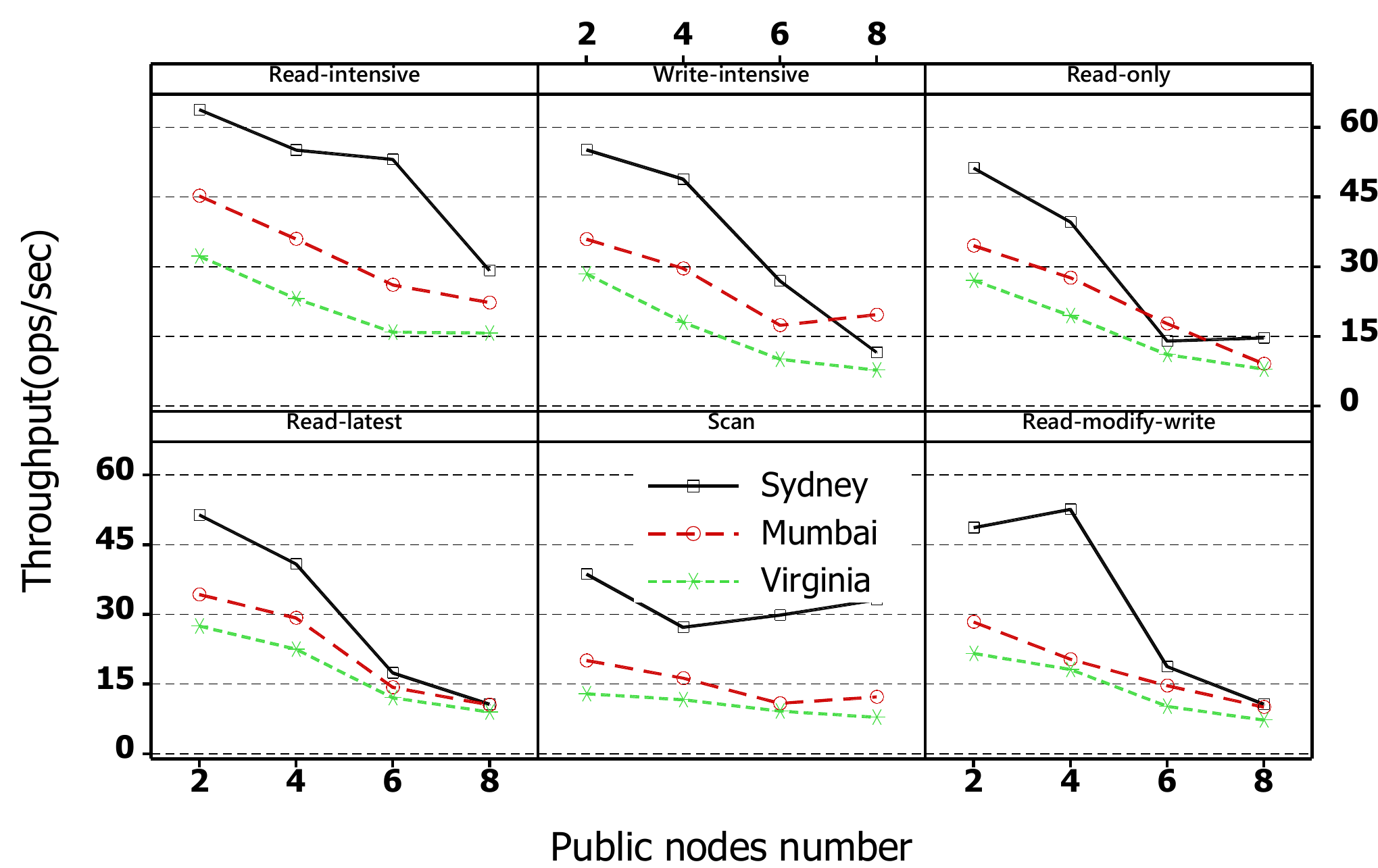}}
	\subfloat[MongoDB]{\label{figur:mongo-hsca}\includegraphics[width=0.5\textwidth]{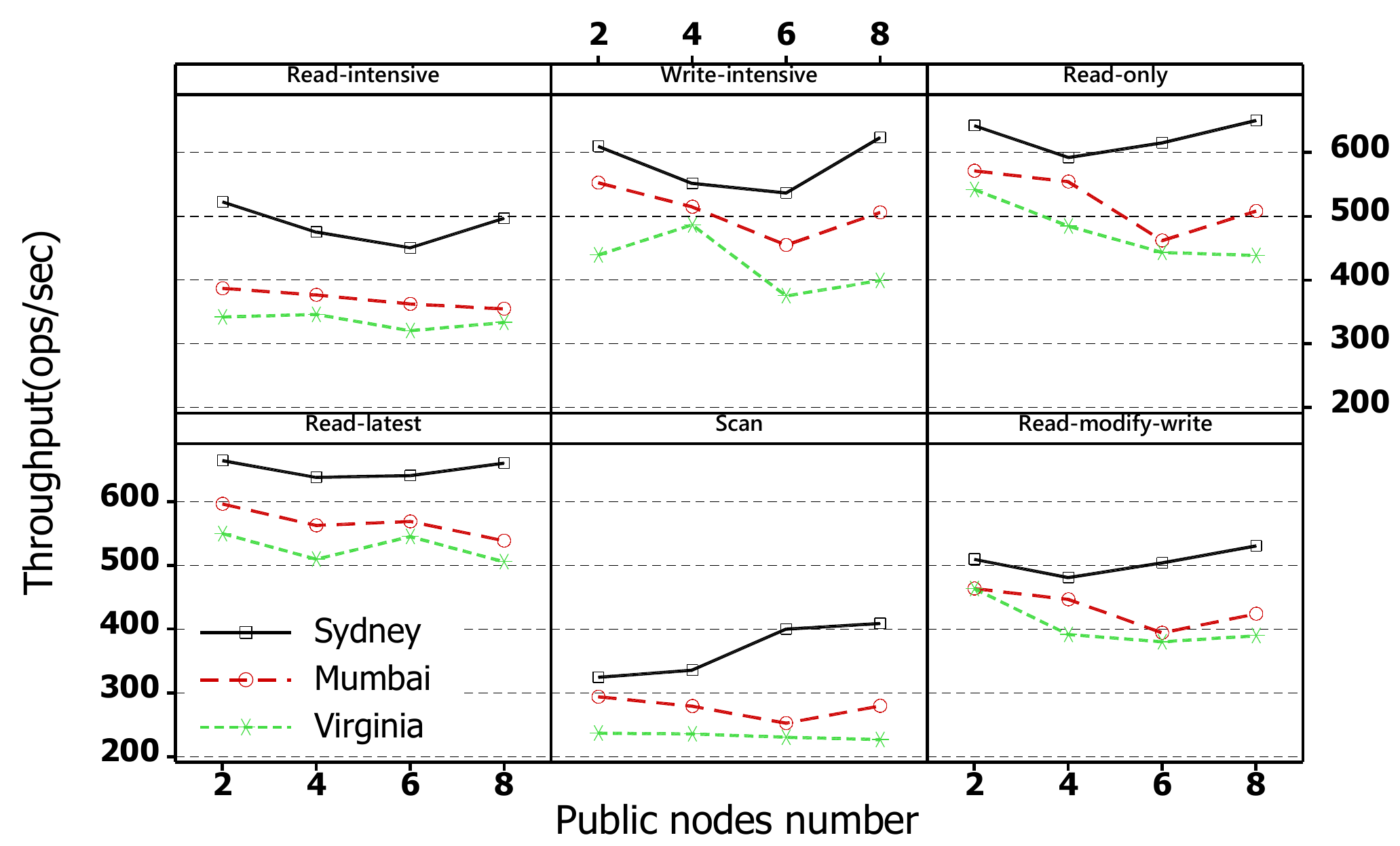}}\\
	\subfloat[Riak]{\label{figur:riak-hsca}\includegraphics[width=0.5\textwidth]{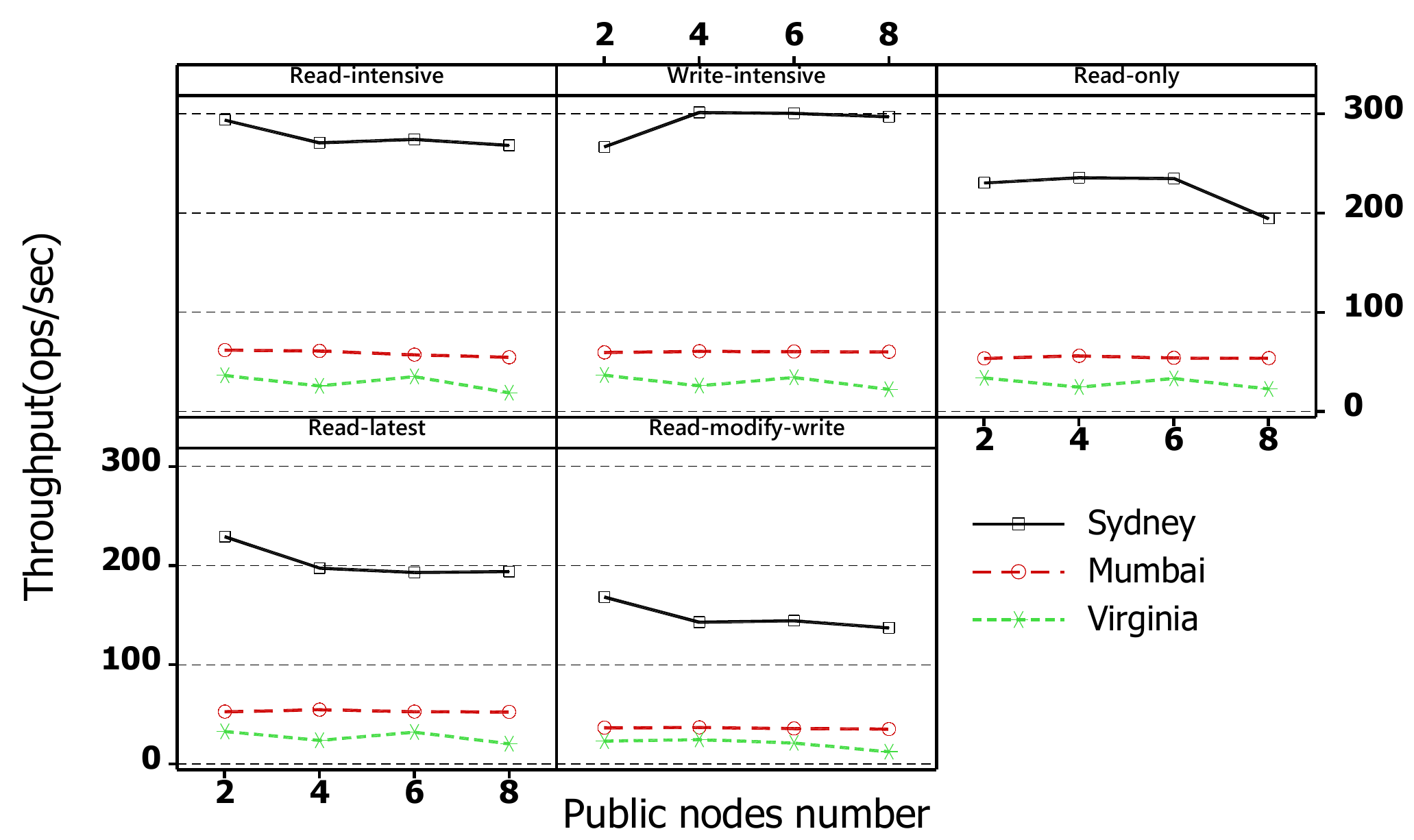}}
	\subfloat[Couchdb]{\label{figur:couchdb-hsca}\includegraphics[width=0.5\textwidth]{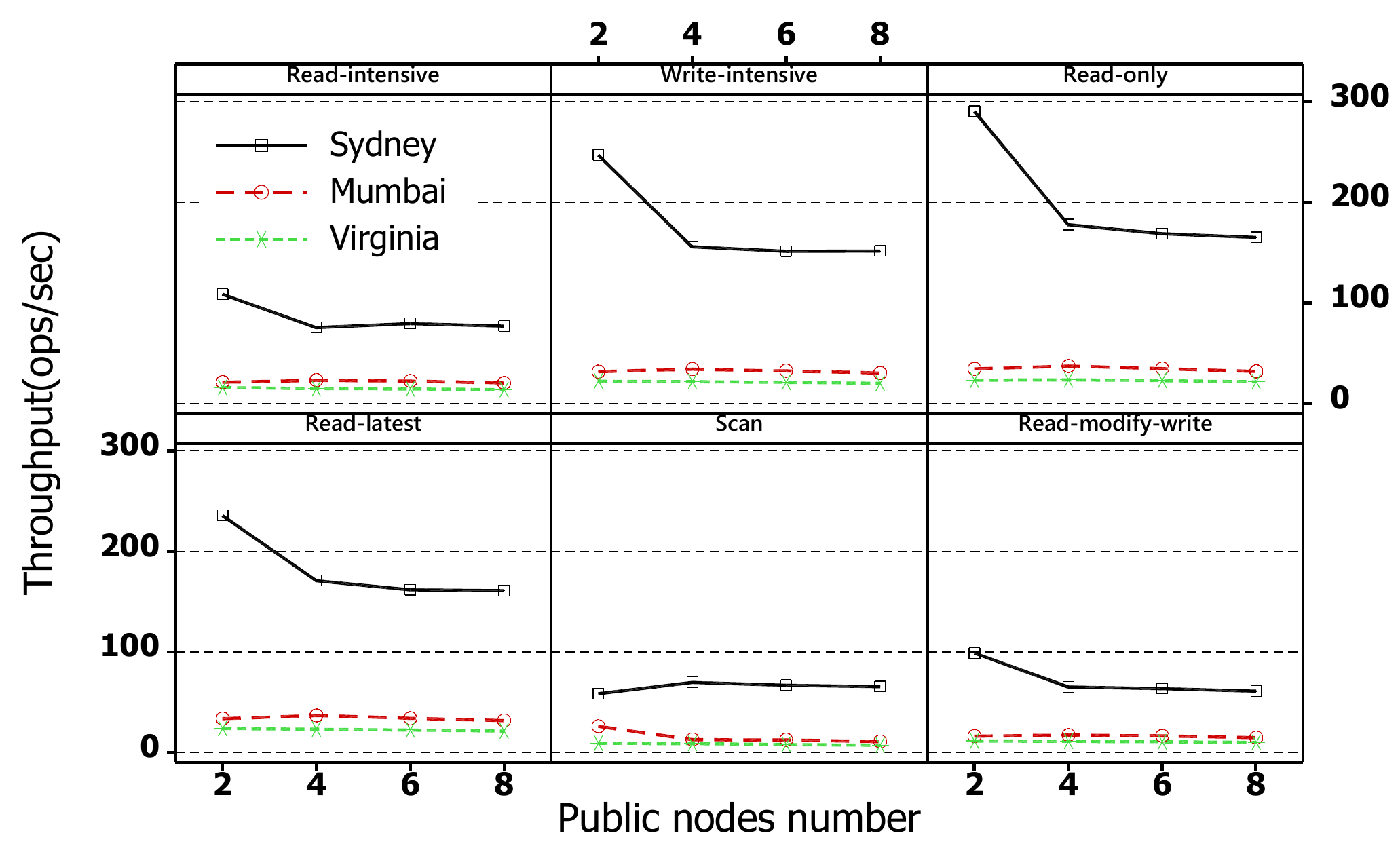}}\\
	\subfloat[Redis]{\label{figur:redis-hsca}\includegraphics[width=0.5\textwidth]{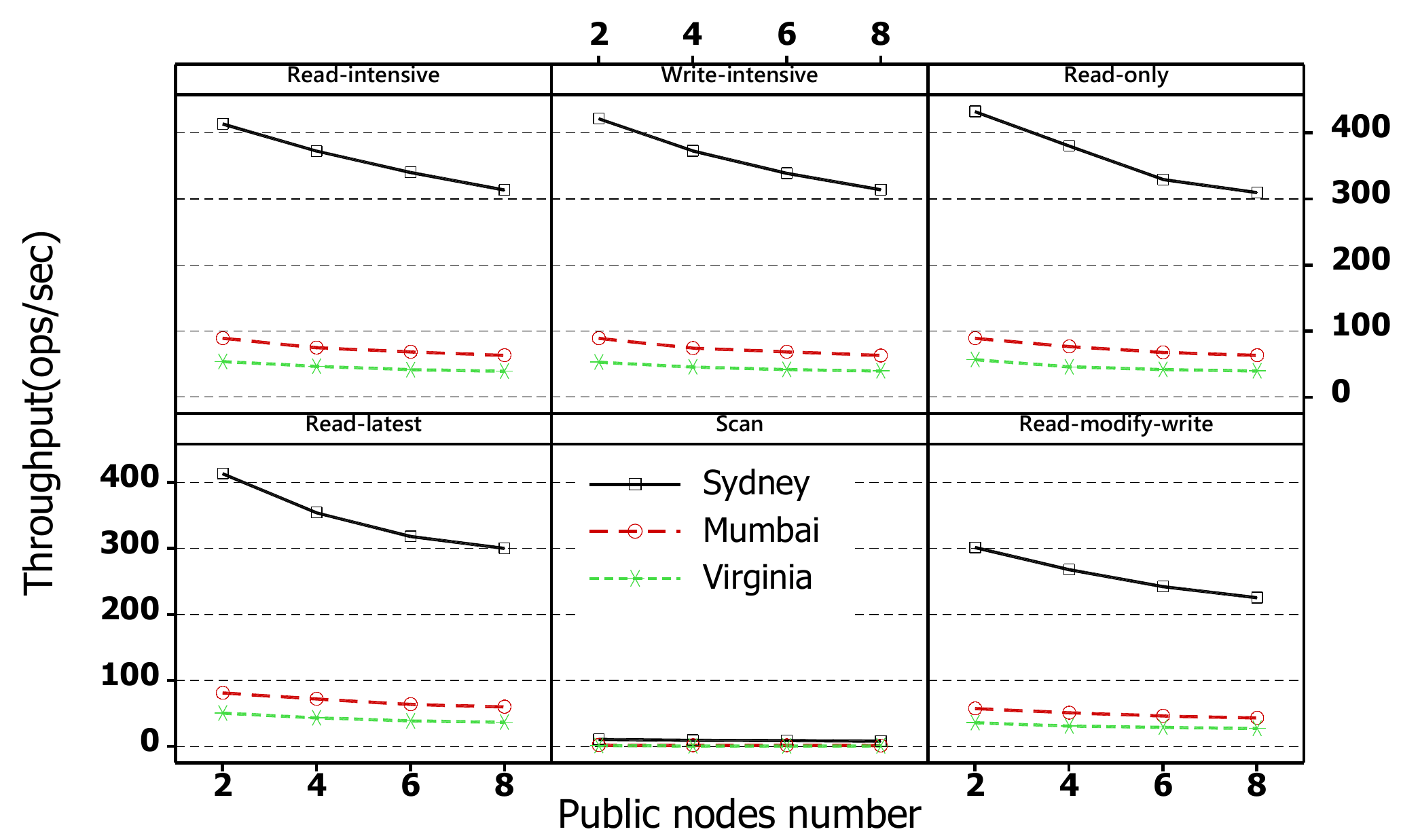}}
	\subfloat[MySQL Cluster]{\label{figur:mysql-hsca}\includegraphics[width=0.5\textwidth]{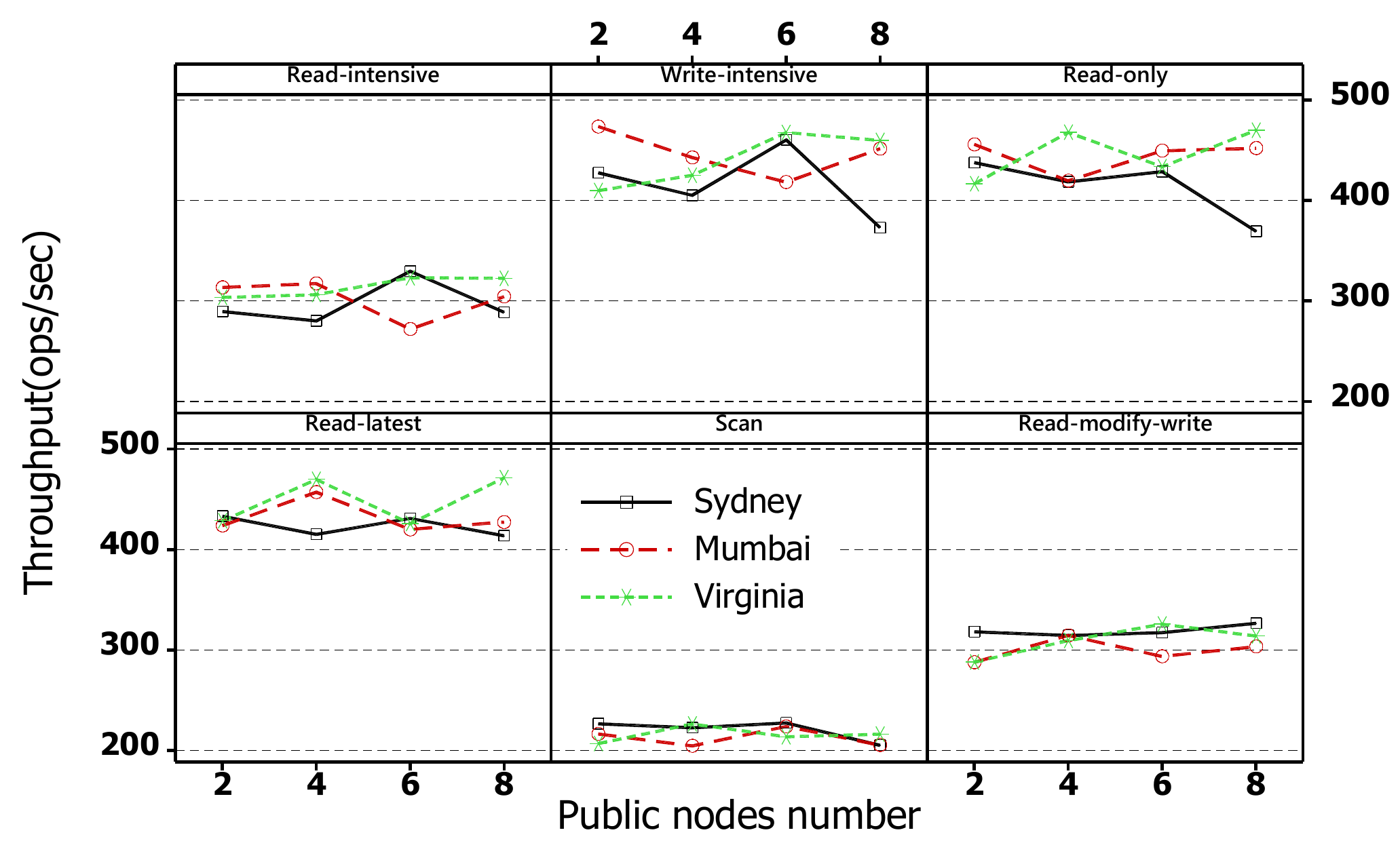}}
	\caption{Horizontal scalability  of MongoDB, Cassandra, Riak, Couchdb, Redis, and MySQL Cluster in Sydney, Mumbai, and Virginia regions.  Axis X represents the number of VM deployed  in a public cloud datacenter.}
	\label{fig:hsca}
\end{figure*}

This section presents the results for RQ2.1, which is related to horizontal scalability of six distributed databases running on a hybrid cloud. In this set of experiments, we investigated the effects of horizontal scalability on the throughput of distributed databases\footnote{Note that in this work, we intended to evaluate horizontal scalability as cloud bursting happened. Thus, we increased the exploitation of more VMs in the public cloud datacenter, not in the private one.}. Horizontal scalability means that adding more computing nodes to the resource pools. Thus, in this experiment, we varied the number of VMs from 2 to 8 with the size of Standard\_B1m (1 vCPU, 2 GB RAM, and 4 GB SSD) in a public cloud datacenter. In compliance with the definition of a hybrid cloud, we also deployed a small VM instance (1 vCPU, 2GB RAM, 10GB HDD) in  private cloud datacenters.    

As shown in Fig. \ref{figur:cass-hsca}, the throughput of Cassandra reduces to half or even more for most of workloads as the number of VMs increased from 2 to 8. This reduction was less for Mumbai and Virginia regions. Thus, bursting more nodes to a public cloud datacenter did not necessarily improve the performance of database because the more VMs there are, the more communication they need with each other to conduct read and write operations in Cassandra (see Section \ref{sec:vmpac}). The throughput of Riak and Redis remained fairly constant as more VMs exploited into the public datacenter in Mumbai and Virginia regions. In Sydney region, Riak slightly incremented in throughput for the write intensive workload, while for Redis's throughput decreased by a factor of $\frac{1}{4}$ (Figs. \ref{figur:riak-hsca} and \ref{figur:redis-hsca}). In the same region, Couchdb's throughput initially reduced with the increment of 2 VMs to 4 VMs, and then its throughput stayed at a constant level (see Fig. \ref{figur:couchdb-hsca}).

In contrast to the four discussed databases above, we observe that an increment in throughput of MongoDB and MySQL in some cases as the number of VMs increased in all regions (Figs. \ref{figur:mongo-hsca} and \ref{figur:mysql-hsca}). However, this performance trend did not increase consistently for both databases since read and write operations transferring over Wide Area Network (WAN) incurred high latency. 

\textbf{Summary}: This set of experiments shows that adding more nodes in a public cloud datacenter to conduct read and write operation over WAN cannot guarantee better throughput especially for RAM-based databases (e.g., Redis) and for quorum-based databases (e.g., Cassandra and Couchdb).

\begin{figure*}[h!]
	\centering
	\subfloat[Cassandra]{\label{figur:cass-vsca}\includegraphics[width=0.5\textwidth]{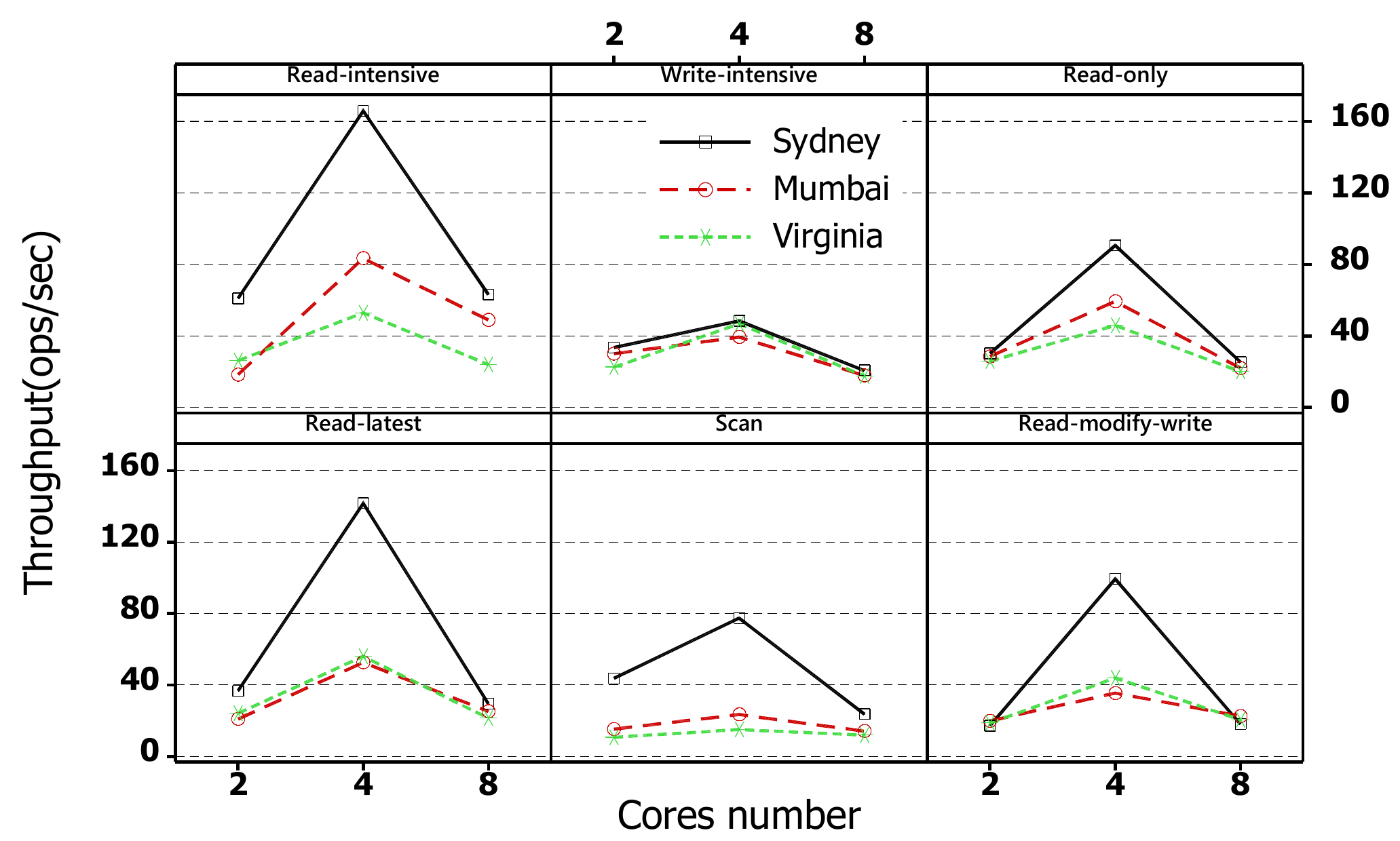}}
	\subfloat[MongoDB]{\label{figur:mongo-vsca}\includegraphics[width=0.5\textwidth]{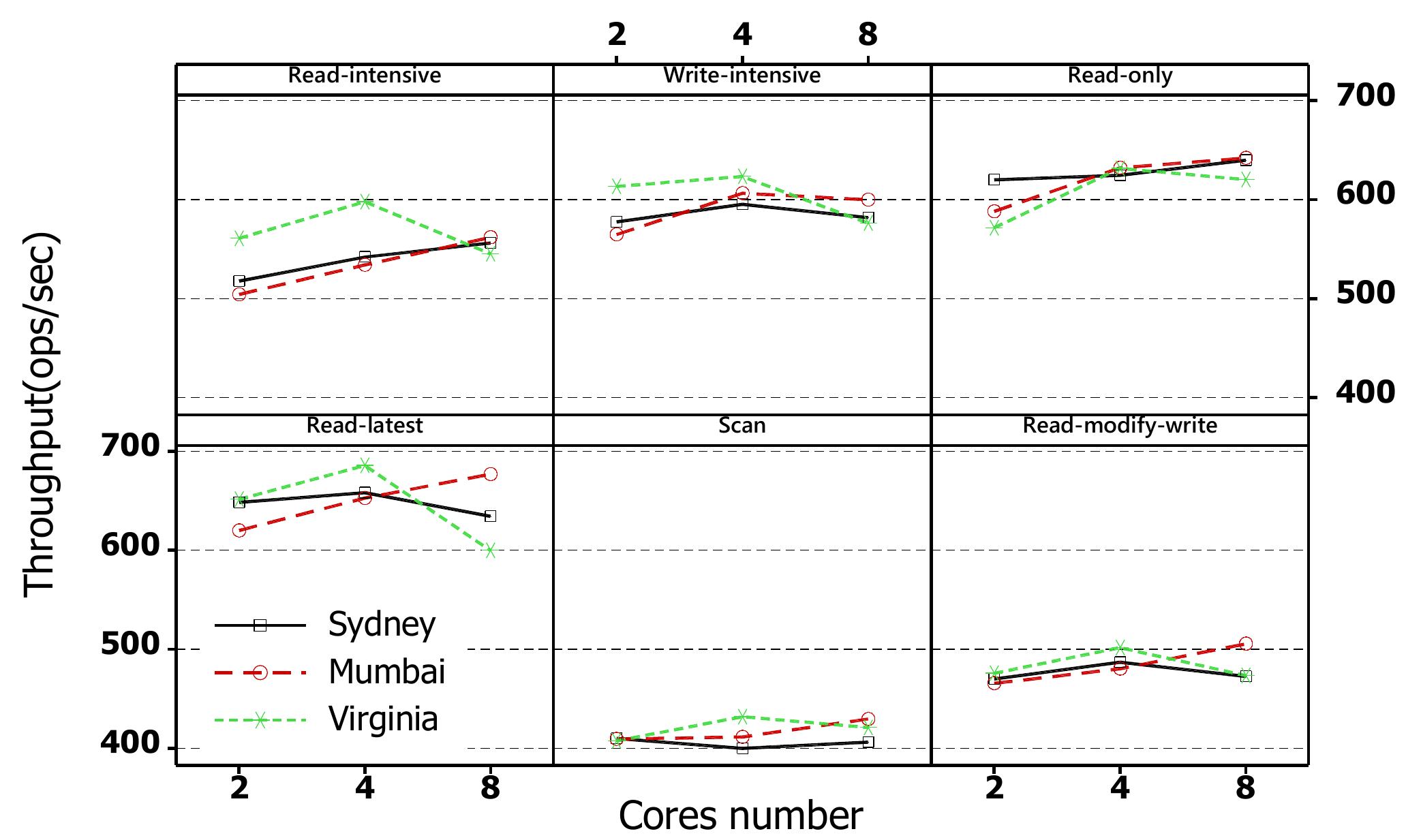}}\\
	\subfloat[Riak]{\label{figur:riak-vsca}\includegraphics[width=0.5\textwidth]{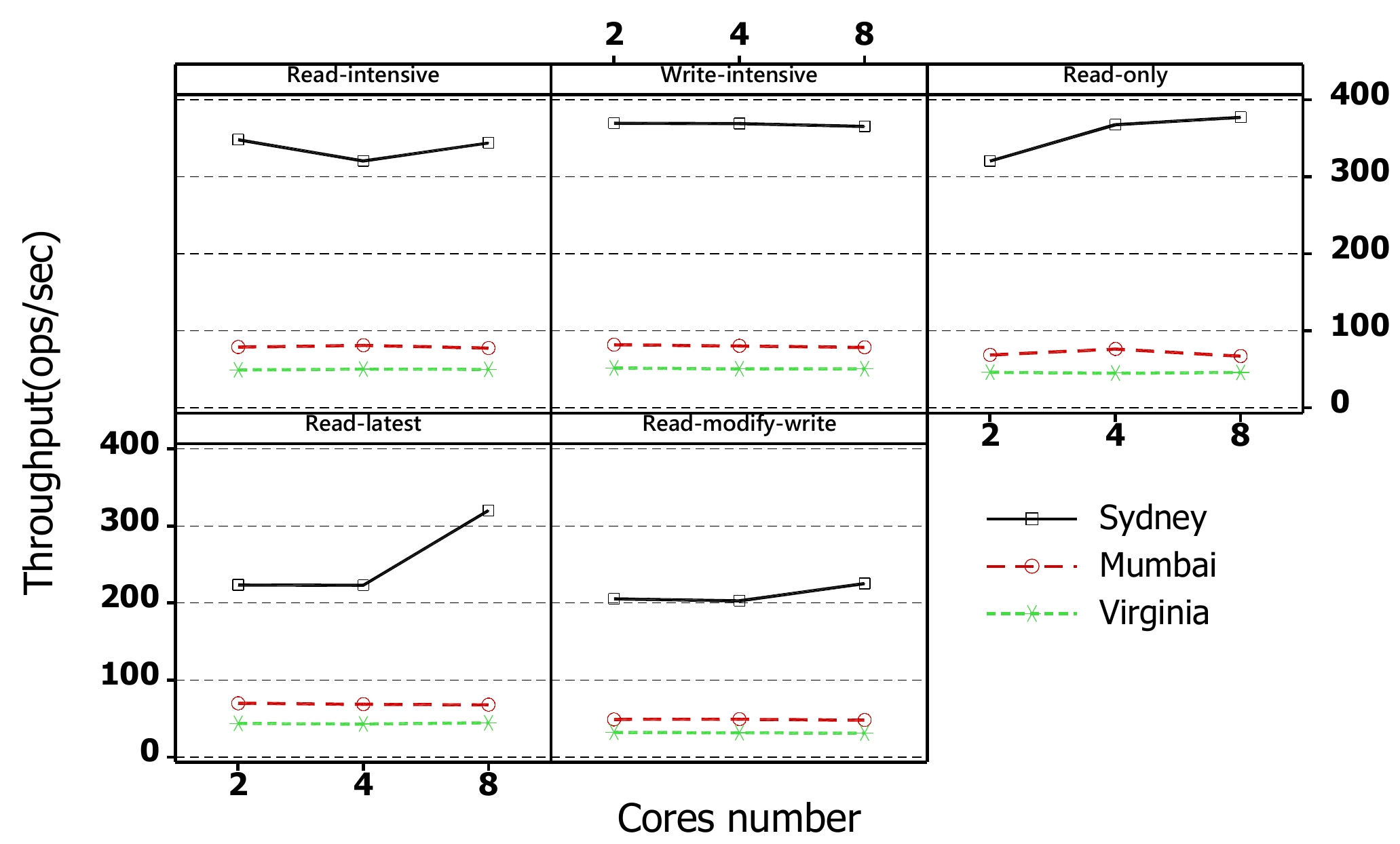}}
	\subfloat[Couchdb]{\label{figur:couchdb-vsca}\includegraphics[width=0.5\textwidth]{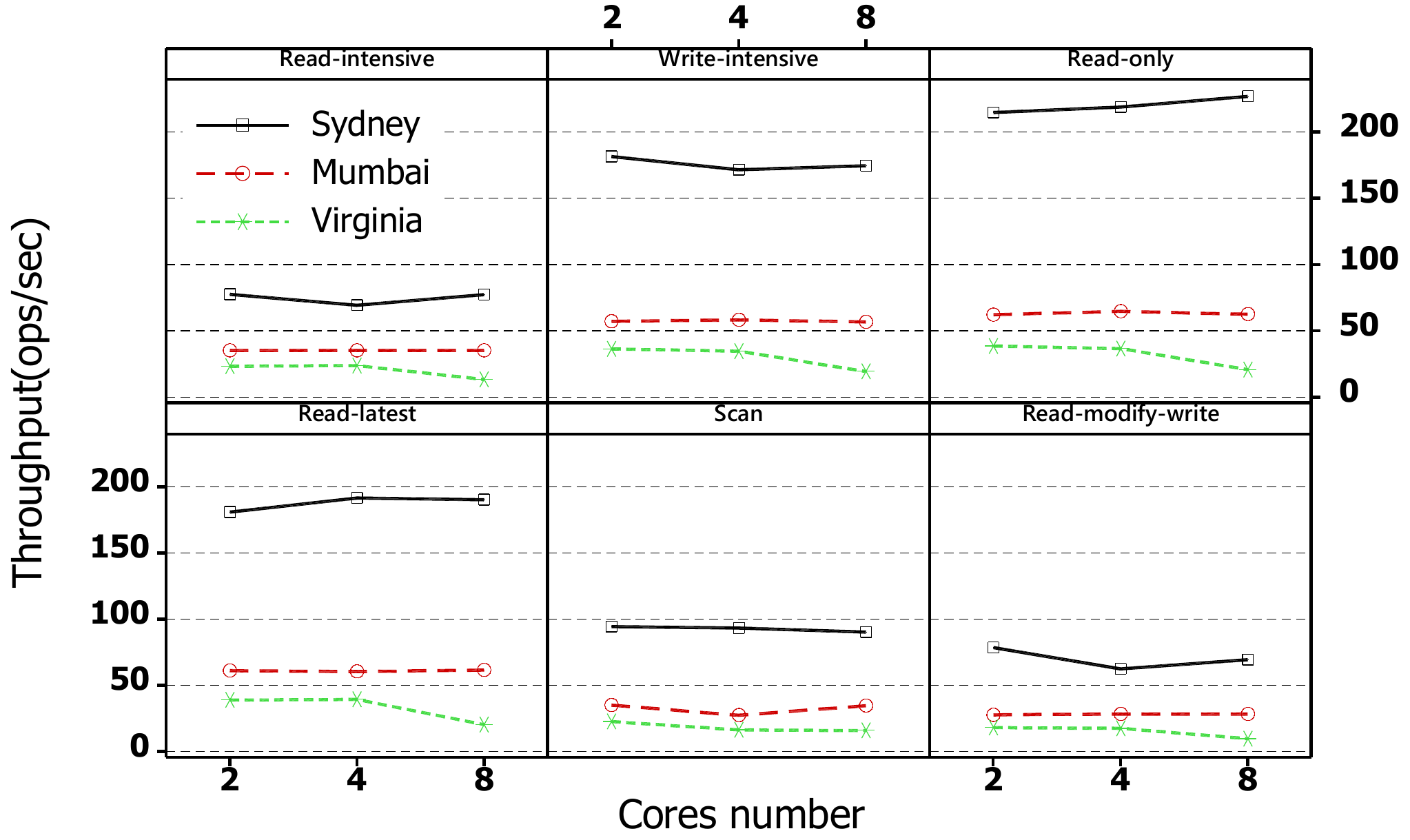}}\\
	\subfloat[Redis]{\label{figur:redis-vsca}\includegraphics[width=0.5\textwidth]{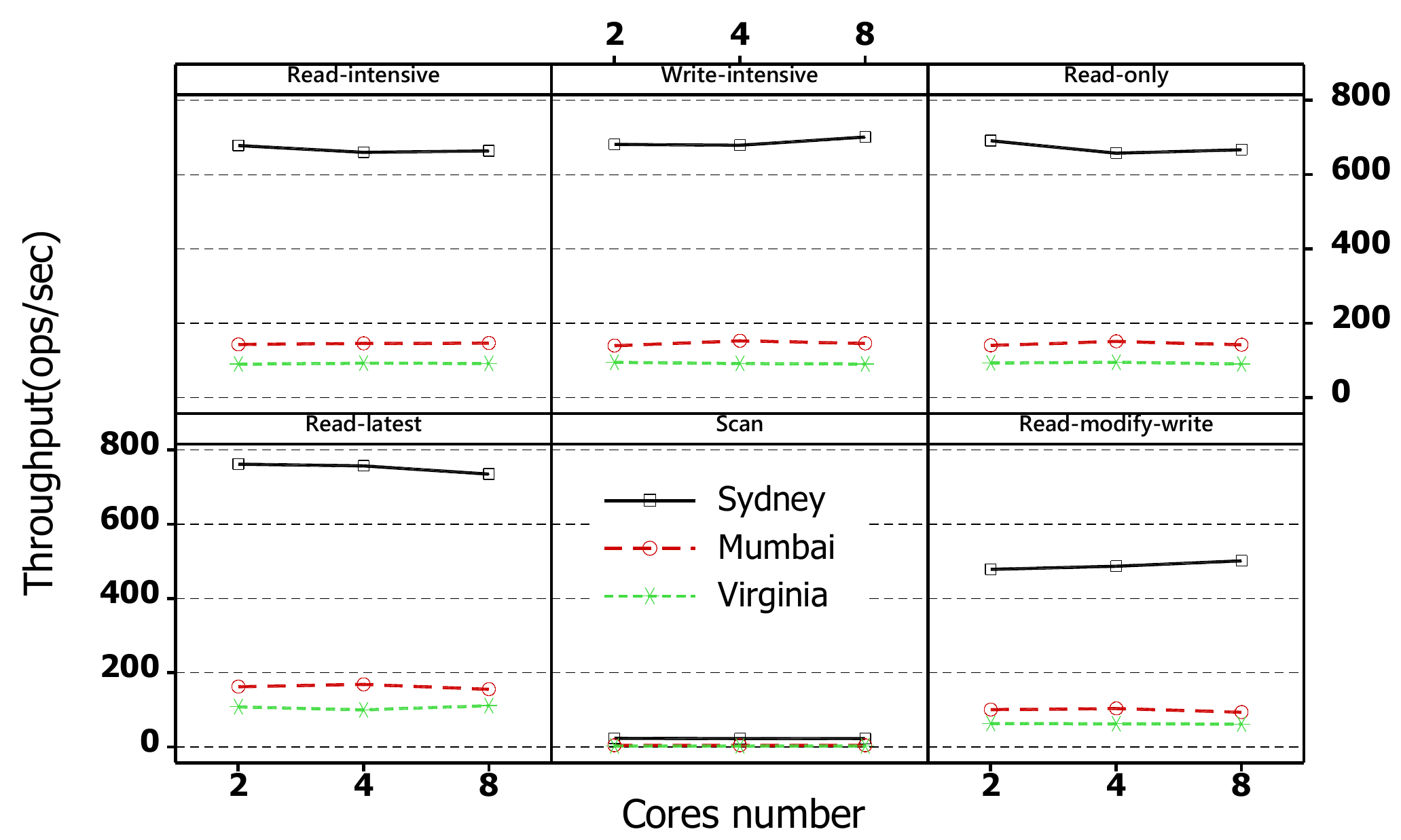}}
	\subfloat[MySQL Cluster]{\label{figur:mysql-vsca}\includegraphics[width=0.5\textwidth]{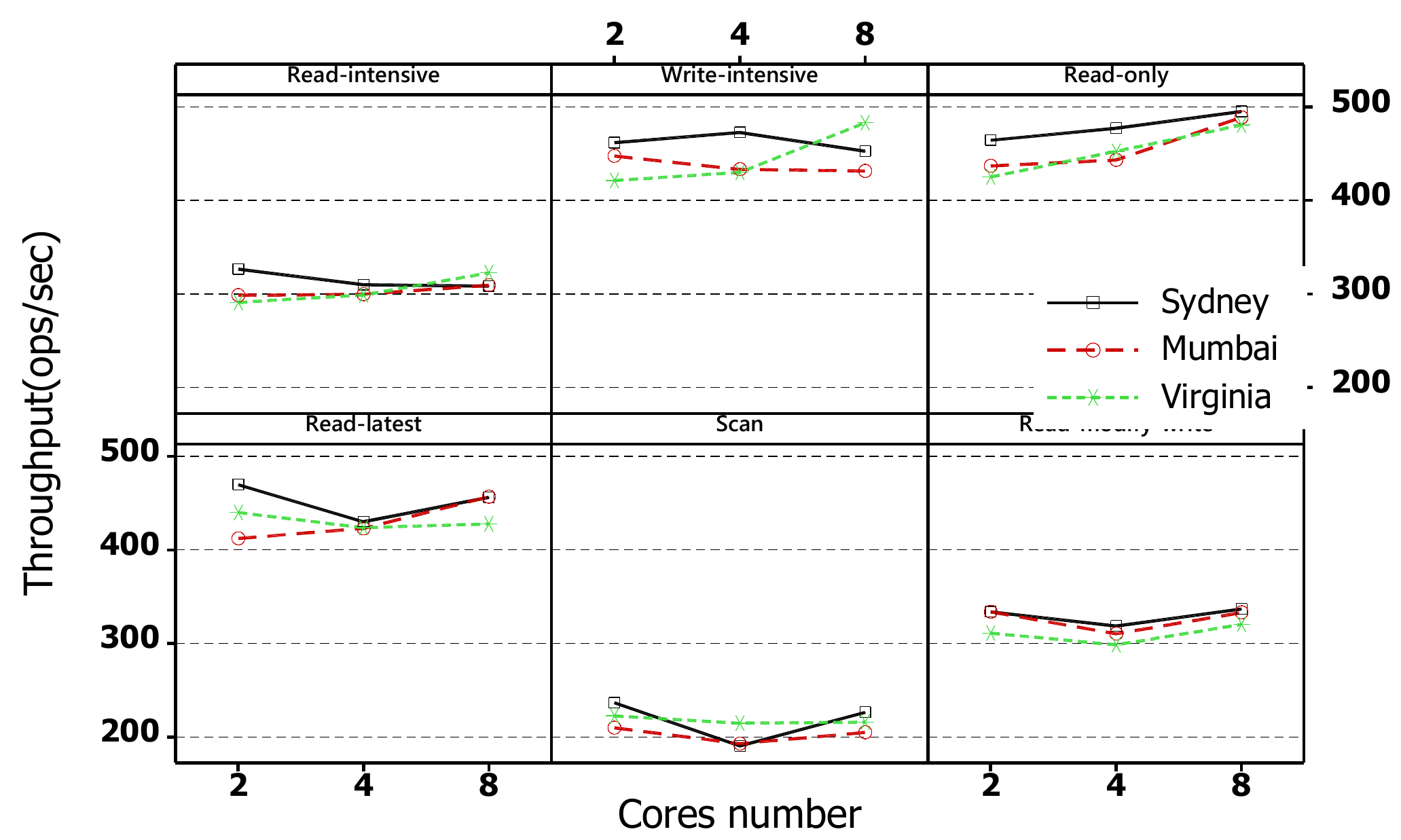}}
	\caption{Vertical scalability of MongDB, Cassandra, Riak, Couchdb, Redis, and MySQL Cluster in Sydney, Mumbai, and Virginia regions. Axis X represents the number of cores for a VM deployed in a public cloud datacenter.}
	\label{fig:vsca}
\end{figure*}

\subsubsection{Vertical Scalability Evaluation}
This set of experiments answers RQ2.2, where the vertical scalability of six distributed databases has been evaluated by varying the number of cores for the VM deployed in the public cloud datacenter in Sydney, Mumbai, and Virginia regions. In these experiments, we exploited two small VMs in private infrastructure resources and one VM in the public cloud datacenter with 2, 4, and 8 cores. Note that we ran three VMs because MySQL and Redis required at least three VMs for cluster configuration. 

Fig. \ref{figur:cass-vsca} shows that the throughput of Cassandra increased 2-4 times when the number of cores varied from 2 to 4 for the VM deployed in Sydney region. Likewise, we observe that the same increment of 1.5-2 times for the VM running in Mumbai and Virginia regions. As depicted in Fig. \ref{figur:riak-vsca}, selection of a larger VM in Sydney region raises the throughput of Riak by 30\% for the read-only and read-latest workloads. The throughput of Redis and Couchdb remained constant or slightly decreased as a larger VM in Mumbai and Virginia regions was exploited (Figs. \ref{figur:redis-vsca} and \ref{figur:couchdb-vsca}). In contrast, a larger VM selection in Sydney region raised the throughput of Couchdb by 10\% -30\% for the read-only and read-latest workloads. 

Figs. \ref{figur:mongo-vsca} and \ref{figur:mysql-vsca} illustrate the effect of a larger VM exploitation on throughput of MongoDB and MySQL. Apart from scan, both databases obtained increment in their throughput in the range of 5\%-20\% in Sydney region, though this incremental, the trend was not linear. This is due to these two databases were unstable in performance over WAN. In contrast to MongoDB, MySQL demonstrated a slight decrement of ~5\% as the number of cores changes from 4 to 8 in Virginia region. 

\textbf{Summary:} The results from this set of experiments demonstrate that deploying a larger VM in a public cloud datacenter can improve the throughput of all databases except Cassandra for most workloads in Sydney region. In this region, Cassndra's throughput significantly increased as the number of cores raised from 2 to 4, and then reduced when the number of cores went up from 4 to 8.  

\subsubsection{VM Packing Evaluation}\label{sec:vmpac} 

\begin{figure*}[h!]
	\centering
	\subfloat[Cassandra]{\label{figur:cass-vmpac}\includegraphics[width=0.5\textwidth]{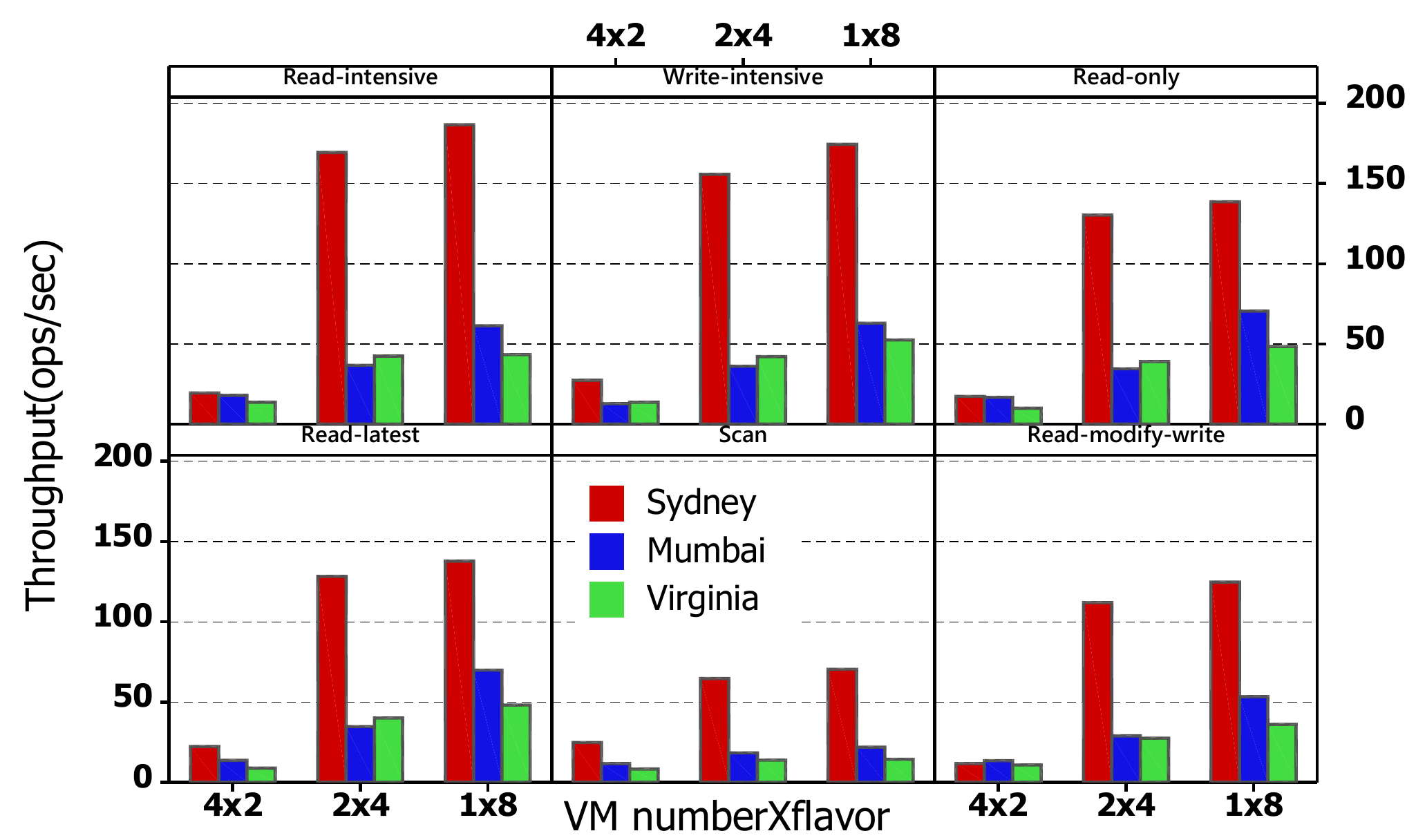}}
	\subfloat[MongoDB]{\label{figur:mongo-vmpac}\includegraphics[width=0.5\textwidth]{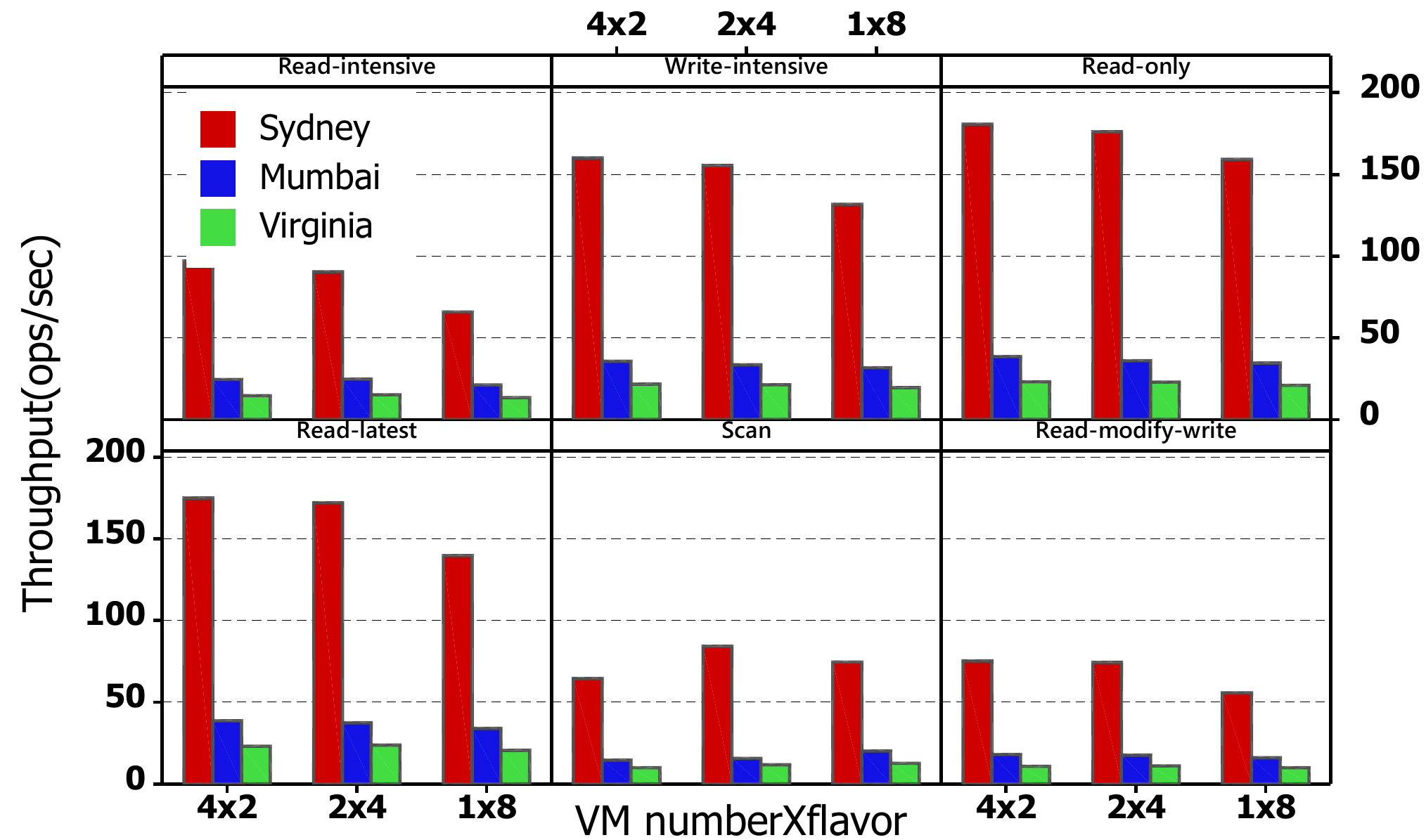}}\\
	\subfloat[Riak]{\label{figur:riak-vmpac}\includegraphics[width=0.5\textwidth]{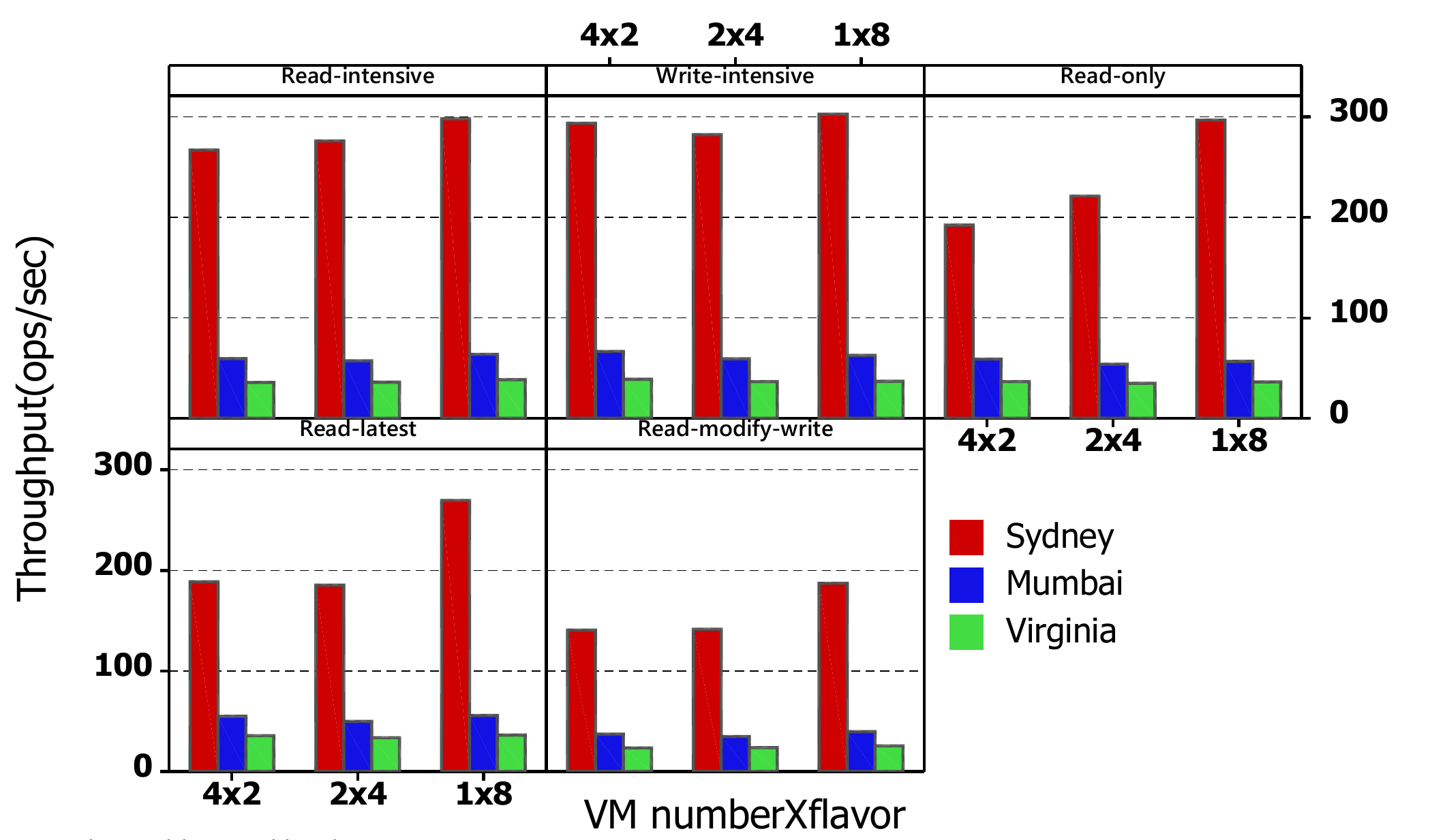}}
	\subfloat[Couchdb]{\label{figur:couchdb-vmp}\includegraphics[width=0.5\textwidth]{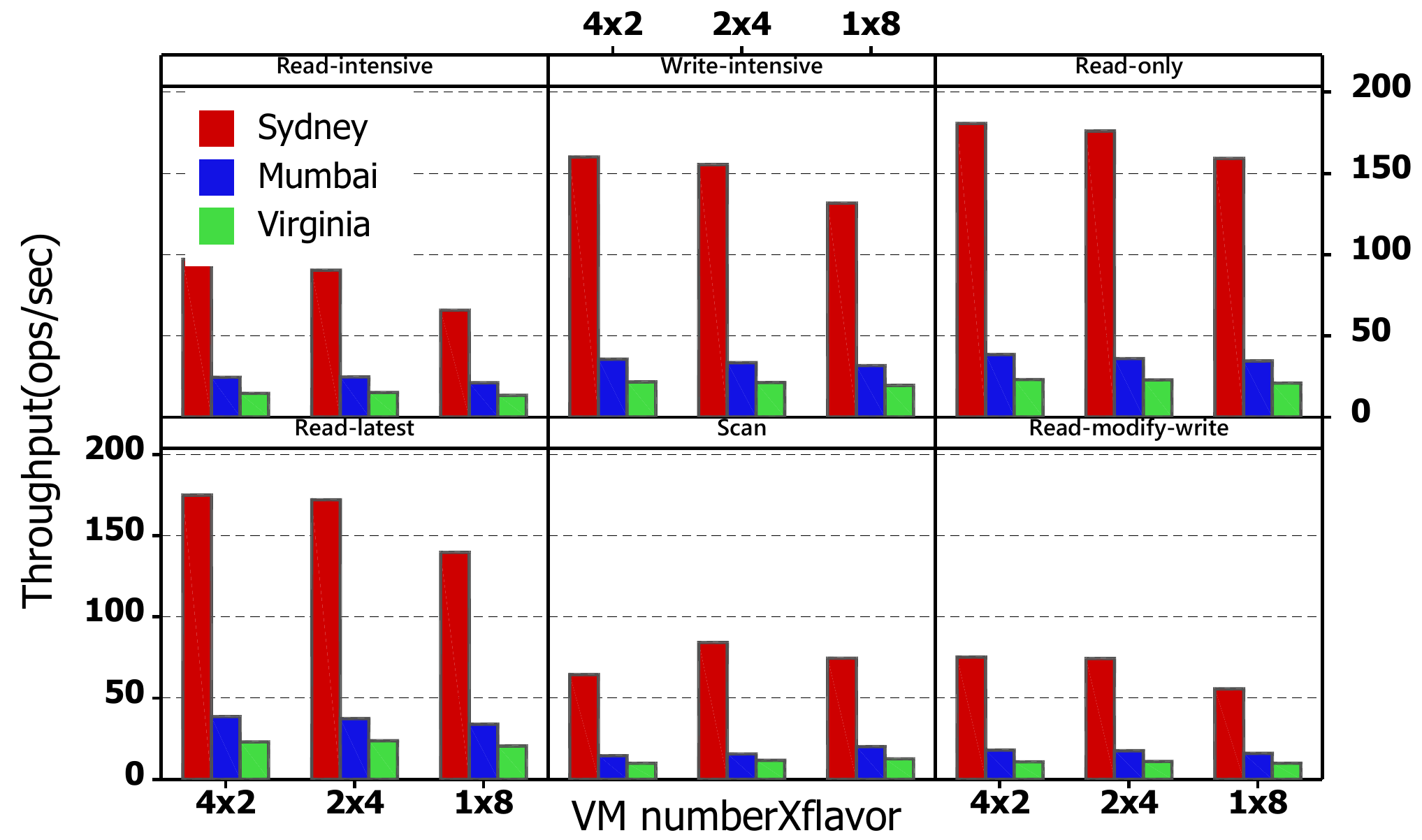}}\\
	\subfloat[Redis]{\label{figur:redis-vmpac}\includegraphics[width=0.5\textwidth]{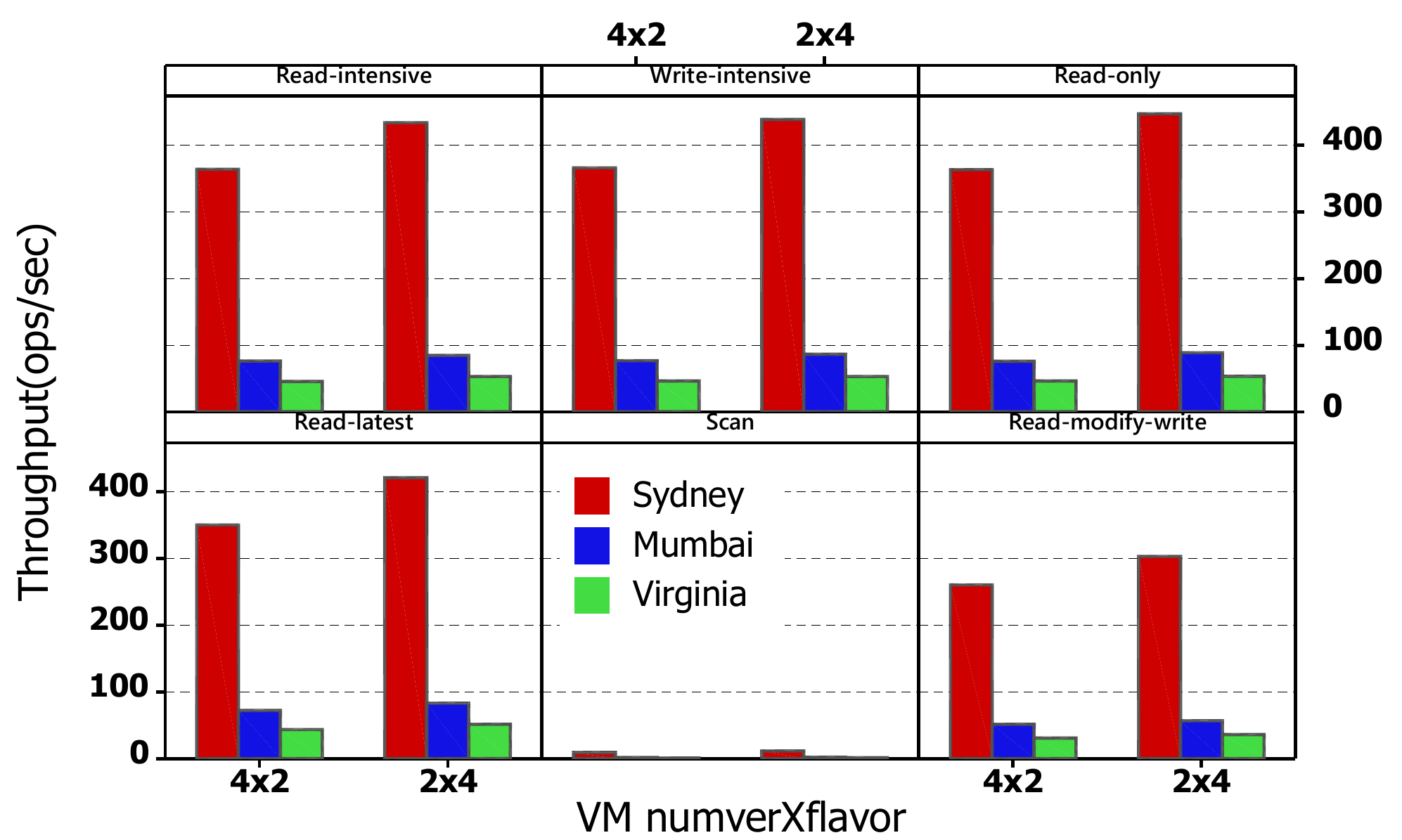}}
	\subfloat[MySQL Cluster]{\label{figur:mysql-vmpac}\includegraphics[width=0.5\textwidth]{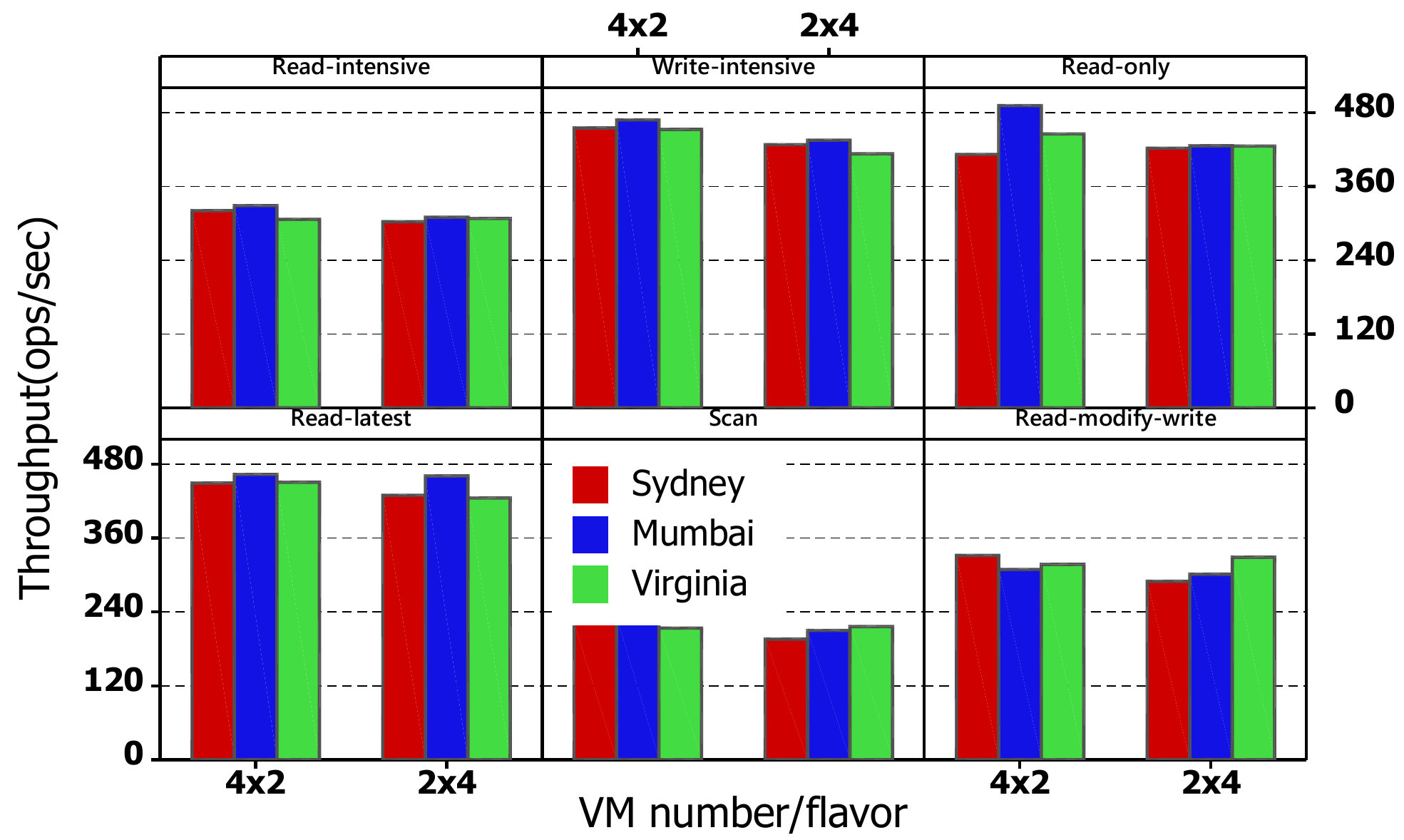}}
	\caption{The effects of VMs number vs. VM size  on throughput for MongDB, Cassandra, Riak, Couchdb, Redis, and MySQL Cluster in Sydney, Mumbai, and Virginia regions. Value $n \times m$  in axis X represents $n$ VMs with size of $m$ cores in a public cloud datacenters.}
	\label{fig:vmpac}
\end{figure*}

This section answers RQ3 related to the performance evaluation of distributed databases as the number and flavor of VM change. In this set of experiments, we fixed one small VM instance in OpenStack (see Table \ref{tab:infras}) and varied the number of VM instances (1,2,4) and their flavor (Standard\_B2ms (2 cores), Standard\_B4ms (4cores), and Standard\_B8ms(8cores)) in the public cloud datacenter so that for each setting VMs number $times$ VMs flavor is a constant value (i.e., 8 cores). This set of experiments helped us to determine whether to select a larger VM in the number of cores or more VM instances with less cores number. Thus, as shown in Fig. \ref{fig:vmpac}, we have 3 configuration settings of $nXm$, where $n$ and $m$ respectively represents the number and flavor of VMs\footnote{Note that in this set of experiments, we have not considered configuration setup of 1X8 for Redis and MySQL because these databases require at least 3 VMs/nodes.}. 

Fig. \ref{figur:cass-vmpac} shows that as the number of cores for a VM increased from 2 to 8, the throughput of Cassandra significantly boosted. In Sydney region, Cassandra's throughput increased by a factor of 8 for read-modify-write, followed by read-intensive (7.2 times), and read-only (7 times). Similarly, for two other regions, we observe an incremental trend for throughput as larger VMs are deployed for all workloads (apart from scan) by a factor of at most 4. Like Cassandra, Riak's throughput increases as the number of cores in a VM increased especially for the read-only and read-latest workload in Sydney region. As shown in Fig. \ref{figur:riak-vmpac}, these two workloads obtained a 30\% improvement of throughput with the increasing number of cores. However, for two other regions, the throughput of Riak remained fairly constant with the increased number of cores. This might be the latency over WAN dominates the latency between VMs in the same cluster. The reason behind such behaviour arising from Cassandra and Riak can be that both databases leveraged a kind of quorum-based mechanism to provide data consistency, which necessitated more communication between VMs. As can be seen in Fig. \ref{fig:vmpac}, as the configuration setting varies from 4X2 to 2X4, Redis's throughput improved at most 10\% for all workload (except scan). Thus, Redis has better performance when it exploited 2 VMs with 4 cores each rather than 4 VMs with 2 cores each. This is due to larger VMs offer bigger RAM, which suits for memory-based databases like Redis.    
          
In contrast to the discussed databases above, Figs. \ref{figur:mongo-vmpac}. \ref{figur:couchdb-vmp}, and \ref{figur:mysql-vmpac} show that the throughput of MongoDB, Couchdb, and MySQL with the default setting \cite{Mansouri2020} did not vary as the configuration setting changed from 4X2 to 1X8. This implies that these databases do not require heavy communication between VMs to conduct read and write operations. No reason to have such communication since  MongoDB provides full replication with eventual consistency, Couchdb offers a local quorum-based mechanism for consistency, and MySQL supports strong consistency between VMs in the same \textit{data node group}. It is worth mentioning that the throughput of MongoDB, Couchdb, and MySQL was not affected as configuration setting (4X2 - 1X8) changed for Mumbai and Virginia regions.  

\textbf{Summary:} The result demonstrate that quorum-based (i.e. Cassandra), RAM-based (i.e., Redis), and Riak  databases can effectively leverage larger VMs in terms of cores and RAM size rather than more VMs with less Cores and RAM size to improve their performance. While the remaining databases cannot exploit larger VMs to improve performance in Mumbai and Virginia regions.

\subsubsection{Lessons Learnt}
From the deployment perspective, construction and deconstruction of a cluster nodes in a private cloud datacenter take less time in comparison to the one in a public cloud datacenter irrespective of its region. The reason might be a public cloud consists of thousands of racks including hundreds of servers, while in our private clouds consists of one rack including two servers. Moreover, as ref elected in the results, we observed that MongoDB and MySQL require less time to run YCSB workload with much less deployment failures. 

In terms of throughput performance under default setting, MongoDB is a clear winner throughout our databases under test in this work. Long distance between the private and public cloud datacenters affects its throughput if more than half of VMs in the cluster configuration are exploited in a public cloud datacenter. Precisely, for the longest distance (i.e., Virginia region), the throughput of MongoDB reduces to at most half with full bursting configuration ($1\_7$) in comparison to the throughput for non bursting $(8\_0)$. We believe that the reason behind such performance of this database is full replication and eventual consistency, which makes close data to data requester (here the broker VM in OpenStack).  

MySQL stands at the second rank in this criterion since it guarantees strong consistency between nodes in the same \textit{data nodes group} rather than guaranteeing Geo-graphical strong consistency required transfer data over WAN. Since the size of data nodes group is equals to two (default value for replicas number), data is close to data requester in all cluster configuration except ($1\_7$). Surprisingly, even for the cluster configuration ($1\_7$), MySQL gains almost the same throughput. This might be that the operations are conducted on the local VM (hosting MySQL) in OpenStack, where data requester and data host are in the same cloud. It might be difficult to achieve such throughput for MySQL if clients issuing data request are geographically located, the number of replicas increases, and  YCSB+T \cite{Dey2014} is leveraged. YCSB+T provides dependency between data which requires strong consistency in case of MySQL. The effect of such aspects should be investigated to better understand the functionality of MySQL at long distance (i.e., Virginia and Mumbai).  

Cassandra deployment demonstrates the most fluctuation in throughput as it is significantly impacted  by distance. This evaluation also confirmed  \cite{sha2014} for three web-based workloads. However, this database provides the best performance for the Read-intensive workload in Sydney region and exposes high throughput for Read-only and Read-latest workloads with hybrid cloud configuration of ($4\_4$) in Sydney region. Based on the obtained results, the default setting for data replication and consistency mechanism should be adapted to achieve better throughput as the distance between data requester node/VM and data host node increases.  

The throughput of Riak, Redis and Couchdb follows semi-parabolic trend line, in which the throughput of these databases reduces as hybrid configuration setup changes from ($8\_0$) to ($4\_4$), and then their throughput gradually increases when more than half of the nodes are exploited in the public cloud in Sydney region. This evaluation demonstrates that the performance of these databases depends on the density of nodes/VMs located in one cloud when the private and public clouds are at close distance. The comparison between three databases shows that Redis outperforms Riak, which, in turn, exposes better throughput in comparison to Couchdb for almost all workload (apart from the scan workload) in Sydney region. In contrast, in Mumbabi and Virginia regions, the throughput of these databases (especially Redis) is significantly impacted as cloud bursting happens.   

%For other databases, Cassandra, Riak, Redis and Couchdb, long distance (e.g., Mumbai and Virginia) significantly affects their performance. In contrast, these databases, apart from Cassandra, their performance  linearly reduces  for half of VM (i.e., $4\_4$) bursting into the public cloud datacenter in Sydney region and then stays at a constant level. 

With respect to horizontal scalability, the results show that MongoDB and MySQL can improve their performance as a greater number of VMs exploited in the Sydney region though this incremental trend is not linear. By contrast, in the Sydney region, the throughput of Cassandra significantly cuts by a factor of $\frac{1}{3}$ as the number of VMs in the public cloud changes from 2 to 8. Likewise, for Riak by a factor of $\frac{1}{4}$. Thus, the results demonstrate that the throughput of Riak, Cassandra and in some cases for MySQL and MongoDB cannot improve as the number of bursting VMs increases. This might be a case because distance does not allow to saturate the link between the private and public clouds  while more data are located far away from the node/VM issued operations.

%For other databases, changes (reduction/increment) on performance is insignificant. Thus, high latency over WAN hinders a well horizontal scalability for these databases. 
With respect to vertical scalanility, MongoDB, MySQL, Riak and Couchdb demonstrate increment in throughput but not significant. For Cassnadra, initially the throughput goes up as the number of cores changes from 2 to 4, and then reduces when the cores number increases from 4 to 8. What we realized from this set of experiments, long distance dominates the performance of these databases with respect to vertical scalability. It seems valuable to investigate the scalability of these databases on shorter distance, namely at distance of several KMs rather than hundreds of KMs.   

%In terms of vertical scalability, most databases especially Cassandra, MongoDB, and Riak databases exhibit increase in performance in Sydney region. For other regions, 
In the last set of experiments, we evaluated the impact of VMs number vs. VMs cores (called \textit{VM packing}) on the performance of modern distributed. This evaluation exposes  latency between VMs in the same cloud, and replacing this latency with one between cores in the same VM. The results demonstrate that less VMs with more cores has significant impact on the performance of Cassandra and then Riak in the Sydney region. This might be that these databases require more communication between VMs. Also, VM packing improves the performance of RAM-based databases such as Redis. In this work, we considered VM packing on the public cloud side since we intended to investigate the impact of distance on it. Otherwise, it is useful to explore the impact of VM packing locally (i.e., on the side of private cloud) on the performance of distributed databases.   

%In contrast, this trade-off has insignificant effect on the performance of Mongo and MySQL. 

\section{Conclusions and Future Work}\label{sec5}
In this paper, we have conducted an extensive evaluation of performance, scalability and VMs size vs. VMs number for six modern and widely used databases (i.e., MongoDB, Cassandra, Riak, Couchdb, Redis, and MySQL Cluster). Unlike the previous studies, we have evaluated these databases on a hybrid cloud spanning on-premises infrastructure resources and public cloud datacenters in the Sydney, Mumbai, and Virginia regions. The selection of these regions reflects the effect of the distance between a pair of private and public datacenters on performance and scalability of these databases. We have observed that MongoDB obtains the first rank among these databases in throughput since it leverages full replication with the eventual mechanism for consistency. MySQL Cluster comes after MongoDB from performance perspective since it uses strong consistency between data nodes in the same \textit{data nodes group}, which avoids guaranteeing geographically strong consistency. In contrast to these two databases, the long distance (i.e., Virginia and Mumbai regions) degrades the performance of other databases (i.e., Cassandra, Riak, Couchdb, and Redis). At close distance (i.e., Sydney region), Riak, Redis and Couchdb exposes at most 50\% reduction in throughput as half of the VMs (i.e., $4\_4$) are burst into a public cloud, and then their performance gradually increases as more than half of the VMs are deployed in the public cloud.

In our experiments, we have observed adding more VMs in a public cloud datacenter can help MongoDB and MySQL to improve their throughput especially in the Sydney region. In contrast, exploiting the larger VM instances in a public cloud datacenter increases the throughput of all the databases (except Cassandra) for the most workloads in the Sydney region. For Cassnadra, in all regions, its throughput increases as the number of cores increases from 2 to 4, and then reduces when the number of cores rises from 4 to 8.  Thus, distance also affects on vertical and horizontal scalability of these databases. Furthermore, in our evaluation, we realized that those databases requiring more communication among VMs to conduct operations (e.g., Cassandra and Riak) as well as the Ram-based database (e.g., Redis) can benefit more from larger VMs compared to more smaller VMs. 

For the future work, this research can be extended in several directions. As all these experiments have been conducted under default settings, it is worth determining the effect of the key setting parameters such as replica number, sharding number (if applicable) and different consistency mechanisms on the performance of NoSQL databases in a hybrid cloud. Since a hybrid cloud spanning two cloud datacenters with one source of issuing read and write operations, it is valuable to extend this architecture to be fully distributed in hardware resources hosting data and issuing operations. Last but not the least, rather than either bursting more or larger VMs in a public datacenter, it might be effective to leverage more larger VMs to improve throughput of all the distributed databases.

%\section*{Acknowledgment}
%We thank Faheem Ullah and anonymous reviewers for their valuable comments in improving the quality of the paper.

\bibliographystyle{elsarticle-num}
\bibliography{references}

\begin{thebibliography}{10}
\expandafter\ifx\csname url\endcsname\relax
  \def\url#1{\texttt{#1}}\fi
\expandafter\ifx\csname urlprefix\endcsname\relax\def\urlprefix{URL }\fi
\expandafter\ifx\csname href\endcsname\relax
  \def\href#1#2{#2} \def\path#1{#1}\fi

\bibitem{orend2010}
O.~K. (Ed.), Analysis and Classification of NoSQL Databases and Evaluation of
  their Ability to Replace an Object-relational Persistence Layer, Technische
  Universität München, 2010.

\bibitem{Chang2008}
F.~Chang, J.~Dean, S.~Ghemawat, W.~C. Hsieh, D.~A. Wallach, M.~Burrows,
  T.~Chandra, A.~Fikes, R.~E. Gruber, Bigtable: A distributed storage system
  for structured data, ACM Trans. Comput. Syst. 26~(2) (2008) 4:1--4:26 (June
  2008).

\bibitem{Leavitt2010}
N.~{Leavitt}, Will nosql databases live up to their promise?, Computer 43~(2)
  (2010) 12--14 (2010).

\bibitem{Schram2012}
A.~Schram, K.~M. Anderson, Mysql to nosql: Data modeling challenges in
  supporting scalability, in: Proceedings of the 3rd Annual Conference on
  Systems, Programming, and Applications: Software for Humanity, SPLASH ’12,
  Association for Computing Machinery, New York, NY, USA, 2012, p. 191–202
  (2012).

\bibitem{Kuznetsov2014}
S.~D. Kuznetsov, A.~V. Poskonin, Nosql data management systems, Program.
  Comput. Softw. 40~(6) (2014) 323–332 (Nov. 2014).

\bibitem{Floratou2012}
A.~Floratou, N.~Teletia, D.~J. DeWitt, J.~M. Patel, D.~Zhang, Can the elephants
  handle the nosql onslaught?, Proc. VLDB Endow. 5~(12) (2012) 1712–1723
  (Aug. 2012).

\bibitem{Parker2013}
Z.~Parker, S.~Poe, S.~V. Vrbsky, Comparing nosql mongodb to an sql db, in:
  Proceedings of the 51st ACM Southeast Conference, ACMSE ’13, Association
  for Computing Machinery, New York, NY, USA, 2013 (2013).

\bibitem{Cooper2010}
B.~F. Cooper, A.~Silberstein, E.~Tam, R.~Ramakrishnan, R.~Sears, Benchmarking
  cloud serving systems with ycsb, in: Proceedings of the 1st ACM Symposium on
  Cloud Computing, SoCC ’10, Association for Computing Machinery, New York,
  NY, USA, 2010, p. 143–154 (2010).

\bibitem{Bokhari2018}
M.~U. Bokhari, Q.~Makki, Y.~K. Tamandani, A survey on cloud computing, in:
  V.~B. Aggarwal, V.~Bhatnagar, D.~K. Mishra (Eds.), Big Data Analytics,
  Springer Singapore, Singapore, 2018, pp. 149--164 (2018).

\bibitem{Patidar2012}
S.~Patidar, D.~Rane, P.~Jain, A survey paper on cloud computing, in:
  Proceedings of the 2012 Second International Conference on Advanced Computing
  \& Communication Technologies, ACCT ’12, IEEE Computer Society, USA, 2012,
  p. 394–398 (2012).

\bibitem{Rimal2009}
B.~P. Rimal, E.~Choi, I.~Lumb, A taxonomy and survey of cloud computing
  systems, in: Proceedings of the 2009 Fifth International Joint Conference on
  INC, IMS and IDC, NCM ’09, IEEE Computer Society, USA, 2009, p. 44–51
  (2009).

\bibitem{Klein2015}
J.~Klein, I.~Gorton, N.~Ernst, P.~Donohoe, K.~Pham, C.~Matser, Performance
  evaluation of nosql databases: A case study, in: Proceedings of the 1st
  Workshop on Performance Analysis of Big Data Systems, PABS ’15, Association
  for Computing Machinery, New York, NY, USA, 2015, p. 5–10 (2015).

\bibitem{Kuhlenkamp2014}
J.~Kuhlenkamp, M.~Klems, O.~R\"{o}ss, Benchmarking scalability and elasticity
  of distributed database systems, Proc. VLDB Endow. 7~(12) (2014) 1219–1230
  (Aug. 2014).

\bibitem{Rabl2012}
T.~Rabl, S.~G\'{o}mez-Villamor, M.~Sadoghi, V.~Munt\'{e}s-Mulero, H.-A.
  Jacobsen, S.~Mankovskii, Solving big data challenges for enterprise
  application performance management, Proc. VLDB Endow. 5~(12) (2012)
  1724–1735 (Aug. 2012).

\bibitem{Li2018}
C.~Li, J.~Tang, Y.~Luo, Towards operational cost minimization for cloud
  bursting with deadline constraints in hybrid clouds, Cluster Computing 21
  (2018) 2013--2029 (12 2018).

\bibitem{Mansouri2020}
Y.~Mansouri, V.~Prokhorenko, M.~A. Babar, An automated implementation of hybrid
  cloud for performance evaluation of distributed databases, Journal of Network
  and Computer Applications (2020) 102740 (2020).

\bibitem{Wu2013}
Z.~Wu, H.~V. Madhyastha, Understanding the latency benefits of multi-cloud
  webservice deployments, SIGCOMM Comput. Commun. Rev. 43~(2) (2013) 13–20
  (Apr. 2013).

\bibitem{Lourenco2015}
J.~Lourenço, B.~Cabral, P.~Carreiro, M.~Vieira, J.~Bernardino, Choosing the
  right nosql database for the job: a quality attribute evaluation, Journal of
  Big Data 2 (2015) 18 (08 2015).

\bibitem{Li2013}
Y.~{Li}, S.~{Manoharan}, A performance comparison of sql and nosql databases,
  in: 2013 IEEE Pacific Rim Conference on Communications, Computers and Signal
  Processing (PACRIM), 2013, pp. 15--19 (Aug 2013).

\bibitem{Abramova2013}
V.~Abramova, J.~Bernardino, Nosql databases: Mongodb vs cassandra, in:
  Proceedings of the International C* Conference on Computer Science and
  Software Engineering, C3S2E ’13, Association for Computing Machinery, New
  York, NY, USA, 2013, p. 14–22 (2013).

\bibitem{Cooper2008}
B.~F. Cooper, R.~Ramakrishnan, U.~Srivastava, A.~Silberstein, P.~Bohannon,
  H.-A. Jacobsen, N.~Puz, D.~Weaver, R.~Yerneni, Pnuts: Yahoo!’s hosted data
  serving platform, Proc. VLDB Endow. 1~(2) (2008) 1277–1288 (Aug. 2008).

\bibitem{vanderVeen2012}
J.~S. {van der Veen}, B.~{van der Waaij}, R.~J. {Meijer}, Sensor data storage
  performance: Sql or nosql, physical or virtual, in: 2012 IEEE Fifth
  International Conference on Cloud Computing, 2012, pp. 431--438 (2012).

\bibitem{Bastiao2014}
L.~Basti{\~a}o, L.~Beroud, C.~Costa, J.~L. Oliveira, Medical imaging archiving:
  A comparison between several nosql solutions, IEEE-EMBS International
  Conference on Biomedical and Health Informatics (BHI) (2014) 65--68 (2014).

\bibitem{Tudorica2011}
B.~G. {Tudorica}, C.~{Bucur}, A comparison between several nosql databases with
  comments and notes, in: 2011 RoEduNet International Conference 10th Edition:
  Networking in Education and Research, 2011, pp. 1--5 (2011).

\bibitem{JingHan2011}
{Jing Han}, {Haihong E}, {Guan Le}, {Jian Du}, Survey on nosql database, in:
  2011 6th International Conference on Pervasive Computing and Applications,
  2011, pp. 363--366 (2011).

\bibitem{Davoudian2018}
A.~Davoudian, L.~Chen, M.~Liu, A survey on nosql stores, ACM Comput. Surv.
  51~(2) (Apr. 2018).

\bibitem{Hecht2011}
R.~{Hecht}, S.~{Jablonski}, Nosql evaluation: A use case oriented survey, in:
  2011 International Conference on Cloud and Service Computing, 2011, pp.
  336--341 (2011).

\bibitem{Lourenco2015a}
J.~R. {Lourenço}, V.~{Abramova}, B.~{Cabral}, J.~{Bernardino}, P.~{Carreiro},
  M.~{Vieira}, No sql in practice: A write-heavy enterprise application, in:
  2015 IEEE International Congress on Big Data, 2015, pp. 584--591 (2015).

\bibitem{Rith2014}
J.~Rith, P.~S. Lehmayr, K.~Meyer-Wegener, Speaking in tongues: Sql access to
  nosql systems, in: Proceedings of the 29th Annual ACM Symposium on Applied
  Computing, SAC ’14, Association for Computing Machinery, New York, NY, USA,
  2014, p. 855–857 (2014).

\bibitem{Toosi]2018}
A.~N. Toosi], R.~Sinnott, R.~Buyya, Resource provisioning for data-intensive
  applications with deadline constraints on hybrid clouds using aneka, Future
  Generation Computer Systems 79 (2018) 765 -- 775 (2018).

\bibitem{Tuli2020}
S.~Tuli, R.~Sandhu, R.~Buyya, Shared data-aware dynamic resource provisioning
  and task scheduling for data intensive applications on hybrid clouds using
  aneka, Future Generation Computer Systems 106 (2020) 595 -- 606 (2020).

\bibitem{Calheiros2012}
R.~N. Calheiros, C.~Vecchiola, D.~Karunamoorthy, R.~Buyya, The aneka platform
  and qos-driven resource provisioning for elastic applications on hybrid
  clouds, Future Generation Computer Systems 28~(6) (2012) 861 -- 870,
  including Special sections SS: Volunteer Computing and Desktop Grids and SS:
  Mobile Ubiquitous Computing (2012).

\bibitem{Vecchiola2012}
C.~Vecchiola, R.~N. Calheiros, D.~Karunamoorthy, R.~Buyya, Deadline-driven
  provisioning of resources for scientific applications in hybrid clouds with
  aneka, Future Generation Computer Systems 28~(1) (2012) 58 -- 65 (2012).

\bibitem{Loreti2015}
D.~{Loreti}, A.~{Ciampolini}, A hybrid cloud infrastructure for big data
  applications, in: 2015 IEEE 17th International Conference on High Performance
  Computing and Communications, 2015 IEEE 7th International Symposium on
  Cyberspace Safety and Security, and 2015 IEEE 12th International Conference
  on Embedded Software and Systems, 2015, pp. 1713--1718 (2015).

\bibitem{zhou2019}
J.~Zhou, T.~Wang, P.~Cong, P.~Lu, T.~Wei, M.~Chen, Cost and makespan-aware
  workflow scheduling in hybrid clouds, Journal of Systems Architecture 100
  (2019) 101631 (2019).

\bibitem{Vulimiri2013}
A.~Vulimiri, P.~B. Godfrey, R.~Mittal, J.~Sherry, S.~Ratnasamy, S.~Shenker, Low
  latency via redundancy, in: Proceedings of the Ninth ACM Conference on
  Emerging Networking Experiments and Technologies, CoNEXT ’13, Association
  for Computing Machinery, New York, NY, USA, 2013, p. 283–294 (2013).

\bibitem{Nishtala2013}
R.~Nishtala, H.~Fugal, S.~Grimm, M.~Kwiatkowski, H.~Lee, H.~C. Li, R.~McElroy,
  M.~Paleczny, D.~Peek, P.~Saab, D.~Stafford, T.~Tung, V.~Venkataramani,
  Scaling memcache at facebook, in: Proceedings of the 10th USENIX Conference
  on Networked Systems Design and Implementation, nsdi’13, USENIX
  Association, USA, 2013, p. 385–398 (2013).

\bibitem{Wu2015}
Z.~Wu, C.~Yu, H.~V. Madhyastha, Costlo: Cost-effective redundancy for lower
  latency variance on cloud storage services, in: Proceedings of the 12th
  USENIX Conference on Networked Systems Design and Implementation, NSDI’15,
  USENIX Association, USA, 2015, p. 543–557 (2015).

\bibitem{Dean2013}
J.~Dean, L.~A. Barroso, The tail at scale, Commun. ACM 56~(2) (2013) 74–80
  (Feb. 2013).

\bibitem{Haughian2016}
G.~Haughian, R.~Osman, W.~J. Knottenbelt, Benchmarking replication in cassandra
  and mongodb nosql datastores, in: S.~Hartmann, H.~Ma (Eds.), Database and
  Expert Systems Applications, Springer International Publishing, Cham, 2016,
  pp. 152--166 (2016).

\bibitem{Chang2006}
F.~Chang, J.~Dean, S.~Ghemawat, W.~C. Hsieh, D.~A. Wallach, M.~Burrows,
  T.~Chandra, A.~Fikes, R.~E. Gruber, Bigtable: A distributed storage system
  for structured data, in: Proceedings of the 7th USENIX Symposium on Operating
  Systems Design and Implementation - Volume 7, OSDI ’06, USENIX Association,
  USA, 2006, p.~15 (2006).

\bibitem{Sivasubramanian2012}
S.~Sivasubramanian, Amazon dynamodb: A seamlessly scalable non-relational
  database service, in: Proceedings of the 2012 ACM SIGMOD International
  Conference on Management of Data, SIGMOD ’12, Association for Computing
  Machinery, New York, NY, USA, 2012, p. 729–730 (2012).

\bibitem{Gudivada2014}
V.~N. {Gudivada}, D.~{Rao}, V.~V. {Raghavan}, Nosql systems for big data
  management, in: 2014 IEEE World Congress on Services, 2014, pp. 190--197
  (2014).

\bibitem{Dey2014}
A.~{Dey}, A.~{Fekete}, R.~{Nambiar}, U.~{Röhm}, Ycsb+t: Benchmarking web-scale
  transactional databases, in: 2014 IEEE 30th International Conference on Data
  Engineering Workshops, 2014, pp. 223--230 (2014).

\bibitem{sha2014}
P.~N. {Shankaranarayanan}, A.~{Sivakumar}, S.~{Rao}, M.~{Tawarmalani},
  Performance sensitive replication in geo-distributed cloud datastores, in:
  2014 44th Annual IEEE/IFIP International Conference on Dependable Systems and
  Networks, 2014, pp. 240--251 (2014).

\end{thebibliography}

\end{document}